\documentclass[11pt,nofootinbib]{article}
\pdfoutput=1

\usepackage{cite}
\usepackage{footmisc}
\usepackage{jheppub,fancyhdr,graphicx,epstopdf,color,amsmath,cases,slashed,hyperref}
\usepackage{makecell}
\usepackage{comment}
\usepackage{caption, subcaption}  
\usepackage{dcolumn}   
\usepackage{amsmath, amssymb}        
\usepackage{color,latexsym, array,multirow,  verbatim, enumerate,cancel}
\usepackage{slashed, soul}
\usepackage{bigcenter}
\usepackage{soul}
\usepackage{footnote}
\makesavenoteenv{tabular}
\makesavenoteenv{table}

\def\issue(#1,#2,#3){{\bf #1}, #2 (#3)}

\catcode`\@=11
\def\lsim{\mathrel{\mathpalette\@versim<}}
\def\gsim{\mathrel{\mathpalette\@versim>}}
\def\@versim#1#2{\vcenter{\offinterlineskip
\ialign{$\m@th#1\hfil##\hfil$\crcr#2\crcr\sim\crcr } }}
\catcode`\@=12

\catcode`@=12

\newcommand{\met}{$\cancel E_T$}

\newcommand{\newc}{\newcommand}
\newc{\wt}{\widetilde}
\newc{\ra}{\rightarrow}

\def\beq {\begin{equation}}
\def\eeq {\end{equation}}
\def\bi {\begin{itemize}}
\def\ei {\end{itemize}}
\def\bea {\begin{eqnarray}}
\def\eea {\end{eqnarray}}

\def \met{\slashed{E}_T}

\newcommand{\br}{\begin{eqnarray}}
\newcommand{\er}{\end{eqnarray}}
\newcommand{\be}{\begin{equation}}
\newcommand{\ee}{\end{equation}}

\newcommand{\ch}{\widetilde \chi^\pm}

\newcommand{\ifb} {\rm {fb}^{-1}}

\def \ch2p {{\wt\chi_2^+}}

\def \ch2m {{\wt\chi_2^-}}

\newc{\dmchi}{\Delta m_{\wt\chi}}

\def\issue(#1,#2,#3){{\bf #1}, #2 (#3)}

\hypersetup{
   colorlinks=true,       
   linkcolor=blue,        
   citecolor=blue,         
   filecolor=magenta,      
   }

\title{Prospects for exotic $h\rightarrow 4 \tau$ decays in single and di-Higgs boson production at the LHC and future hadron colliders}

\author[a,b]{Amit Adhikary}
\author[c]{Shankha Banerjee}
\author[d]{Rahool Kumar Barman}
\author[e]{Brian Batell}
\author[b]{Biplob Bhattacherjee}
\author[b]{Camellia Bose}
\author[f]{Zhuoni Qian}
\author[g]{Michael Spannowsky}

\affiliation[a]{Institute of Theoretical Physics, Faculty of Physics, University of Warsaw, Pasteura 5, PL 02-093, Warsaw, Poland}
\affiliation[b]{Centre for High Energy Physics, Indian Institute of Science, Bangalore 560012, India}
\affiliation[c]{CERN, Theoretical Physics Department, CH-1211 Geneva 23, Switzerland}
\affiliation[d]{Department of Physics, Oklahoma State University, Stillwater, Oklahoma, 74078, USA}
\affiliation[e]{Pittsburgh Particle Physics, Astrophysics, and Cosmology Center, Department of Physics and Astronomy, University of Pittsburgh, Pittsburgh, USA}
\affiliation[f]{Hangzhou Normal University, Hangzhou, Zhejiang 311121, China}
\affiliation[g]{Institute for Particle Physics Phenomenology, Durham University, South Road, Durham, DH1 3LE}

\emailAdd{amit.adhikary@fuw.edu.pl}
\emailAdd{shankha.banerjee@cern.ch}
\emailAdd{rahool.barman@okstate.edu}
\emailAdd{batell@pitt.edu}
\emailAdd{biplob@iisc.ac.in}
\emailAdd{camelliabose@iisc.ac.in}
\emailAdd{zhuoniqian@hznu.edu.cn}
\emailAdd{michael.spannowsky@durham.ac.uk}

\date{\today}

\abstract
{We study the prospects for observing exotic decays of the Standard Model Higgs boson $h$ into light beyond the Standard Model scalars $a$ with mass $m_{a} \lesssim m_{h}/2$ in the single Higgs and Higgs pair production channels at the high luminosity run of the Large Hadron Collider (HL-LHC). Discovery prospects for single Higgs production in the gluon-gluon fusion and vector boson fusion modes with the Higgs boson decaying via the exotic mode $h \to aa \to 4\tau$ are analyzed at the HL-LHC. The projected sensitivity for exotic Higgs decays in the nonresonant Higgs pair production channel $pp \to hh \to (h \to b\bar{b})(h \to aa \to 4\tau) \to 2b4\tau$ at the HL-LHC and a future $\sqrt{s}=100~$TeV hadron collider (FCC-hh) are also estimated. Furthermore, we study HL-LHC's potential reach for the Higgs-strahlung process in the $2b4\tau$ channel, taking into account the contamination from nonresonant Higgs pair production. 
Finally, the potential reach for resonant Higgs pair production in the $2b4\tau$ channel at the HL-LHC is also explored for several choices of $\{m_{H},m_{a}\}$.
Our studies suggest that significant improvements over existing bounds are achievable in several production channels, motivating new dedicated searches for $h \rightarrow aa \rightarrow 4 \tau$ at the HL-LHC and future colliders.
} 

\begin{document}
\maketitle

\newpage
\section{Introduction}
\label{sec:intro}

Despite bearing a high degree of consistency with the SM predictions, the current Higgs measurements still allow significant room for the Higgs boson to have non-standard decays. The ATLAS collaboration has analyzed the $\sqrt{s}=13~\mathrm{TeV}$ data collected at $\mathcal{L}=139~\mathrm{fb}^{-1}$ and has derived upper limits on the branching ratios of the Higgs boson to undetected $\sim 19\%$ and invisible particles $\sim 11\%$, at $95\%$ CL, through combined measurements of single Higgs boson production in $ggF$, $VBF$, $Vh$ and $t\bar{t}h$ modes, and decay in $h \to b\bar{b}, \gamma\gamma, ZZ^{\star}, W^{+}W^{-},\tau^{+}\tau^{-}, \mu^{+}\mu^{-}$ channels~\cite{ATLAS-CONF-2020-027}. Thus, non-standard or exotic decays of the Higgs boson are still allowed with appreciable branching rates and remain a well-motivated and exciting opportunity to probe new physics~\cite{Curtin:2013fra, Cepeda:2021rql}. 
Given the generic expectation of novel event topologies and decay kinematics in a variety of well-motivated exotic Higgs decay channels, dedicated search strategies are required to fully discern their discovery potential.

Exotic decays of the Higgs boson can be realized in various beyond the SM (BSM) frameworks. One typical example is the Higgs portal scenario where the Higgs field weakly couples to a light hidden sector 
~\cite{Silveira:1985rk,Burgess:2000yq,Draper:2010ew,Englert:2011yb, Bhattacherjee:2013jca, Robens_2015, Robens_2016, Robens_2020, Bauer_2017, Alves:2021puo}. Similarly, BSM theories with extended Higgs sectors, including supersymmetric extensions, often predict exotic Higgs decays. A widely studied example from the latter category is the Next-to-Minimal Supersymmetric SM where the Higgs sector consists of two Higgs doublets and a singlet~\cite{Ellwanger_2010, NILLES1983346,PhysRevD.39.844}. Here, the lightest (pseudo)scalar Higgs can be singlet-like with mass $\leq m_{h}/2$, leading to exotic decays of the SM-like Higgs boson. The ATLAS and CMS collaborations have explored exotic Higgs decays in single Higgs production channels \textit{viz.}, $pp \to h \to aa \to X,~ X= 4b$~\cite{ATLAS:2018pvw, ATLAS:2020ahi}, $2b 2\tau$~\cite{Sirunyan_2018}, $2b 2\mu$~\cite{ATLAS:2018emt, CMS:2018nsh, ATLAS:2021hbr}, $4\mu$~\cite{Sirunyan_2019, ATLAS:2018coo}, $2\mu 2\tau$~\cite{Sirunyan_2020, Sirunyan_2020_2, CMS:2018qvj, CMS:2017dmg, ATLAS:2015unc}, $4\gamma$~\cite{ATLAS:2015rsn} and $2\gamma 2j$ ($j$=jets)~\cite{ATLAS:2018jnf}. Here, $a$ is a new spin-0 boson that couples with the SM Higgs boson $h$ with mass $m_{a} \leq m_{h}/2$ such that it is kinematically possible for $h$ to decay via $h \to aa$. These analyses assume several extensions of the Two Higgs Doublet Model (2HDM) while maintaining SM-like Higgs production cross-sections such as Ref~\cite{CMS:2018nsh, Sirunyan_2018, CMS:2018qvj, CMS:2017dmg} by CMS and Ref~\cite{ATLAS:2018emt, ATLAS:2018pvw, ATLAS:2020ahi} by ATLAS collaboration, where Type III 2HDM+S model have been incorporated. Assuming SM production cross section, upper limits have been derived on $Br(h \to aa \to X)$ at $95\%$ CL, as presented in Table~\ref{tab:limits}. 

\begin{table}[t!]
\centering
 \begin{tabular}{|c | c | c | c |} 
 \hline
 Channel         & Mass of a, $m_a$    & \multicolumn{2}{c|}{$95\%$ CL upper limit on} \\ \cline{3-4} 
 (X)             & (GeV)    & $\sigma\times Br(h\to aa\to X)$ (fb)     & $Br(h\to aa\to X)$ \\\hline\hline
 
 $b\bar{b}b\bar{b}$~\cite{ATLAS:2018pvw, ATLAS:2020ahi}     & $[20,60]$  & $[3000,1300]$  & $-$        \\\hline

 $2b 2\tau$~\cite{Sirunyan_2018}     & $[15,60]$  & $-$     & $[0.03,0.12]$    \\\hline

 $2b 2\mu$~\cite{ATLAS:2018emt}  & $[20,60]$    & $-$     & $(1.2-8.4)\times 10^{-4}$    \\\hline

 $2b 2\mu$~\cite{CMS:2018nsh}    & $[20,62.5]$  & $-$     & $(1-7)\times 10^{-4}$    \\\hline 

 $2b 2\mu$~\cite{ATLAS:2021hbr}  & $[16,62]$  & $-$     & $(0.2-4)\times 10^{-4}$    \\\hline 
 
 $4\mu$~\cite{Sirunyan_2019, ATLAS:2018coo}   & $[0.25,8.5]$  & $[0.15,0.39]$     & $-$    \\\hline

 $4\tau$~\cite{Cai:2020lao}   & $[15,60]$  & $-$     & $[0.30,0.10]]$   \\\hline
 
 $2\mu 2\tau$~\cite{Sirunyan_2020}   & $[3.6,21]$  & $-$     & upto $1.5\times 10^{-4}$    \\\hline
 
 $4\tau/2\mu 2\tau$~\cite{Sirunyan_2020_2}   & $[4,15]~(9)$  & $-$     & $[0.23,0.16]~(0.022)$    \\\hline

 $2\mu 2\tau$~\cite{CMS:2018qvj, CMS:2017dmg, ATLAS:2015unc}   & $[15,62.5]$  & $-$     & upto $1.2\times 10^{-4}$    \\\hline
 
 $4\gamma$~\cite{ATLAS:2015rsn}   & $[10,62]$  & $[3\times 10^{-4},4\times 10^{-4}]\times \sigma_{SM}$     & $-$    \\\hline
 
$2\gamma 2j$~\cite{ATLAS:2018jnf}   & $[20,60]$  & $[3100,9000]$     & $-$    \\\hline
 
\end{tabular}
\caption{\it Model-independent upper limits at $95\%$ CL on the branching ratio of Higgs boson to pseudoscalars with a further decay into various four-particle final states. 
}
\label{tab:limits}
\end{table}

In this paper, we study the prospects for probing the exotic Higgs decay
\begin{equation}
\label{eq:h4tau}
h \rightarrow  a a \rightarrow 4 \tau \, 
\end{equation}
at the HL-LHC and a future 100 TeV hadron collider (FCC-hh) from several directions. This channel is particularly well-motivated for a couple of reasons.
From a theoretical perspective, if $a$ has preferential couplings to leptons with an interaction strength that is proportional to the lepton mass, as is often the case, then it is natural to expect $a\rightarrow \tau \bar \tau$ to dominate over other possible $a$ decay channels. 
For simplicity, we will take a phenomenological approach in this work, assuming that $a$ decays solely via $a\rightarrow \tau \bar \tau$ and the branching ratios to any other possible channels are negligible. 
Besides being theoretically motivated, the current experimental bounds on the exotic decay branching ratio  $Br(h \to aa \to 4\tau)$ are relatively weak, particularly for $m_a$ larger than about 15 GeV, allowing for the possibility of large event rates depending on the Higgs production channel under consideration. 
Let us summarize the current experimental status for this channel. 

A CMS search in the $h\to aa\to 4\tau/2\mu 2\tau$ mode has placed $95\%$ CL upper limits on $Br(h \to aa \to 4\tau)$ in the low $m_a$ range of 4 - 15 GeV~\cite{Sirunyan_2020_2}. The limits vary from $0.23$ at $m_{a}$ = 4 GeV to $0.16$ at $m_{a}$ = 15 GeV, being strongest at $m_{a}$ = 9 GeV with a limit of $0.022$.
On the other hand, for heavier exotic scalars, $Br(h \to aa \to 4\tau)$ is only weakly constrained by existing experimental analyses~\cite{CMS:2012pxt,CMS-PAS-SUS-13-010,Curtin:2013fra}. Upper limits on $Br(h \to aa \to 4\tau)$ in the $m_{a}\geq 15~$GeV regime has been derived in a recent ATLAS search which focuses on the $gg \to h \to aa \to 4\tau$ mode, considering the $4\tau$ final state with two same-sign~(SS) charged leptons and two SS $\tau$ jets~\cite{Cai:2020lao}. This search excludes $Br(h \to aa \to 4\tau)$ up to $0.3$~($0.1$) for $m_a=15~(60)~$GeV in the SS $\mu^{\pm}\mu^{\pm}$ + SS $\tau$ jets channel at $95\%$ CL.

If the exotic decay (\ref{eq:h4tau}) is present, it is natural to expect it to first be detected in the single Higgs production channels with the largest rate. Therefore, our first investigations in this work focus on the discovery prospects for $h \to aa \to 4\tau$ in the $ggF$ and $VBF$ induced single Higgs production channels at the HL-LHC.

Observing Higgs boson pair production, and in turn measuring the Higgs self-coupling and studying the scalar potential, is one of the major goals of the HL-LHC and future colliders. 
New physics can significantly impact the discovery prospects of the di-Higgs channel.

As is well-known, the rates of di-Higgs production can be enhanced in the presence of heavy resonances, as well as by new particles in the loops. 
Besides new physics which modifies di-Higgs production, it is also worthwhile to consider the impact of exotic Higgs decays in Higgs pair production. 

Such a case where one of the Higgs decays to a pair of invisible particles was first studied in Ref.~\cite{Banerjee:2016nzb} and later in Refs.~\cite{Arganda:2017wjh,Alves:2019emf}. 
Probing di-Higgs production in the SM is extremely challenging, with the projected constraints on the Higgs trilinear self-coupling at the FCC-hh being $\sim 5\%$~\cite{FCC:2018byv}. Thus additional final states could in principle be helpful in probing the di-Higgs production and the Higgs self-coupling further. With this motivation, we will study the prospects for probing non-resonant Higgs pair production at the HL-LHC and FCC-hh, with one of the Higgs bosons decaying exotically via $h \to aa \to 4\tau$ (Eq.~(\ref{eq:h4tau})) and the other decaying via $h \to b\bar{b}$, thus, culminating in the $gg \to hh \to (h \to b\bar{b})(h \to aa \to 4\tau)$ final state.

It is worth reiterating here that one generically expects to first observe such exotic decays in single Higgs production channels and only later in $hh$ production, owing to the lower rates of the latter.
At $\sqrt{s}=13~\mathrm{TeV}$ LHC, the $hh$ production cross section in the $ggF$ mode is $\sigma_{hh}^{ggF} = 31.05^{+2.2\%}_{-5.0\%}~\mathrm{fb}$~\cite{Dawson:1998py,Borowka:2016ehy,Baglio:2018lrj,deFlorian:2013jea,Shao:2013bz,deFlorian:2015moa,Grazzini:2018bsd,Baglio:2020ini} at next-to-next-to-LO (NNLO), which is roughly three orders of magnitude smaller than $ggF$ induced single Higgs production. $\sigma_{hh}^{ggF}$ improves only to $36.69_{-4.9\%}^{+2.1\%}~\mathrm{fb}$ and $1224^{+0.9\%}_{-3.2\%}~\mathrm{fb}$
~\cite{Dawson:1998py,Borowka:2016ehy,Baglio:2018lrj,deFlorian:2013jea,Shao:2013bz,deFlorian:2015moa,Grazzini:2018bsd,Baglio:2020ini} at $\sqrt{s}=14$ and $100~\mathrm{TeV}$, respectively. Besides $ggF$, $hh$ production at the LHC can also proceed via vector boson fusion ($VBF$), associated production with a vector boson ($Vhh$, $V=W^\pm/Z$), and associated production with a top-anti-top pair ($t\bar{t}hh$). However, the latter processes have comparatively smaller cross-sections, making them more challenging to probe. Despite the lower expected production rates in the di-Higgs channel, the possibility of exploiting novel final states, such as $gg \to hh \to (h \to b\bar{b})(h \to aa \to 4\tau)$, to probe the Higgs self coupling and scalar potential warrants further investigation.

It must be noted that both ATLAS and CMS collaborations have explored non-resonant Higgs pair production in numerous final states with a significant focus on the scenario where the Higgs boson decays via SM modes \textit{viz.} $4b$~\cite{Aaboud:2018knk}, $b\bar{b}\tau^{+}\tau^{-}$~\cite{Aaboud:2018sfw,Sirunyan:2017djm}, $b\bar{b}\gamma\gamma$~\cite{Aaboud:2018ftw,CMS:2017ihs}, $b\bar{b}WW^{*}$~\cite{Aaboud:2018zhh}, $WW^{*}\gamma\gamma$~\cite{Aaboud:2018ewm} and $4W$~\cite{Aaboud:2018ksn}. However, due to the non-observation of any substantial excess over the SM expectation in these channels, upper limits have been derived on the di-Higgs production cross-section ($\sigma_{hh}$) times SM branching ratio. In principle, Beyond Standard Model (BSM) physics can impact the Higgs pair production cross-section. Reference \cite{deFlorian:2227475} discusses benchmark BSM scenarios like the Higgs Singlet model and the 2HDM model containing a heavy Higgs boson that can enhance the di-Higgs production rate. Moreover, deviations in the Higgs self-coupling can modify the $hh$ production cross-section. Reference \cite{Durieux:2022hbu} studies the custodial weak quadruplet extension and the Gegenbauer's Twin model in detail that predict large Higgs self-coupling deviations. The current measurements from CMS in the combined $b\bar{b}ZZ$, $multilepton$, $b\bar{b}\gamma\gamma$, $b\bar{b}\tau\tau$, and $b\bar{b}b\bar{b}$ channels~\cite{CMS:2022dwd} and ATLAS in the combined $b\bar{b}\gamma\gamma$, $b\bar{b}\tau\tau$, and $b\bar{b}b\bar{b}$ channels~\cite{ATLAS:2022jtk} have constrained the di-Higgs signal strength ($\mu_{hh}$) within $\mu_{hh} < 3.4$ and $2.4$ at $95\%$ CL, respectively. Discovery prospects for non-resonant di-Higgs production and its potential sensitivity to probe $\lambda_{h}$ at the future hadron colliders have also been widely studied in the literature (see Refs.~\cite{Dolan:2012rv, Kim:2018cxf, Kim:2019wns, Barr:2013tda, Barger:2013jfa, Kling:2016lay, Alves:2017ued, Adhikary:2017jtu, Amacker:2020bmn, Abdughani:2020xfo, Heinrich:2019bkc, Arganda:2018ftn, Chang:2018uwu, Cao:2015oxx, Mangano:2020sao, Banerjee:2019jys, Banerjee:2018yxy, Kling:2016lay, Bizon:2018syu, Goncalves:2018qas, Barr:2014sga, Chang:2018uwu, Contino:2016spe, Park:2020yps, Adhikary:2020fqf} and references therein), and they do not exhibit much promise at the HL-LHC. As discussed previously, the major bottleneck in non-resonant di-Higgs searches at the LHC is the low production rates, rendering them weaker than the single Higgs production channels, despite bearing a richer phenomenology. However, the rate bottleneck can be alleviated in various new physics scenarios. New physics models with extended Higgs sectors, modified top quark Yukawa interaction, or heavy color-charged states as in supersymmetric or extra-dimension theories, are a few typical examples~\cite{Liu:2004pv,Dib:2005re,Pierce:2006dh,Wang:2007zx,Kanemura:2008ub,Contino:2010mh,Grober:2010yv,Sun:2012zzm,Contino:2012xk,Kribs:2012kz,Dolan:2012ac,Nishiwaki:2013cma,Ellwanger:2013ova,Cao:2013si,Dawson:2012mk,No:2013wsa,Goertz:2014qta,Liu:2014rba,Chen:2014xra,Baglio:2014nea,Hespel:2014sla,Barger:2014taa,Dawson:2015oha,Azatov:2015oxa,Lu:2015qqa,Lu:2015jza,Carvalho:2015ttv,Cao:2015oaa,Costa:2015llh,Batell:2015koa,Cao:2016zob,Kotwal:2016tex,Grober:2016wmf,Bian:2016awe,Crivellin:2016ihg,Gorbahn:2016uoy,Carvalho:2016rys,Huang:2017nnw,Nakamura:2017irk,Gao:2019uco,Huang:2019bcs,Barducci:2019xkq,Basler:2019nas,Alves:2019igs,Englert:2019eyl,Bauer:2017cov,Babu:2018uik,Basler:2018dac,Alves:2018jsw,Adhikary:2018ise,Borowka:2018pxx,Chen:2018uim,Alves:2018oct,Buchalla:2018yce,Heng:2018kyd,Kim:2018uty,Flores:2019hcf,Englert:2019xhz,DiMicco:2019ngk,Alasfar:2019pmn,Capozi:2019xsi,Li:2019uyy,Cheung:2020xij}. In new physics scenarios with an extended Higgs sector, a heavier Higgs boson $H$ which decays to two SM-like Higgs bosons $H \to hh$ can be resonantly produced $(gg \to H \to hh)$, which can potentially increase the di-Higgs production rate. Studies on the future collider prospects of resonant di-Higgs production, with both SM-like Higgs bosons decaying via SM decay channels, can be found in Refs.~\cite{Adhikary:2018ise,Kon:2018vmv,Arhrib:2018qmw,Barducci:2019xkq,DiMicco:2019ngk,Bahl:2020kwe,Barman:2020ulr,Cheung:2022bhx,Arroyo-Urena:2022oft,Bhaskar:2022ygp,Kanemura:2022ldq}. In light of the possible presence of the exotic decay mode of the SM-like Higgs boson $h \to aa \to 4\tau$, the projected sensitivity for resonant di-Higgs searches warrants a thorough evaluation. Accordingly, we perform a detailed collider study to evaluate the projected reach of resonant di-Higgs production at the HL-LHC, in the $gg \to H \to hh \to (h \to b\bar{b})(h \to aa \to 4\tau)$ channel, for various combinations of heavy and light Higgs masses, $m_{H}$ and $m_{a}$, respectively.

Another channel of interest is Higgs-Strahlung production $pp \to Zh$, a background in di-Higgs searches. The $Zh$ production mode can also lead to the $2b4\tau$ final state with $Z \to b\bar{b}$ and $h$ decaying via the exotic mode $h \to aa \to 4\tau$, thus, providing a supplementary channel in the search for exotic Higgs decays. The backgrounds of this channel include the $2b4\tau$ final state arising from non-resonant di-Higgs mode. In this work, we analyze the potential sensitivity for probing exotic Higgs decays in the $pp \to Zh \to (Z \to b\bar{b})(h \to aa \to 4\tau)$ channel at the HL-LHC and contrast the results with that from non-resonant di-Higgs searches.

The plan of the paper is as follows. In Sec.~\ref{sec:hto4tau}, we study the projected reach for single Higgs production at the HL-LHC, with the Higgs boson decaying exotically $h \to aa \to 4\tau$. Both $ggF$~(Sec.~\ref{sec:ggF}) and $VBF$~(Sec.~\ref{sec:VBF}) production modes are considered, and our results are presented as projected upper limits on the Higgs boson signal strength $\mu_{h}^{ggF~(VBF)} = \sigma_{h}^{ggF(VBF)}/\sigma_{h_{SM}}^{ggF(VBF)}$ as a function of $Br(h \to aa \to 4\tau)$ at $95\%$ CL. Sec.~\ref{sec:non-res} delves into the general kinematic features of the non-resonant di-Higgs production in the $ggF$ channel, $gg \to hh\to b\bar{b}4\tau$, and explores its prospects at the HL-LHC and FCC-hh. We further discuss the possibility of probing exotic Higgs boson decay in the Higgs-strahlung channel $pp \to Z(\to b\bar{b})h(\to aa\to 4\tau)$ in Sec.~\ref{sec:zh}. The case of resonant di-Higgs production in the $gg \to H \to hh \to (h \to b\bar{b})(h \to aa \to 4\tau)$ channel is studied in Sec.~\ref{sec:res}. Finally, we summarize in Sec.~\ref{sec:summary}.

\section{The $h\to 4\tau$ channel} 
\label{sec:hto4tau}

The $ggF$ channel is the dominant mode for single Higgs production at the LHC, with a cross-section of ${49.68}^{+8.2\%}_{-8.7\%}~\mathrm{pb}$ at NNLO+NNLL QCD and NLO EW~\cite{htwiki} at $\sqrt{s}=14~\mathrm{TeV}$. The VBF production mode is the second largest, with a cross-section of ${4.260}^{+2.2\%}_{-2.1\%}$ pb at NNLO QCD and NLO EW~\cite{htwiki}. The ATLAS and CMS collaboration probed these production modes in various final states, \textit{viz.} $ h \to b\bar{b}$~\cite{ATLAS:2020bhl,ATLAS:2021tbi}, $\gamma\gamma$~\cite{ATLAS:2022tnm,ATLAS:2022qef}, $\tau\tau$~\cite{ATLAS:2022yrq}, $ZZ^*$~\cite{ATLAS:2022net, ATLAS:2022qef}, $WW^*$~\cite{ATLAS:2022ooq, ATLAS:2021pkb} and $\mu\mu$~\cite{ATLAS:2020fzp}. As discussed previously, current Higgs measurements at the LHC still have enough uncertainties to allow exotic decays of the SM-like Higgs boson, and such exotic Higgs decays are expected to appear in single Higgs search channels sooner than in non-resonant Higgs pair production channels due to larger rates. Correspondingly, the experimental collaborations have analyzed exotic decays of the Higgs boson $pp \to h \to aa$ at the current LHC in several final states, as discussed in Sec.~\ref{sec:intro}. Extending this search to HL-LHC is pertinent, which is precisely this section's goal. We study the projected reach for single Higgs production in $ggF$ and $VBF$ modes, with the Higgs decaying via $h \to aa \to 4\tau$, at the HL-LHC. The $ggF$ channel is viable due to its large cross-section, while the $VBF$ channel offers a unique final state topology. On the other hand, the choice for the decay channel $h \to aa \to 4\tau$ is motivated by these three factors: (i) a modest signal production rate at the future LHC along with tractable background rates, (ii) a rich phenomenology offered by the multiple $\tau$s which could decay either leptonically~($\tau_{l}$) or hadronically~($\tau_{h}$) leading to a wide array of potential final states, and (iii) the absence of dedicated studies on this particular channel in the literature. We perform the analysis for several benchmark scalar masses, $m_a = 20, 30, 40, 50$ and 60~GeV, and combine both leptonic and hadronic decay modes of the $\tau$ lepton.

\subsection{$ggF$ production: $gg \to h \to aa \to 4\tau$}
\label{sec:ggF}

\begin{figure}[htb!]
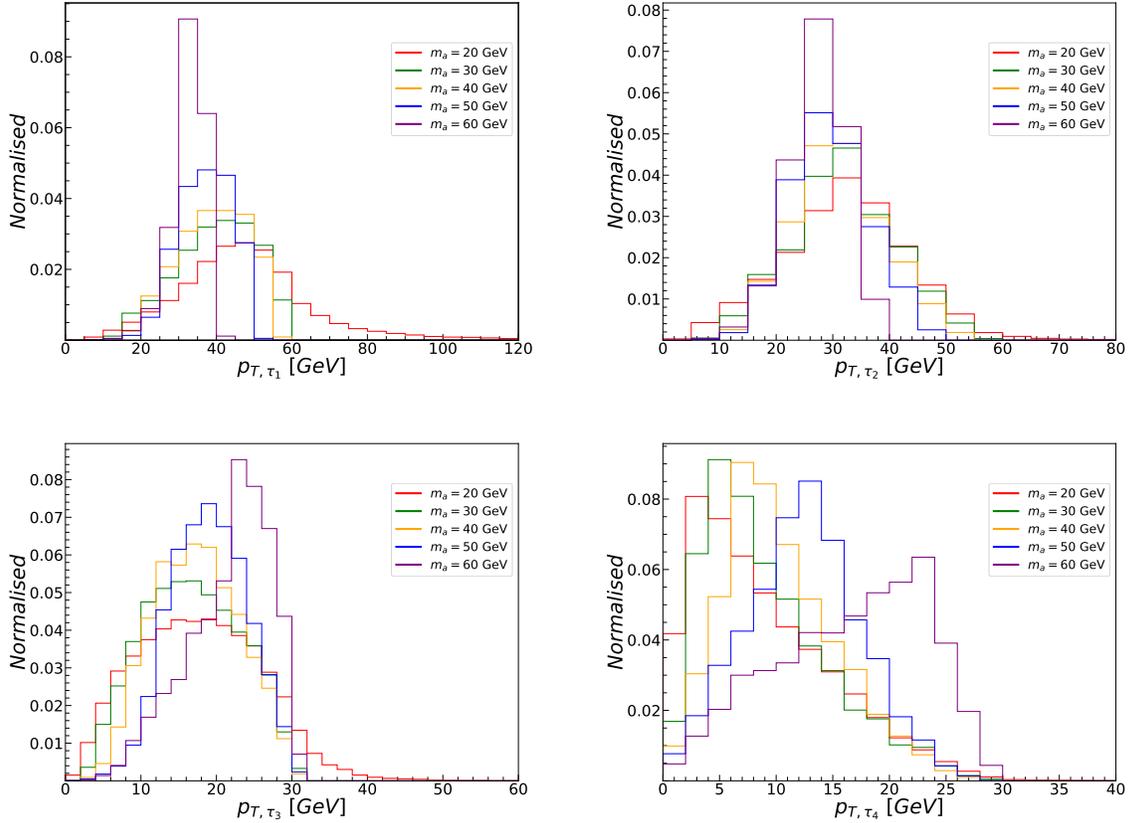

    \centering
    \includegraphics[width=0.5\textwidth]{./pt_tau1_ggF.pdf}~
    \includegraphics[width=0.5\textwidth]{./pt_tau2_ggF.pdf}\\
    \includegraphics[width=0.5\textwidth]{./pt_tau3_ggF.pdf}~
    \includegraphics[width=0.5\textwidth]{./pt_tau4_ggF.pdf}
    \caption{\it Distributions for the transverse momentum $p_T$ of the $\tau$ leptons at the truth level in the $ggF$ induced single Higgs production channel $gg \to (h \to aa \to 4\tau)$, at $\sqrt{s}=14~$TeV LHC.}
    \label{fig:pt_4tau_ggF}
\end{figure}

The kinematic features of the $\tau$ leptons in the process $gg \to h \to aa \to 4\tau$ have substantial dependence on the mass of the exotic scalar $a$. For illustrative purposes, we present the distributions for transverse momentum $p_{T}$ of the four $\tau$ leptons at the truth level in Fig.~\ref{fig:pt_4tau_ggF}. The $\tau$ leptons are $p_T$ ordered as  $\tau_i$ [$i=1-4$], 1 being the hardest. For smaller values of $m_{a}$, \textit{viz.} $m_{a} = 20~\mathrm{GeV}$, the exotic scalars $a$ are considerably boosted, which eventually translates to and largely regulates the boost associated with the $\tau$ leptons. This leads to a wide variation between the distributions for $p_{T,\tau_{1}}$ and $p_{T,\tau_{4}}$. On the other hand, in the $m_{a}=60~\mathrm{GeV}$ scenario, the $a$s are produced almost at rest in the centre of mass~(c.o.m) frame of $h$. In this case, the boost associated with the $\tau$ leptons is largely governed by the mass difference $m_{a} - m_{\tau}$. Hence, the $p_{T,\tau_{1}}$ and $p_{T,\tau_{4}}$ distributions for exotic scalars with larger mass manifest relatively close to each other. Implications from the dependence of $p_{T,\tau_{i}}~ (i = 1-4)$ distributions on $m_{a}$ would become evident in the estimation of projected limits in the latter part of this section.

The main source of backgrounds are the inclusive $4\ell$ and $h\rightarrow Z Z^* \rightarrow 4\ell$~($\ell = e,\mu,\tau$) processes. Subdominant contributions arise from QCD-QED $4\ell2\nu$, $4\ell2b$, $t\bar{t}Z$, $t\bar{t}h$, $t\bar{t}ZZ$ and $t\bar{t}WW$. Signal and background events have been generated at LO with \texttt{MadGraph5\_aMC@NLO}~\footnote{The generation level cuts are tabulated in Appendix~\ref{sec:appendixA}~(see Table~\ref{app1:1})}~\cite{Alwall:2014hca} at $\sqrt{s}=14~\mathrm{TeV}$. Showering and hadronization effects in the signal and $h\rightarrow Z Z^* \rightarrow 4\ell$ background have been simulated with \texttt{Pythia6}~\cite{Sjostrand:2001yu} along with \texttt{CTEQ6L1} Parton distribution functions (PDF) set. We use \texttt{Pythia8}~\cite{Sjostrand:2014zea} with the \texttt{NN23LO PDF} set for the rest of the backgrounds. Jets have been reconstructed using the anti-$kT$~\cite{Cacciari:2008gp} algorithm with jet reconstruction parameters $R = 0.4$, and transverse momentum $p_T > 20~\mathrm{GeV}$ within the  {\tt FastJet}~\cite{Cacciari:2011ma} framework. Detector response has been simulated with \texttt{Delphes-3.4.1}~\cite{deFavereau:2013fsa} using the detector card for ATLAS with the following modifications: $b$-tagging efficiency, as well as the efficiency of a light jet or $c$ jet being mistagged as a $b$-tagged jet, are defined as functions of the jet $p_{T}$ considering the medium working point~(see Fig.~17 in Ref.~\cite{Sirunyan:2017ezt}. For example, the b-tagging efficiency is about $62\%$ for $p_T\sim 20$ GeV, while the $c$ (light) jet mistagging efficiency is 12$\%$ (2$\%$). Considering $p_T\sim 50$ GeV, the b-tag efficiency and $c$, light jet mistag efficiency is about $67\%$ and $12\%$, $0.8\%$, respectively.
$\tau$-tagging efficiencies for the 1-prong and 3-prong $\tau$-tagged jet ($\tau_h$) are fixed at $55\%$ and $50\%$, respectively, and $j \to \tau_h$ fake rate is set to $0.35\%$ (see Fig.3 in \cite{CMS-PAS-TAU-16-002}). 

The presence of exotic decay channels also modifies the total decay width of $h$~($\Gamma_{h}$) with respect to its SM value. In the present scenario, the modified total decay width of $h$ is given by $\Gamma_{h} = \Gamma_{h_{SM}} + \Gamma_{h \to aa \to 4\tau}$. Correspondingly, the SM branching ratios of $h$ are scaled by the factor, $\sim \left(1 - Br(h \to aa \to 4\tau)\right)$.

Events with exactly four $\tau$ objects and zero $b$-tagged jets are selected. We consider both leptonic and hadronic decay modes of the $\tau$s and include these $\tau$ decay scenarios in the present analysis: (a) four $\tau$-tagged jets ($4\tau_{h}$), (b) three $\tau_{h}$ and one $\ell$~($\ell=e^{\pm},\mu^{\pm}$), (c) two $\tau_{h}$ and two $\ell$, (d) one $\tau_{h}$ and three $\ell$~\footnote{Unless otherwise specified, $\tau_{h}$ and $\ell$ will be collectively referred to as $\tau$ in the rest of the analysis. Here, we do not include the four $\ell$ final states as the $t\bar{t}+X$ backgrounds become very large with minor improvement in signal efficiency.}. We want to mention that fake backgrounds can arise from a light jet faked as a hadronic $\tau$. However, their contribution is suppressed because the mistagging efficiency is very low, and we demand four $\tau$ objects in the final state. We do not include these backgrounds in our analysis. The $\tau_{h}$ and $\ell$ must satisfy $p_{T}> 20$ GeV and $>10$ GeV, respectively. The minimum distance in the $\eta-\phi$ plane between the $\tau$ objects is as follows, $\Delta R(\tau_h,\tau_h)>0.4$, $\Delta R(\tau_h,\ell)>0.4$ and $\Delta R(\ell,\ell)>0.1$~\footnote{We choose the separation in the $\eta-\phi$ plane between two leptons to be $>0.1$ because the leptons coming from boosted $\tau$s are highly collimated and demanding a greater $\Delta R$ separation decreases the signal efficiency.}. The $\tau$ objects must be within $|\eta| <3.0$. The veto on $b$ jets is applied to reduce backgrounds like QCD-QED $4l2b$, $t\bar{t}Z$, $t\bar{t}h$, $t\bar{t}ZZ$ and $t\bar{t}WW$. Furthermore, events are required to satisfy the generation level cuts described in Table~\ref{app1:1}.

Several kinematic observables are reconstructed in order to discriminate the signal from the background through a multivariate analysis, 
\begin{equation}
\begin{split}
 \Delta R_{\tau_i \tau_j} (i,j = 1-4; i \neq j),~\Delta R_{\tau\tau}^{min},~\Delta R_{\tau\tau}^{max},
 ~p_{T,4\tau},~m_{4\tau}^{vis},~m_{T,h},~\met.
\label{eqn:var_hto4tau_ggf}
\end{split}
\end{equation}
where, $\Delta R_{\tau\tau}^{min~(max)}$ represents the minimum (maximum) separation between any pair of $\tau$ objects, $p_{T,4\tau}$ and $m_{4\tau}^{vis}$ are the visible transverse momentum and invariant mass of the Higgs boson $h$ respectively, $m_{T,h}$ is the transverse mass of $h$, defined as $m_{T,h}^2=(\sum_i E_{T,i})^2 - (\sum_i \vec{p}_{T,i})^2$ where $i$ runs over the visible $\tau$ decay products and $\met$, the missing transverse momentum. The training observables considered in Eq.~\ref{eqn:var_hto4tau_ggf} are chosen due to their sensitivity to the mass of the exotic scalar, which dictates the final state kinematics. The multivariate analysis is performed using the eXtreme Gradient Boosted, or XGBoost \cite{Chen:2016:XST:2939672.2939785} technique, which is a decision tree-based machine learning algorithm. The multi-class classification algorithm is adopted through the \texttt{multi:softprob} objective function and three network hyper-parameters are optimized, \texttt{max depth}, \texttt{$\eta$} and \texttt{$\lambda$}. Here, \texttt{max depth} is the maximum depth of a tree that XGBoost constructs, \texttt{$\eta$} is the learning rate, and \texttt{$\lambda$} is the L2 regularization applied to weights. Furthermore, the signal and background events in the training dataset are weighted according to their relative cross-sections. The output from the trained network is a set of probability scores corresponding to each class. In other words, the model predicts, for every event, the probability of the event belonging to the signal and the different background classes. 

\begin{figure}[htb!]
    \centering
    \includegraphics[width=0.48\textwidth]{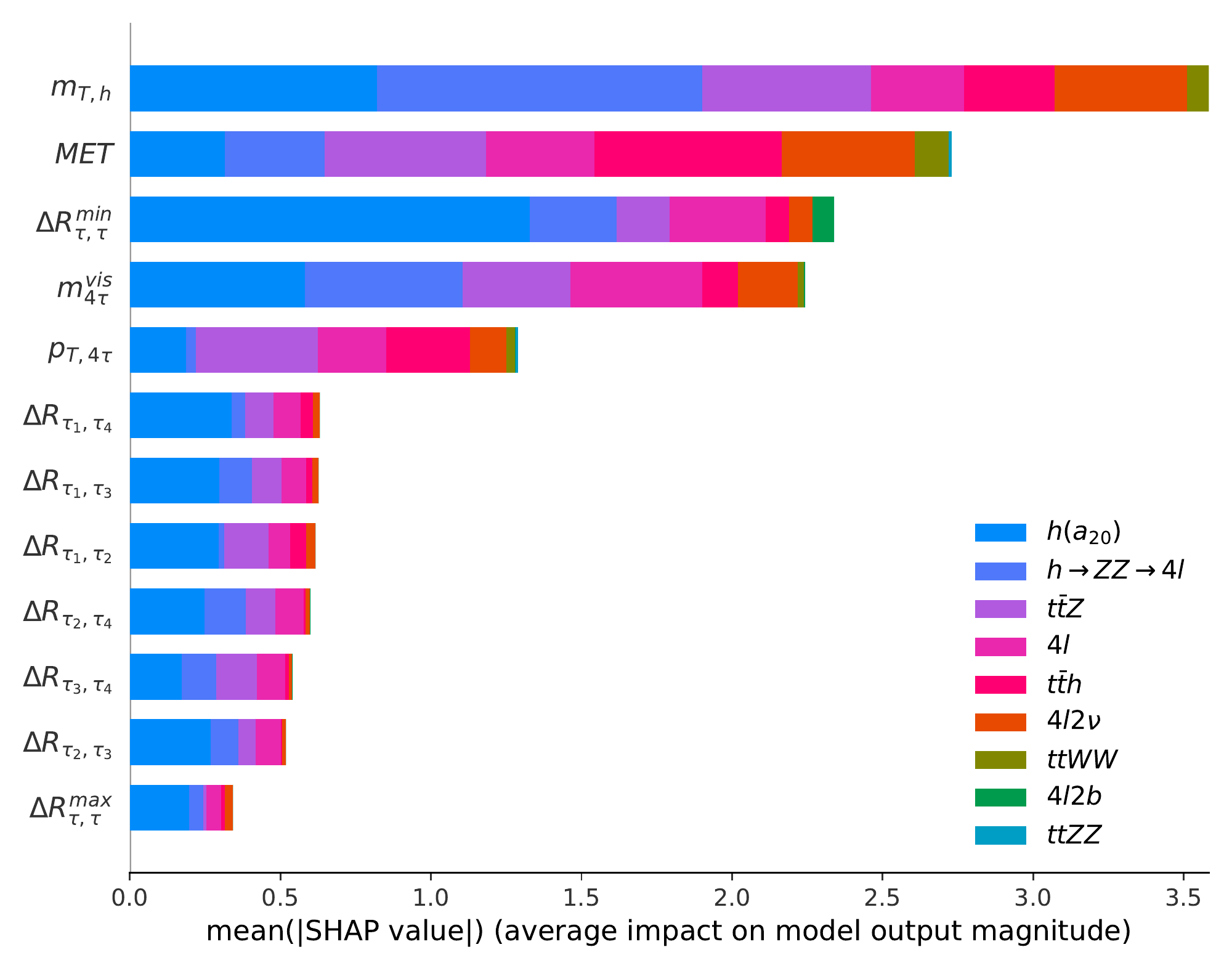}~
    \includegraphics[width=0.48\textwidth]{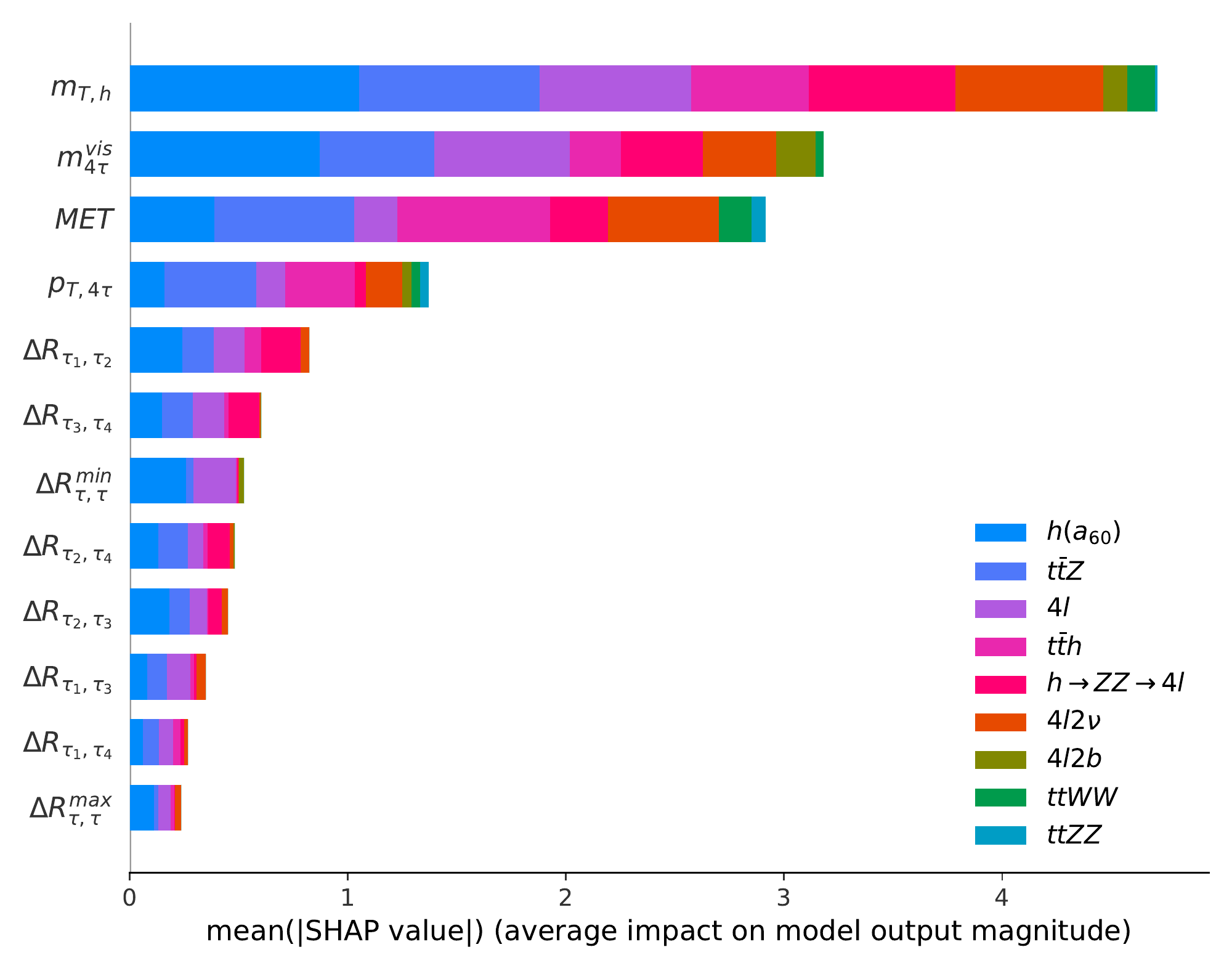}
    \caption{\it Mean of the absolute SHAP values for the kinematic variables used to perform the XGBoost analysis for $m_a$ = 20 GeV (left) and 60 GeV (right) in the $gg \to h\to aa\to 4\tau$ channel, at $\sqrt{s}=14~\rm{TeV}$ LHC with $\mathcal{L}=3~\rm{ab}^{-1}$. A higher absolute SHAP value indicates a higher rank.}
    \label{fig:shap_ggF}
\end{figure}

We rank the kinematic variables using SHAP (SHapley Additive exPlanations) \cite{Lundberg17, Lundberg18} based on how well they performed in the XGBoost analysis. SHAP is an individualized feature attribution method that uses the concept of Shapley values \cite{Shapley+2016+307+318} to determine how much each feature contributed to the model's output. SHAP has emerged as a popular tool for interpreting machine learning results in collider studies \cite{Cornell_2022, Alvestad:2021sje, Grojean:2020ech, CMS:2021qzz, Grojean:2022mef}. The classification result for each event is equal to the total SHAP values of all the features in that specific event. We obtain a mean of the events' absolute SHAP values by averaging over all of the events. The influence of a variable in categorizing an event as a signal or one of the backgrounds increases with the SHAP value. The mean SHAP values of all the kinematic variables for exotic scalar of mass 20 GeV (left) and 60 GeV (right), along with backgrounds, are summarized in  Fig.~\ref{fig:shap_ggF}. The variables, in this case, are ranked according to their feature importance. Signals and backgrounds are labelled with a class in multi-class classification. The different colors next to each feature on the y-axis correspond to a class of signal or one of the backgrounds. The length of each colored bar for a given feature shows the contribution of that feature to classifying events into that class. It should be noted that $\Delta R^{min}_{\tau\tau}$ has a more significant impact on the signal class for the 20 GeV scalar than it does for the 60 GeV case. This is because when the scalars have a mass of 20 GeV, the enhanced boost causes the $\tau$ objects to become collimated, reducing the minimum $\Delta R$ between $\tau$s and distinguishing the signal more from all other backgrounds. As the mass increases, the $\tau$s become more dispersed, and the signal loses the advantage of having a low $R^{min}_{\tau\tau}$ in comparison to the backgrounds.

\begin{figure}[htb!]
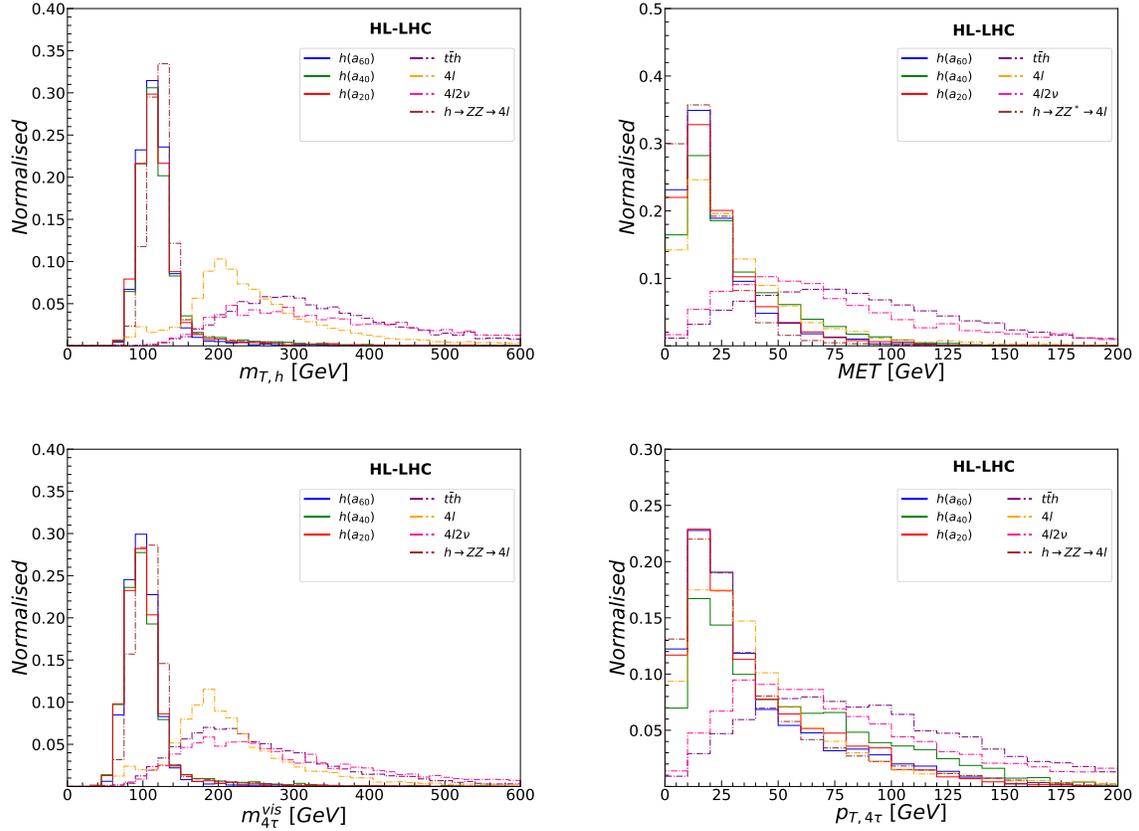

\centering
\includegraphics[width=0.5\textwidth]{./MTh_14_ggF.pdf}~
\includegraphics[width=0.5\textwidth]{./ptmis_14_ggF.pdf}\\
\includegraphics[width=0.5\textwidth]{./mvis4tau_14_ggF.pdf}~
\includegraphics[width=0.5\textwidth]{./pt4tau_14_ggF.pdf}
\caption{\it  Distributions for the transverse mass of the reconstructed Higgs boson $m_{T,h}$, missing transverse momentum $\met$, invariant mass and transverse momentum for the visible $4\tau$ system, $m_{4\tau}^{vis}$ and $p_{T,4\tau}$, respectively, for signal benchmarks corresponding to $m_{a}=20,~40,~{60~\rm GeV}$, and relevant backgrounds, in the $gg \to h \to aa \to 4\tau$ channel at $\sqrt{s}=14~${TeV} LHC with $\mathcal{L}=3~\rm{ab}^{-1}$.}
\label{fig:h_ggf_imp_obs}
\end{figure}

The training observables with the highest feature importance scores are $m_{T,h}$, $m_{4\tau}^{vis}$, $\met$ and  $p_{T,4\tau}$. In Fig.~\ref{fig:h_ggf_imp_obs}, we show the distributions of these observables at the detector level for signal benchmarks corresponding to $m_{a} = 20, 40, 60~$GeV, and the relevant backgrounds. 
The signal efficiency and background yields $B$ at the HL-LHC from the XGBoost analysis are shown in Table~\ref{tab:ggF_h_XGB}. We also compute the signal yields $S$ and signal significance at the HL-LHC, assuming $Br(h \to aa \to 4\tau) = 0.1\%$. Here, signal significance is defined as $\mathcal{S}=S/\sqrt{S+B}$. The signal efficiency is almost similar for $m_{a} = 20, 30$ and $40~$GeV. In Fig.~\ref{fig:pt_4tau_ggF}, we observe that the distributions for $p_{T,\tau_{4}}$ for smaller $m_a$ at the truth level peaks at $p_{T} \lesssim 10~$GeV. However, we apply stronger selection cuts on the visible decay products of these $\tau$ leptons at the detector level, $p_{T,\ell} > 10~$GeV and $p_{T,\tau_{h}} > 20~$GeV, which selects events only from the tail of the $p_{T,\tau_{4}}$ distributions in the $m_a = 20, 30 $ and 40~GeV scenario, thereby, leading to lower signal efficiencies. On the other hand, the background yield increases with $m_{a}$. Overall, the significance falls from 6.0 at $m_{a} = 20~$GeV to 5.2 at $m_{a} = 40~$GeV. For higher values of $m_{a}$ $viz$ $m_a = 50$ and 60~GeV, the signal efficiency registers an improvement, leading to higher signal significance. The dominant contributions to the background yield arise from the $4\ell$ and $h\rightarrow Z Z^* \rightarrow 4\ell$~($\ell = e,\mu,\tau$) processes, as discussed previously. For example, in the $m_a = 60~$ GeV scenario, the total background yield is 237, out of which 54$\%$  are inclusive $4\ell$ events, 44$\%$ are $h\rightarrow Z Z^* \rightarrow 4\ell$ events and the other sub-dominant backgrounds comprise the remaining 2$\%$. In Table~\ref{tab:ggF_h_XGB}, we also show the signal significance given a systematic uncertainty of $\sigma_{unc} = 5\%$, using $\mathcal{S}=S/\sqrt{S+B+((S+B)\times \sigma_{unc})^2}$. Including systematic uncertainties barely reduces the significance due to relatively large $S/B \sim \mathcal{O}(1)$. It lowers only by $3-10\%$ for $\sigma_{unc} = 5\%$.  

\begin{table}[htbp!]
\centering
\scalebox{0.9}{%
\begin{tabular}{|c|c|c|c|c|c|}\hline
$\sqrt{s}$ & $m_a$ & Total Background & Signal Efficiency  & Signal & Significance \\

(TeV)      & (GeV) & Yield, B  & ($\times 10^{-4}$) & Yield, S  & (5$\%$ sys.)\\\hline

\multirow{5}{*}{14} & 20 & $50$ & $4.3$ & $64$ & $6.0 (5.8)$ \\ \cline{2-6}

 & 30 & $65$ & $4.1$ & $61$  & $5.4 (5.2)$  \\ \cline{2-6}
 
 & 40 & $87$ & $4.2$ & $63$  & $5.2 (4.9)$  \\ \cline{2-6}
 
 & 50 & $137$ & $6$ & $87$  & $5.8 (5.3)$  \\ \cline{2-6}

 & 60 & $237$  & $15$ & $221$ & $10.3 (9.0)$ \\ \hline
 
\end{tabular}}
\caption{\it  Signal efficiency and background yields in the $gg \to h \to aa \to 4\tau$ channel at the HL-LHC from the XGBoost analysis. Signal yields and signal significance at the HL-LHC are also shown, under the assumption $Br(h \to aa \to 4\tau) = 0.1\%$. Signal significance for $5\%$ systematic uncertainty are shown in parenthesis.}
\label{tab:ggF_h_XGB}
\end{table}

For comparison, we use two more models as multivariate methods. First, we perform a multivariate analysis using the Boosted Decision Tree Decorrelated (BDTD) algorithm in the TMVA framework. We optimize the following parameters while training the signal and background events, \textit{viz.} number and length of decision trees, \texttt{NTrees} and \texttt{MaxDepth}, respectively; minimum number of events (in $\%$) in each leaf, \texttt{MinNodeSize} and \texttt{NCuts}. We utilize \texttt{Adaptive Boost} for boosting the weak classifiers. We maintain a stable Kolmogorov-Smirnov (KS) test score of $> 0.1$ to avoid overtraining signal and background samples. The signal, background yields, and signal significance after the BDTD analysis are tabulated in Table~\ref{tab:ggF_h_BDT}. The signal efficiency is poor compared to XGBoost results. For a scalar of mass, $m_a$ = 20 (60) GeV, the XGBoost yields a signal significance of 6.0 standard deviations (s.d.) (10.3 s.d.). In contrast, BDTD results in 3.6 s.d. (2.9 s.d.), a decrease by 40 (70) $\%$.

\begin{table}[htbp!]
\centering
\scalebox{0.9}{%
\begin{tabular}{|c|c|c|c|c|c|}\hline
$\sqrt{s}$ & $m_a$ & Total Background & Signal Efficiency  & Signal & Significance \\

(TeV)      & (GeV) & Yield, B  & ($\times 10^{-4}$) & Yield, S  & (5$\%$ sys.)\\\hline

\multirow{5}{*}{14} & 20 & $58$ & $2.4$ & $35$ & $3.6~(3.5)$ \\ \cline{2-6}

 & 30 & $95$ & $1.8$ & $27$  & $2.5 (2.2)$  \\ \cline{2-6}
 
 & 40 & $105$ & $2.4$ & $36$  & $3.0 (2.8)$  \\ \cline{2-6}
 
 & 50 & $137$ & $2.3$ & $34$  & $2.6 (2.3)$  \\ \cline{2-6}

 & 60 & $48$  & $1.7$ & $25$ & $2.9 (2.8)$ \\ \hline
 
\end{tabular}}
\caption{\it  Signal efficiency and background yields in the $gg \to h \to aa \to 4\tau$ channel at the HL-LHC from the BDTD-TMVA analysis. Signal yields and signal significance at the HL-LHC are also shown, under the assumption $Br(h \to aa \to 4\tau) = 0.1\%$. Signal significance for $5\%$ systematic uncertainty are shown in parenthesis.}
\label{tab:ggF_h_BDT}
\end{table}

Next, we use a Deep Neural Network (DNN) model. The network consists of 1 input layer, 3 hidden layers with 16, 32, and 16 nodes, respectively, and one output layer with 9 nodes for 1 signal and 8 backgrounds. Each layer except the output layer uses the \texttt{ReLu} activation function. The final layer uses the \texttt{Softmax} function to output probabilities to each class label. The model minimizes categorical cross-entropy loss with a learning rate of 0.1, and the \texttt{Adam} optimizer. The model stops learning when the performance does not improve after 5 epochs to avoid overfitting. 

Using the DNN, the results are listed in Table \ref{tab:ggF_h_DNN}. There is a 5-6$\%$ improvement in performance using DNN. Since XGBoost and DNN have similar classification performances, we continue using XGBoost as our model for the rest of the analysis. 

\begin{table}[htbp!]
\centering
\scalebox{0.9}{%
\begin{tabular}{|c|c|c|c|c|c|}\hline
$\sqrt{s}$ & $m_a$ & Total Background & Signal Efficiency  & Signal & Significance \\

(TeV)      & (GeV) & Yield, B  & ($\times 10^{-4}$) & Yield, S  & (5$\%$ sys.)\\\hline

\multirow{5}{*}{14} & 20 & $56$ & $4.5$ & $66$ & $6.0 (5.8)$ \\ \cline{2-6}

 & 30 & $66$ & $4.4$ & $66$  & $5.7 (5.5)$  \\ \cline{2-6}
 
 & 40 & $80$ & $4.4$ & $67$  & $5.4 (5.1)$  \\ \cline{2-6}
 
 & 50 & $150$ & $6.4$ & $96$  & $6.1 (5.5)$  \\ \cline{2-6}

 & 60 & $210$  & $15$ & $228$ & $10.9 (9.8)$ \\ \hline
 
\end{tabular}}
\caption{\it  Signal efficiency and background yields in the $gg \to h \to aa \to 4\tau$ channel at the HL-LHC from the DNN analysis. Signal yields and signal significance at the HL-LHC are also shown, under the assumption $Br(h \to aa \to 4\tau) = 0.1\%$. Signal significance for $5\%$ systematic uncertainty are shown in parenthesis.}
\label{tab:ggF_h_DNN}
\end{table}

\begin{figure}[htb!]
\centering
\includegraphics[width=0.5\textwidth]{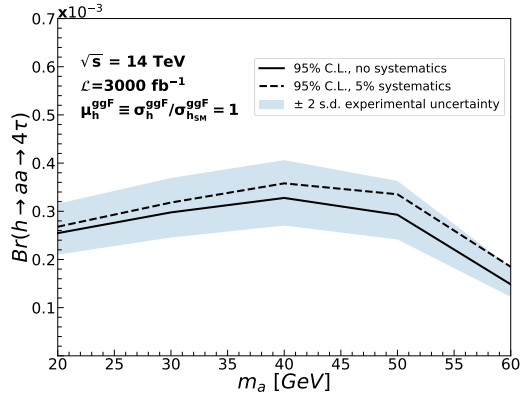}
\caption{\it Upper limit projection for $Br(h\to aa\to 4\tau)$ at 95$\%$ C.L., as a function of exotic scalar mass when $\mu_{h}^{ggF}$ is unity at $\sqrt{s}=14$ TeV. The blue band represents the variation in $Br(h\to aa\to 4\tau)$ within 2 s.d. interval of $\mu_{h}^{ggF}$ as measured by CMS \cite{CMS:2022dwd} and ATLAS \cite{ATLAS:2022vkf} collaborations. The solid and dashed lines refer to adding zero and $5\%$ systematic uncertainty, respectively.
}
\label{fig:muh_14}
\end{figure}

The measured Higgs signal strength, denoted as $\mu^{ggF}_{h} \equiv \sigma^{ggF}_{h}/\sigma^{ggF}_{h_{SM}}$, is consistent with the SM predictions~\cite{CMS:2022dwd, ATLAS:2022vkf}. The observed $\sigma^{ggF}_{h}$ is constrained within $\sim$ 20$\%$ of $\sigma^{ggF}_{h_{SM}}$, at 2 s.d. uncertainty \footnote{The errors on the signal strength measurement by CMS and ATLAS have been combined in quadrature to obtain the approximate error on $\mu^{ggF}_{h}$.}. In Fig.~\ref{fig:muh_14}, we present the upper limit projections for $Br(h \to aa \to 4\tau)$ as a function of $m_a$ for the SM scenario $\sigma^{ggF}_{h} = \sigma^{ggF}_{h_{SM}}$, or $\mu^{ggF}_{h} = 1$. We observe that the HL-LHC would be able to probe exotic Higgs decays up to $Br(h \to aa \to 4\tau) \sim 0.025\%$~($0.015\%$) for $m_{a} = 20~$(60) GeV assuming SM production rates for $h$. With a 5$\%$ systematic uncertainty, the upper limits on $Br(h \to aa \to 4\tau)$ become $\sim 0.027\%$~($0.018\%$) for $m_{a} = 20~$(60) GeV. The blue band illustrates the variation in the upper limit within a 2 s.d. interval of the current signal strength measurements obtained by the CMS \cite{CMS:2022dwd} and ATLAS \cite{ATLAS:2022vkf} collaborations.

As mentioned earlier, our present analysis does not consider backgrounds where a light jet might fake as a $\tau_h$. Ref.~\cite{Cai:2020lao} analyzed the SS $\mu^{\pm}\mu^{\pm} +$~SS $\tau$ jets channel considering the dominant background from the fake $j\to\tau_h$ events while using a medium working point for $\tau$ identification. It must be noted that in prior studies~(for example, see Fig. 9 in \cite{ATL-PHYS-PUB-2015-045}),
the medium working point with one-prong $\tau$ tagging efficiency of 55$\%$ corresponds to 1$\%$ $j\to\tau_h$ fake rate. In our analysis, we use $\tau$ tagging efficiency of an MVA-based tagger \cite{CMS-PAS-TAU-16-002}, with 0.35$\%$ $j\to \tau_h$ fake rate for 55$\%$ one-prong $\tau$-tagging efficiency. With an improved machine learning based $\tau$ tagger~\cite{Tumasyan_2022}, the fake rate is expected to further reduce in the future. However, the backgrounds from $j\to\tau_h$ fakes might still play an important role in contaminating our signal, which can be estimated by choosing a tighter working point for $\tau$ identification. For example, choosing a $\tau$-tagging efficiency of 30$\%$ for $m_a=$ 60 GeV and assuming that the total background yield doubles upon considering the contribution of additional backgrounds from $j\to \tau_h$ fakes, we obtain an upper limit on $Br(h \to aa \to 4\tau) \sim  0.06\%$ (0.07$\%$), assuming $0\%$ (5$\%$) systematic uncertainty. This limit is roughly 4 times weaker than the upper limit presented in Fig.~\ref{fig:muh_14}.

Our current analysis methodology yields subpar signal efficiency for scalars with  $m_a < $ 20 GeV. In this scenario, the decay products of $\tau$s are collimated and hard to isolate. The requirement of four $\tau$ objects at the detector level becomes too strict to achieve decent signal efficiency. For instance, even with a large $ggF$ production cross-section at HL-LHC, we expect to observe only five events for $\mathcal{L}=3~\rm{ab}^{-1}$, for an exotic scalar of mass $m_a$ = 10 GeV, leading to a signal significance of only 1.33. The upper limit on the branching ratio for the exotic Higgs decay considering the SM case is almost an order of magnitude weaker than the $m_{a}\geq20~$GeV scenarios and comes about $Br(h \to aa \to 4\tau) \sim 0.168\%$. Due to the weaker sensitivity, we do not consider very low-mass exotic scalars in the present analysis. However, it must be noted that the projection above for $m_{a} = 10~$GeV are stronger than the current limits on $Br(h \to aa \to 4\tau) \sim 3\%$~\cite{Sirunyan_2020_2} by roughly an order of magnitude. 

The analysis done in this paper utilizes no kinematic observables which depend on the exotic scalar mass, which is a free parameter. However, we have performed an alternate analysis by reconstructing the exotic scalars from the final state $\tau$s. The analysis depends on the choice of $m_a$, and we see that the results are comparable. At the detector level, the presence of multiple missing particles~($\nu$) produced from $\tau$ decays makes it challenging to reconstruct $a$. To circumvent this issue, we adopt the Collinear Mass Approximation~(CMA) technique~\cite{Elagin:2010aw}. This technique is based on two assumptions: visible and invisible components from $\tau$ decays are nearly collinear~($\theta_{vis} \sim \theta_{\nu}$ and $\phi_{vis} \sim \phi_{\nu}$), and neutrinos are the only source of missing energy. In the present study, the validity of the first assumption is determined by the difference between $m_{a}$ and $m_{\tau}$, and the boost carried by $m_{a}$. Low mass exotic scalars, $m_{a}\sim 20$~GeV, are considerably boosted due to the large mass gap with $m_{h}$ while heavier values of $m_{a}$~($\sim 60$~GeV) are produced almost at rest in the rest frame of $h$. However, as discussed previously, the $\tau$ lepton is boosted in the latter case due to the relatively large mass gap between $a$ and $\tau$. These conditions validate the first approximation of the CMA technique. The second approximation naturally holds since the only missing energy source is the neutrinos produced from the $\tau$ decays. One drawback of this technique is its high sensitivity to $\met$ resolution, leading to overestimating the reconstructed Higgs mass with long tails in the distribution.

Since the final state decay products of exotically decaying Higgs, $\tau$s, are identical at the detector level, the goal is to choose the right pair of $\tau$s to reconstruct the two exotic scalars. We first check this at the parton level where the full kinematics is known. The four visible $\tau$ decay objects can be grouped into two pairs corresponding to the two equal mass exotic scalars, $a_{1}$ and $a_{2}$, in three independent ways, with one of them being the correct $\tau$ pair. First, we choose the following variables to see their potential in having similar features across most of the events for the correct pairing, $\Delta R_{\tau\tau}^1$, $\Delta R_{\tau\tau}^2$, $m_{4\tau}^{col}$, ($m_{\tau\tau}^{a1_{col}}$/$m_{\tau\tau}^{a2_{col}}$), ($m_{\tau\tau}^{a1_{col}}$ - $m_{\tau\tau}^{a2_{col}}$) and $\chi^2_{min}$. Here, $\Delta R_{\tau\tau}^1$ and $\Delta R_{\tau\tau}^2$ are $\Delta R$ separations between $\tau$s in a pair which are used to construct light scalars. The $m_{\tau\tau}^{a1_{col}/a2_{col}}$ and $m_{4\tau}^{col}$ are the collinear mass of $a_{1}/a_{2}$ and exotically decaying 125 GeV Higgs boson, respectively, as defined in the previous paragraph. In case of $\chi^2_{min}$, we choose the combination of $\tau$s that minimises the function

\begin{eqnarray}
\chi^2_{min} &=&  \left [ 
\frac{\left ( (m_{\tau\tau}^{a1_{col}})^2- m^2_a \right )^2}{\sigma_{a1}^4} \,   +
\frac{\left ( (m_{\tau\tau}^{a2_{col}})^2- m^2_a \right )^2}{\sigma_{a2}^4}  \, \right ], 
\label{eqn:chi2_nonres}
\end{eqnarray}

where, $\sigma_{a1/a2} = 0.1\times m_{\tau\tau}^{a1_{col}/a2_{col}}$~\cite{ATLAS:2018rnh}. Among these six variables, over most events, the $\chi^2_{min}$ performs better with a minimum value for the correct $\tau$ pair. So, we choose the $\chi^2_{min}$ to reconstruct the exotic scalars in our analysis. However, this method has some additional complications,

\begin{itemize}
\item For a selected pair of $\tau$s, they must contain oppositely charged $\tau$s to reconstruct the neutral exotic scalars. This information is absent in the $\chi^2$ minimisation procedure. While this can be easily implemented for a leptonically decaying $\tau$, there is an ambiguity in defining the charge of a hadronic $\tau$ or $\tau$-jet. We explicitly checked this in Delphes, and a significant number of events contain $\tau$-jets with charges other than $\pm 1$. In those cases, assigning $\tau$ pairs to an exotic scalar and constructing variables using them might be misleading.
\item We are using the mass of the exotic scalar while calculating $\chi^2_{min}$. But we do not have that knowledge in actual experiments. Hence, we choose to continue with our methodology, which is not dependent on exotic scalar mass information.
\end{itemize}

In continuation of the CMA technique, we do not remove those events having different charges in a $\tau$ pair. Instead, we choose the $\tau$ pairs from the second $\chi^2$ minimum. This helps in improving signal efficiency. The $\tau\tau$ pairing derived from Eq.~\ref{eqn:chi2_nonres} is also used to reconstruct the visible invariant mass of the exotic scalars, $m_{\tau\tau}^{a1_{vis}}$ and $m_{\tau\tau}^{a2_{vis}}$, and we include them in XGBoost analysis. After performing the analysis, we found no significant improvement in the final results using these mass-dependent variables. For scalars of mass $m_a = 20~$ GeV, this methodology gives a signal significance of 6. As a result, we do not employ this method further in this study, but we acknowledge that it has room for improvement.

\subsection{$VBF$ production: $pp \to (h \to aa \to 4\tau)jj$}
\label{sec:VBF}

Having discussed the HL-LHC prospects for $ggF$ induced single Higgs production channel, we now focus on single Higgs production in the $VBF$ mode $pp \to hjj \to (h \to aa \to 4\tau)jj$. The dominant background stems from the QCD-QED $4\ell 2j$ process. Subdominant background contributions can arise from QCD-QED $4\ell 2b$, $t\bar{t}Z,~t\bar{t}h,~t\bar{t}ZZ$ and $t\bar{t}WW$ processes. We select events containing exactly four $\tau$ objects with at least one $\tau$-tagged jet and at least two light jets in the final state. Fig. \ref{fig:pt_4tau_vbf} in
Appendix B shows the $p_{T,\tau}$ distributions at the parton-level for the five signal benchmarks. The choice for $\tau$ objects is similar to that in Sec.~\ref{sec:ggF}. The trigger cuts for the final state objects are, 
\begin{equation}
    p_{T,\tau_{\ell}} > ~10~{\rm{GeV}}, ~p_{T,\tau_{h}} > 20~{\rm{GeV}},
    ~p_{T, j_1/j_2} > ~30~{\rm{GeV}},~|\eta_{\tau_{\ell}/\tau_{h}}| < 3.0,|\eta_{\tau_{j_1/j_2}}| < 4.0,
\end{equation}
where $j_{1}$ and $j_{2}$ are the hardest-$p_{T}$ light jets in the final state. The $VBF$ Higgs production channel leads to a unique topology with the $VBF$ jets produced back to back in the forward regions of the detector. Prompted by the large pseudorapidity difference between the $VBF$ jets and a large invariant mass, we require events to satisfy $\eta_{j_1}.\eta_{j_2} < 0$ and $m_{j_1j_2} > 500~$GeV. Furthermore, the pseudorapidity difference between these $VBF$ jets $\Delta \eta_{j_1j_2}$ is used as a training observable in the XGBoost analysis. We also veto events containing any $b$-tagged jet with $p_{T} > 30~$GeV and $|\eta| < 3$ in order to suppress the $t\bar{t} + X$ backgrounds. Kinematic cuts that are specific to the $VBF$ topology $viz.~ m_{j_1 j_2} > 500$ GeV is also imposed at the event generation level to improve the population of events in the phase space of our interest. Similar to Sec.~\ref{sec:ggF}, we also apply the $\Delta R$ cuts, $\Delta R_{\tau_{h},\tau_{h/\ell}} > 0.4$, and $\Delta R_{\tau_{\ell},\tau_{\ell}} > 0.1$. We also apply the generation level cuts, which are summarized in Appendix~\ref{sec:appendixA}, along with the signal and background cross-sections.

We next turn our attention to the multivariate XGBoost analysis to discriminate the exotically decaying $VBF$ Higgs signal from the SM backgrounds. The optimization is performed for the 5 signal benchmarks considered in Sec.~\ref{sec:ggF}. The following kinematic observables are used to perform the training,
\begin{equation}
\begin{split}
\Delta R_{\tau_i \tau_j} (i,j = 1-4; i \neq j),~\Delta R_{\tau\tau}^{min},~\Delta R_{\tau\tau}^{max},\\
~H_{T},~m_{T,h},~m_{4\tau}^{vis},~
p_{T,jj},~m_{jj},~\Delta \eta_{jj},~\Delta R_{jj,4\tau},~\met.
\label{eqn:h_VBF}
\end{split}
\end{equation}

\begin{figure}[htb!]
\centering
\includegraphics[width=0.46\textwidth]{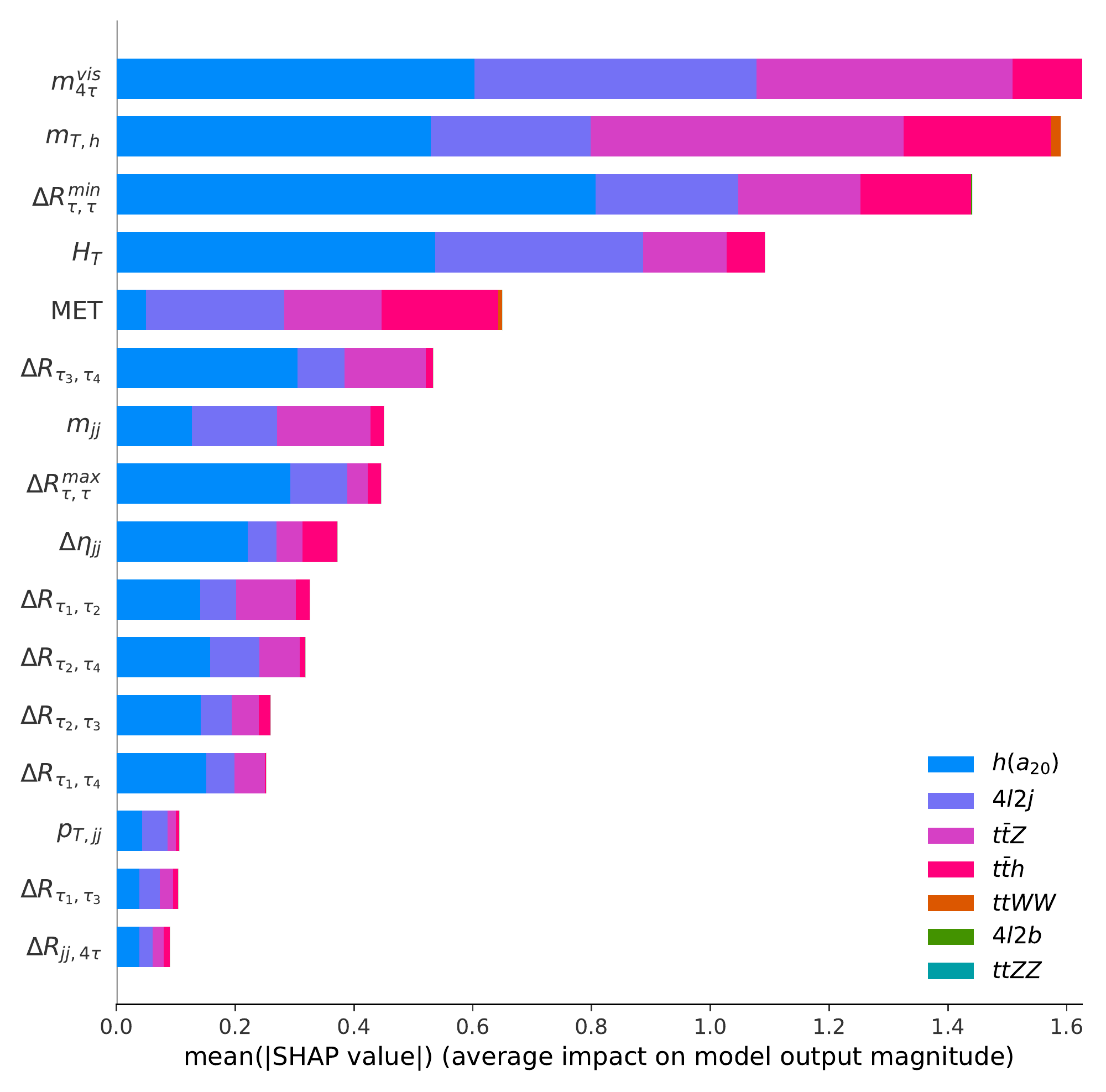}~
\includegraphics[width=0.46\textwidth]{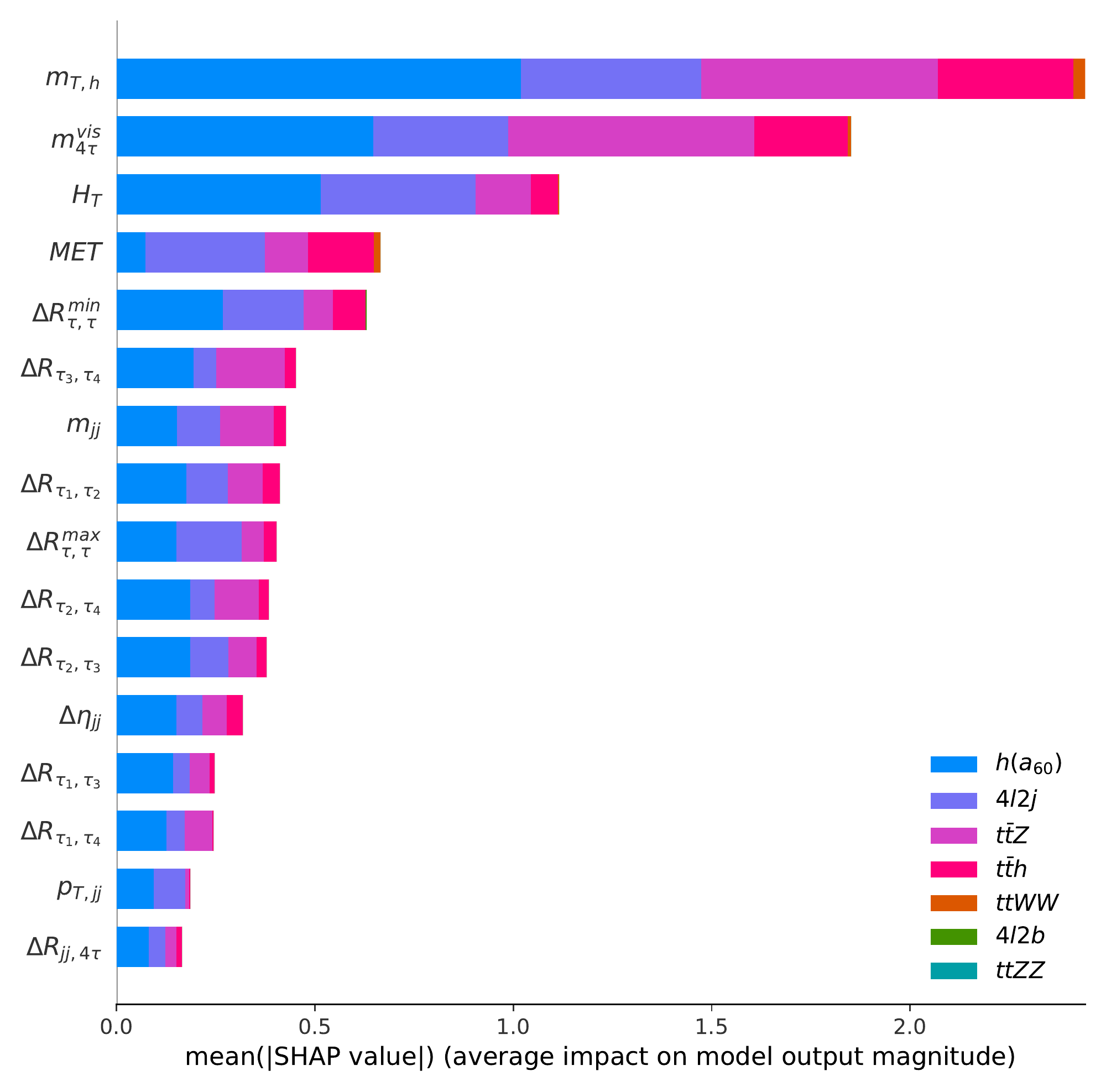}

\caption{\it Mean of absolute SHAP values for the kinematic observables~(Eq.~\ref{eqn:h_VBF}) used to perform the XGBoost analysis for $m_{a}=20~$GeV~(left) and $60~$GeV~(right) in the $VBF$ $pp \to hjj \to (h \to aa \to 4\tau)jj$ channel, at the HL-LHC. A higher absolute SHAP value indicates a higher rank.} 
\label{fig:VBF_shap}
\end{figure}
Here, $H_T$ is the scalar $p_T$ sum of the visible $\tau$ objects and the hardest two VBF jets in the final state, 
$p_{T,j_1j_2}$ is the transverse momentum for the pair of $VBF$ tagged jets, $\Delta R_{jj,4\tau}$ is the $\Delta R$ between the $VBF$ dijet system and the visible Higgs system reconstructed from the visible $\tau$s. The other kinematic observables in Eq.~\ref{eqn:h_VBF} have their usual meanings. The relative importance of the training observables in Eq.~\ref{eqn:h_VBF} in the XGBoost analysis is measured using SHAP analysis. We present the results for the $m_{a}=20$ and $60~$GeV scenarios in Fig.~\ref{fig:VBF_shap}. The top four observables with highest SHAP scores for the $m_{a}=60~$GeV scenario are $m_{4\tau}^{vis}$, $m_{T,h}$, $H_{T}$, and $\met$.

\begin{figure}[htb!]
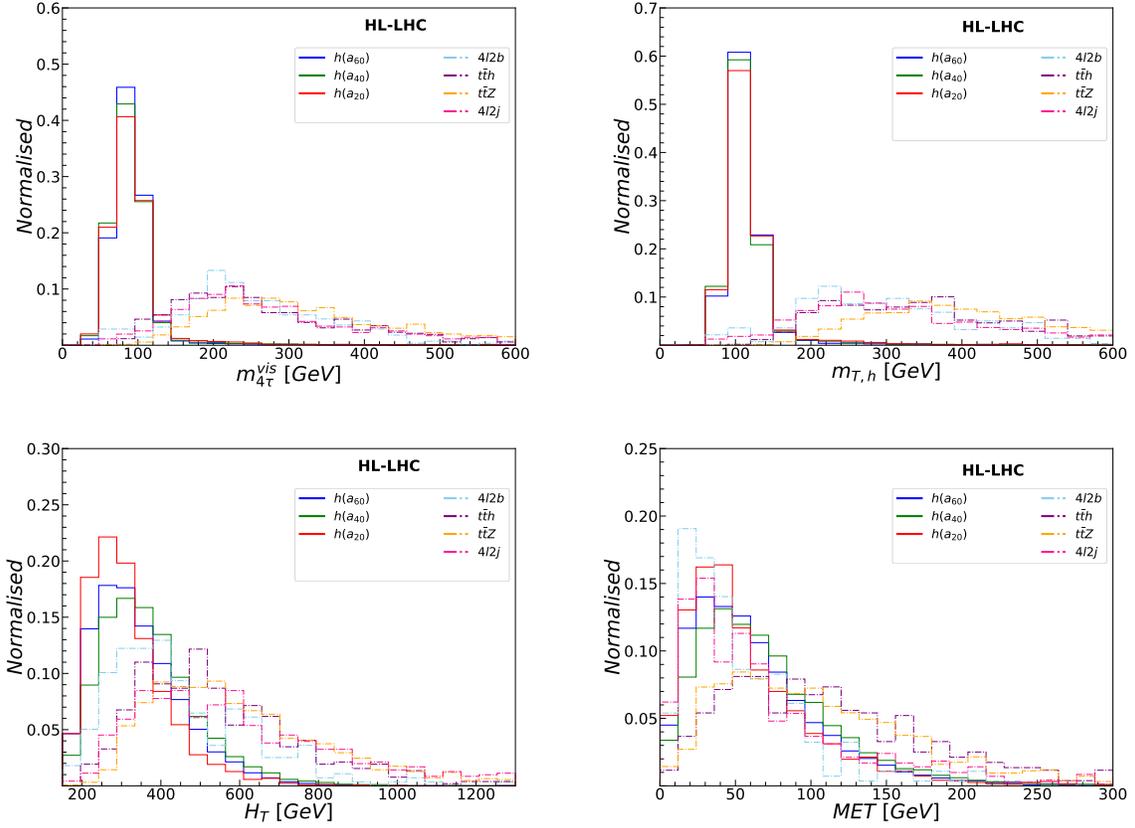

\centering
\includegraphics[width=0.5\textwidth]{./mvis4tau_14_VBF.pdf}~
\includegraphics[width=0.5\textwidth]{./MTh_14_VBF.pdf}\\
\includegraphics[width=0.5\textwidth]{./HT_14_VBF.pdf}~
\includegraphics[width=0.5\textwidth]{./ptmis_14_VBF.pdf}\\
\caption{\it  Distributions for $m_{4\tau}^{vis}$, $m_{T,h}$, $H_{T}$ and $\met$ at the detector level for different signal benchmarks with exotic scalar masses $m_{a}=20,~40$ and ${60~\rm GeV}$ in the VBF single Higgs production channel $pp \to hjj \to (h \to aa \to 4\tau)jj$. Distributions for the QCD-QED $4\ell 2j$, $4\ell 2b$, $t\bar{t}Z$ and $t\bar{t}h$ background processes, are also shown. We consider $\sqrt{s}=14~$TeV LHC with $\mathcal{L}=3~\rm{ab}^{-1}$.}
\label{fig:VBF_most_senstive}
\end{figure}

For illustrative purposes, we show their distributions at the detector level in Fig.~\ref{fig:VBF_most_senstive}. We observe that $m_{4\tau}^{vis}$ and $m_{T,h}$ distributions for the signal benchmarks peak at $\lesssim m_{h}$ due to the decay products from Higgs resonance comprising of missing energy from the neutrinos. On the other hand, the background distributions peak at higher values and are relatively flatter, thereby leading to excellent signal-to-background discrimination. The $H_{T}$ distributions for the signal benchmarks and background processes also exhibit a similar trend. In the distributions for $\slashed{E}_{T}$, the signal benchmarks lead to peaks at $\slashed{E}_{T} \sim 30-40~$GeV. Here, the missing energy arises mainly due to neutrinos from the decay of $\tau$ leptons. We observe that the $\slashed{E}_{T}$ distributions for the $4\ell 2b$ background process overlap considerably with the signal benchmark. However, the rest of the backgrounds are relatively flatter, with peaks at higher values. 

\begin{table}[htbp!]
\centering
\scalebox{0.9}{%
\begin{tabular}{|c|c|c|c|c|c|}\hline
$\sqrt{s}$ & $m_a$ & Total Background & Signal Efficiency  & Signal & Significance \\

(TeV)      & (GeV) & Yield, B  & ($\times 10^{-4}$) & Yield, S  & (5$\%$ sys.)\\\hline

\multirow{5}{*}{14} & 20 & $5$ & $3.8$ & $5$ & $1.5 (1.5)$ \\ \cline{2-6}

 & 30 & $6$ & $8.4$ & $11$  & $2.6 (2.6)$  \\ \cline{2-6}
 
 & 40 & $7$ & $10$ & $13$  & $2.9 (2.9)$  \\ \cline{2-6}
 
 & 50 & $8$ & $12$ & $15$  & $3.1 (3.1)$  \\ \cline{2-6}

 & 60 & $9$  & $15$ & $20$ & $3.7 (3.6)$ \\ \hline
 
 \end{tabular}}
\caption{\it  The signal, background yields, and signal significance at the HL-LHC after the XGBoost analysis in the VBF single Higgs production where Higgs decays via $h \to aa \to 4\tau$. Here, we assume Br($h\to aa\to 4\tau)\sim 0.1\%$. }
\label{tab:VBF_h_XGB}
\end{table}

\begin{figure}[htb!]
\centering
\includegraphics[width=0.5\textwidth]{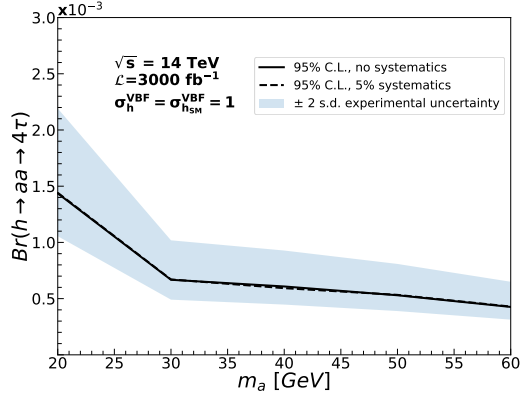}
\caption{\it Upper limit projection for $Br(h\to aa\to 4\tau)$ at 95$\%$ C.L., as a function of exotic scalar mass when $\mu_{h}^{VBF}$ is unity at $\sqrt{s}=14$ TeV. The blue band represents the variation in $Br(h\to aa\to 4\tau)$ within 2 s.d. interval of $\mu_{h}^{VBF}$ as measured by CMS \cite{CMS:2022dwd} and ATLAS \cite{ATLAS:2022vkf} collaborations. The solid and dashed lines refer to adding zero and $5\%$ systematic uncertainty, respectively.
}
\label{fig:muh_14_VBF}
\end{figure}

The signal efficiency and background yields at the HL-LHC for the five signal benchmarks considered in Sec.~\ref{sec:ggF} are shown in Table~\ref{tab:VBF_h_XGB}. Keeping in line with the discussion in Sec.~\ref{sec:ggF}, we also compute the signal yields and the signal significance under the assumption $Br(h \to aa \to 4\tau) = 0.1\%$. For a given signal benchmark, the significance from the $VBF$ channel is lower than that from the $ggF$ channel~(c.f. Table~\ref{tab:ggF_h_XGB}) due to a smaller production rate for the former. Signal efficiency and background yields increase with $m_{a}$. However, the relative growth for signal efficiency is larger than the background yields leading to an overall improvement for the signal significance. In Table~\ref{tab:VBF_h_XGB}, the significance values shown within parentheses correspond to a systematic uncertainty of $5\%$. We observe that the signal significance values remain almost unaffected by systematic uncertainties due to relatively large $S/B \sim \mathcal{O}(1)$ values. The results shown in Table~\ref{tab:VBF_h_XGB} can be translated into upper limits in the plane of $Br(h \to aa \to 4\tau)$ as a function of $m_{a}$ as shown in Fig.~\ref{fig:muh_14_VBF}, under the assumption of $\mu_{h}^{VBF}=1$, where $\mu_{h}^{VBF} = \sigma_{h}^{VBF}/\sigma_{h_{SM}}^{VBF}$. For $m_{a}=20~$GeV~(60~GeV), the HL-LHC would be able to probe exotic Higgs branching ratios as small as $Br(h \to aa \to 4\tau) \lesssim 0.14\%$~($0.043\%$) through searches in the VBF Higgs production channel at 2 s.d. uncertainty.  The observed $\sigma_{h}^{VBF}$ at the LHC is constrained within $\sim 40\%$ of $\sigma_{h_{SM}}^{VBF}$ at 2 s.d. uncertainty~\cite{CMS:2022dwd, ATLAS:2022vkf}. The blue band displays the variation in the upper limit within 2 s.d. uncertainty of $\mu_{h}^{VBF}$ measurements. 

As previously discussed in Sec.~\ref{sec:intro}, signatures for exotic Higgs decay are expected to be observed in single Higgs production channels before in non-resonant di-Higgs searches. Nonetheless, it must be noted that Higgs pair production offers a far richer phenomenology, which can be utilized to complement the potential sensitivity from single Higgs production channels. Therefore, in the following section, we perform a detailed collider study to estimate the future potential at the HL-LHC to probe exotic Higgs decays in the non-resonant di-Higgs production.

\section{Non-resonant Higgs pair production: $gg \to hh \to (h \to b\bar{b})(h \to aa \to 4\tau)$}
\label{sec:non-res}

In this section, our focus is the non-resonant di-Higgs production channel with one Higgs boson decaying into $b\bar{b}$ and the other decaying into a pair of light exotic scalars: $gg \to hh \to (h \to b\bar{b})(h \to aa \to 4\tau)$, at $\sqrt{s} = 14~{\rm TeV}$ HL-LHC.

The major sources of background are $t\bar{t}h$, $t\bar{t}Z$, and $Zh \to (Z \to b\bar{b})(h \to aa \to 4\tau)$ processes. In addition, QCD-QED $4\ell 2b$, $t\bar{t}ZZ$, and $t\bar{t}WW$ contribute subdominantly.  
The signal and background events are generated using the simulation chain considered in Sec.~\ref{sec:hto4tau}. The jet reconstruction parameters are fixed at $R = 0.4$ and $p_{T} > 20~{\rm GeV}$. Other event reconstruction parameters remain unchanged from Sec.~\ref{sec:hto4tau}.

We first perform a traditional cut-and-count collider analysis, optimising the cuts on selected kinematics observables. Afterwards, we follow a machine-learning-based approach using the XGBoost algorithm. Here again, we consider five signal benchmarks corresponding to different exotic scalar masses $m_{a}=20,30,40,50$ and 60~GeV. Then, we translate our results into projected upper limits on $Br(h \to aa \to 4\tau)$ with the condition that $\sigma_{hh}^{ggF} = \sigma_{hh,SM}^{ggF}$. Under this assumption, we also estimate the projected upper limits for the di-Higgs signal strength factor $\mu_{hh}^{ggF} = \sigma_{hh}^{ggF}/\sigma_{hh,SM}^{ggF}$ as a function of $Br(h \to aa \to 4\tau)$ for different values of $m_a$. 

We select events containing exactly two $b$-tagged jets with $p_{T} > 30$~GeV and $|\eta|<3.0$, and four $\tau$ objects. Both leptonic and hadronic $\tau$ decay modes are considered. Similar to the analysis in Sec.~\ref{sec:hto4tau}, the fully leptonic final state from exotic Higgs decay is ignored due to the small production rate and complex combinatorial ambiguity. The selected events are also required to pass the $\tau$ trigger cuts (see Sec.~\ref{sec:ggF}) and the generation level cuts, summarized in Table~\ref{app1:2}. Furthermore, $p_{T, b_1} > 40$ GeV ($b$-jets are $p_T$ sorted), and and $0.4 < \Delta R_{b_1 b_2} < 2$.

\begin{figure}[htb!]
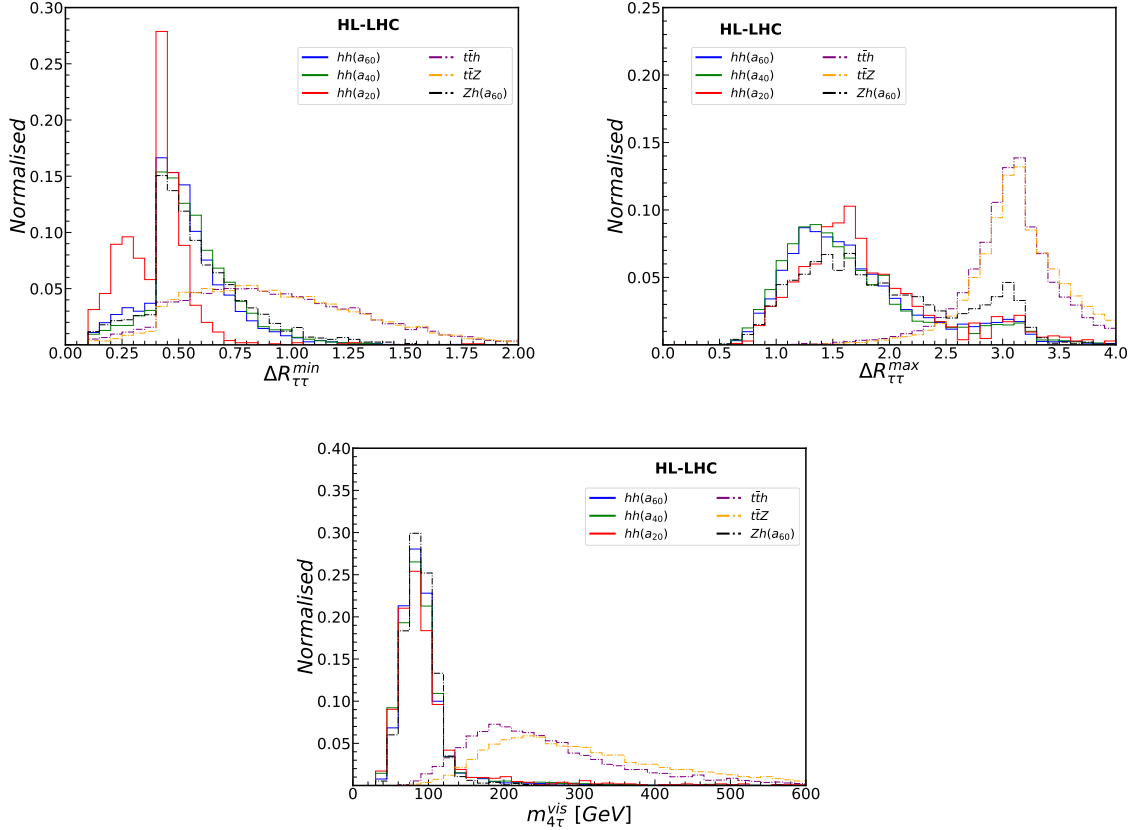

\centering
\includegraphics[width=0.5\textwidth]{./drmin_14_hh.pdf}~
\includegraphics[width=0.5\textwidth]{./drmax_14_hh.pdf}\\
\includegraphics[width=0.5\textwidth]{./mvis4tau_14_hh.pdf}~
\caption{\it  Distributions of $\Delta R_{\tau\tau}^{min}$, $\Delta R_{\tau\tau}^{max}$, and $m_{4\tau}^{vis}$ for signal benchmarks $m_{a}=20,~40,~{60~\rm GeV}$, and dominant backgrounds, in the $pp \to hh \to (h \to b\bar{b})(h \to aa \to 4\tau)$ channel at $\sqrt{s}=14~${TeV} LHC.}
\label{fig:nonres_cutbased_optimised_14TeV}
\end{figure}

Having discussed the event selection cuts, we next turn our attention towards reconstructing the di-Higgs system. One of the Higgses is reconstructed upon requiring 90 GeV $< m_{bb} < 130$ GeV following the optimization in Ref.~\cite{Adhikary:2017jtu}.  In addition, we require $p_{T,b\bar{b}} > 80$ GeV. However, reconstructing the exotically decaying $h$ is challenging due to the presence of multiple invisible particles in the final state from the decay of the $\tau$ leptons. Bypassing this complexity, we reconstruct the visible invariant mass of the exotically decaying $h$, $m_{4\tau}^{vis}$, similar to single Higgs production. In Fig.~\ref{fig:nonres_cutbased_optimised_14TeV}, we illustrate the $m_{4\tau}^{vis}$ distributions for the signal benchmarks and the dominant backgrounds at $\sqrt{s}=14$ TeV. For both c.o.m energies, we observe that the peak for $m_{4\tau}^{vis}$ distributions for the signal process and the $Zh$ background falls close to $80~$GeV. On the other hand, the dominant $t\bar{t} + X$ backgrounds peak at a higher value and are flatter. Overall, $m_{4\tau}^{vis}$ proves to be an excellent discriminator against the associated $t\bar{t}$ backgrounds.  

Another category of observables of considerable interest is the angular correlations between the $\tau$s. In Fig.~\ref{fig:nonres_cutbased_optimised_14TeV}, we present the $\Delta R_{\tau\tau}^{min(max)}$ distributions, which corresponds to the smallest (largest) $\Delta R$ separation between any pair of visible components from the decay of $\tau$ leptons. Both observables exhibit excellent discrimination against the $t\bar{t}h$ and $t\bar{t}Z$ backgrounds. We would like to point out the following observations,
\begin{itemize}
    \item Two distinct peaks are observed in the $\Delta R_{\tau\tau}^{min}$ distributions for the signal process, $0.1<\Delta R_{\tau\tau}^{min}<0.4$ and $\Delta R_{\tau\tau}^{min}>0.4$. This segregation is an implication of the $\Delta R$ selection cuts. While the minimum distance between the visible components of two leptonically decaying $\tau$ leptons~($\tau_{\ell}$) is $\Delta R \gtrsim 0.1$, the minimum separation between $\tau_{h}$ pairs or $\{\tau_{h},\tau_{\ell}\}$ pair must be greater than $\Delta R \gtrsim 0.4$. Hence the $\Delta R_{\tau\tau}^{min}<0.4$ region is associated which those events where the least separated $\tau$s are leptonic. 
    \item The $\Delta R$ separation between the $\tau$ leptons produced from the decay of an exotic scalar $a \to \tau\tau$ is inversely correlated to the mass gap between $h$ and $a$. As a result, the $\tau$ siblings with identical mother $a$ would be more collimated in the $m_{a}=20~$GeV scenario compared to the $m_{a}=60~$GeV scenario. Let us take the example of $m_{a}=20$ GeV. Here, the light exotic Higgs boson is produced with a relatively larger boost owing to its small mass, leading to highly collimated decay products. This further implies a narrower peak for smaller $m_{a}$. As a result, the smallest $\Delta R$ separation is exhibited by the $\tau$ pairs produced from the same light scalar. On the other hand, at relatively higher $m_{a}$ values, $m_{a} \gtrsim 30$ GeV, $\tau$ decay products from different light scalars start constituting $\Delta R_{\tau\tau}^{min}$. 
   \item The peak for the $\Delta R_{\tau\tau}^{max}$ distributions exhibit a mild shift towards larger values with decreasing $m_{a}$. At smaller $m_{a}$, the $\tau$ siblings are highly collimated with their three momentum vectors roughly pointing in the same direction as their parent scalar $a$. Therefore, the $\tau$ pair candidates with the largest $\Delta R$ separation are more likely to arise from different scalar parents at smaller values of $m_a$.
\end{itemize}

Taking into account these observations, we perform a cut-based collider analysis by optimising the selection cuts on $m_{4\tau}^{vis}$, $\Delta R_{\tau\tau}^{min}$ and $\Delta R_{\tau\tau}^{max}$. We analyse five signal benchmarks considered in Sec.~\ref{sec:hto4tau} at the HL-LHC and FCC-hh. Our goal is to maximise the signal significance $\mathcal{S} = S/\sqrt{S+B}$.

\begin{table}[htb!]
\begin{bigcenter}\scalebox{0.9}{
\begin{tabular}{|c|}\hline 
Basic selection cuts \\ \hline
$p_{T,b_{1}(b_{2})} > 40 \; (30)~{\rm GeV}$, $0.4 < \Delta R_{bb} < 2.0$\\ 
$90~{\rm GeV} < m_{bb} < 130~{\rm GeV}$ \\
$p_{T,bb}>80$ GeV \\\hline \hline
Optimised cuts for 14 TeV analysis \\ \hline
$m_{4\tau}^{vis}< [140,130,120,120,120]$ GeV for $m_a = [20,30,40,50,60]$ GeV \\
$\Delta R_{\tau\tau}^{min}<[0.6,0.9,1.2,1.2,1.2]$ and $\Delta R_{\tau\tau}^{max}<[2.4,2.6,2.6,2.8,2.8]$  for $m_a = [20,30,40,50,60]$ GeV\\\hline

\end{tabular}}
\end{bigcenter}
\caption{\it Basic selection and optimised cuts imposed in the cut-and-count analysis for $pp \to hh \to (h \to b\bar{b})(h \to aa \to 4\tau)$ channel at HL-LHC.}
\label{tab:cuts_hh_14TeV}
\end{table}

In Table~\ref{tab:cuts_hh_14TeV}, we summarise the basic selection cuts and the optimised cuts on $m_{4\tau}^{vis}$, $\Delta R_{\tau\tau}^{min}$ and $\Delta R_{\tau\tau}^{max}$. We observe that the optimised cuts on $m_{4\tau}^{vis}$ get slightly stronger with larger values of $m_{a}$. For example, for $m_{a} = 20$ GeV, the signal significance is maximised at $m_{4\tau}^{vis} < 140$ GeV while at $m_{a}=60$ GeV, the signal significance is maximised at $m_{4\tau}^{vis} < 120$ GeV. 
We further observe that the optimised selection cuts get weaker with increasing $m_{a}$ since the peak of $\Delta R_{\tau\tau}^{min}$ distributions shift to lower values with decreasing $m_{a}$. A similar variation is also exhibited by the optimised cuts on $\Delta R_{\tau\tau}^{max}$.

\begin{table}[htb!]
\begin{bigcenter}\scalebox{0.9}{
\begin{tabular}{|c||c|c|c|c|c|c|c|c|c|}
\hline
\multirow{3}{*}{Cut flow} & \multicolumn{7}{c|}{Event rates at $\sqrt{s}=14$ TeV with $\mathcal{L} = 3$ ab$^{-1}$} & \multirow{3}{*}{$\frac{S}{(S+B)}$} & \multirow{3}{*}{Significance}\\ \cline{2-8}

  & Signal & \multicolumn{6}{c|}{Backgrounds} &&\\\cline{3-8} 
 
 &  & $t\bar{t}h$ & $t\bar{t}Z$ & $Zh$ & $4\ell 2b$ & $t\bar{t}ZZ$ & $t\bar{t}WW$ && \\\hline\hline

\multicolumn{10}{|c|}{$m_a=20$ GeV}\\\hline 

\hline
$p_{T,b}$, $\Delta R_{bb}$        & 2.8 & 34 & 81 & 2.3 & 2.0 & 0.72 & 1.4 & 0.023 & 0.25 \\ \hline
$m_{bb}$                          & 2.0 & 12 & 26 & 0.40 & 0.66 & 0.24 & 0.46 & 0.047 & 0.30 \\ \hline
$p_{T,bb}$                        & 1.9 & 11 & 25 & 0.38 & 0.63 & 0.23 & 0.44 & 0.048 & 0.30 \\ \hline
$m_{4\tau}^{vis}$                 & 1.8 & 1.3  & 1.0  & 0.36 & 0.074 & 0.007 & 0.012 & 0.39 & 0.8 \\ \hline
$\Delta R_{\tau\tau}^{min~(max)}$ & 1.7 & 0.24  & 0.22  & 0.32 & 0.042 & 0.002 & 0.001 & 0.67 & 1.1 \\ \hline

\multicolumn{10}{|c|}{$m_a=30$ GeV}\\\hline 
 
\hline
$p_{T,b}$, $\Delta R_{bb}$        & 8.4 & 34 & 81 & 6 & 2 & 0.72 & 1.4 & 0.063 & 0.72 \\ \hline
$m_{bb}$                          & 5.8 & 12 & 26 & 1.0 & 0.66 & 0.24 & 0.46 & 0.12 & 0.85 \\ \hline
$p_{T,bb}$                        & 5.8 & 11 & 25 & 1.0 & 0.63 & 0.23 & 0.44 & 0.13 & 0.86 \\ \hline
$m_{4\tau}^{vis}$                 & 5.3 & 1.0  & 0.64  & 1.0 & 0.067 & 0.004 & 0.006 & 0.658 & 1.9 \\ \hline
$\Delta R_{\tau\tau}^{min~(max)}$ & 5.2 & 0.38  & 0.24  & 0.95 & 0.045 & 0.002 & 0.002 & 0.762 & 2.0 \\ \hline

\multicolumn{10}{|c|}{$m_a=40$ GeV}\\\hline

\hline
$p_{T,b}$, $\Delta R_{bb}$        & 10 & 34 & 81 & 7.3 & 2.0 & 0.72 & 1.4 & 0.074 & 0.87 \\ \hline
$m_{bb}$                          & 7.0  & 12 & 26 & 1.4 & 0.66 & 0.24 & 0.46 & 0.15 & 1.0 \\ \hline
$p_{T,bb}$                        & 6.9  & 11 & 25 & 1.4 & 0.63 & 0.23 & 0.44 & 0.15 & 1.0 \\ \hline
$m_{4\tau}^{vis}$                 & 6.3  & 0.73  & 0.42  & 1.4 & 0.061 & 0.002 & 0.003 & 0.71 & 2.1 \\ \hline
$\Delta R_{\tau\tau}^{min~(max)}$ & 6.1  & 0.29  & 0.22  & 1.3 & 0.042 & 0.001 & 0.001 & 0.77 & 2.2 \\ \hline

\multicolumn{10}{|c|}{$m_a=50$ GeV}\\\hline

\hline
$p_{T,b}$, $\Delta R_{bb}$        & 11 & 34 & 81 & 8.4 & 2.0 & 0.72 & 1.4 & 0.082 & 0.96 \\ \hline
$m_{bb}$                          & 7.9  & 12 & 26 & 1.8 & 0.66 & 0.24 & 0.46 & 0.16 & 1.1 \\ \hline
$p_{T,bb}$                        & 7.8  & 11 & 25 & 1.7 & 0.63 & 0.23 & 0.44 & 0.16 & 1.1 \\ \hline
$m_{4\tau}^{vis}$                 & 7.1  & 0.73  & 0.42  & 1.6 & 0.061 & 0.002 & 0.003 & 0.716 & 2.2 \\ \hline
$\Delta R_{\tau\tau}^{min~(max)}$ & 6.8  & 0.31  & 0.26  & 1.5 & 0.048 & 0.002 & 0.002 & 0.759 & 2.3 \\ \hline

\multicolumn{10}{|c|}{$m_a=60$ GeV}\\\hline

\hline
$p_{T,b}$, $\Delta R_{bb}$        & 13 & 34 & 81 & 10 & 2.0 & 0.72 & 1.4 & 0.094 & 1.1 \\ \hline
$m_{bb}$                          & 9.4  & 12 & 26 & 1.8  & 0.66 & 0.24 & 0.46 & 0.18 & 1.3 \\ \hline
$p_{T,bb}$                        & 9.3  & 11 & 25 & 1.8  & 0.63 & 0.23 & 0.44 & 0.19 & 1.3 \\ \hline
$m_{4\tau}^{vis}$                 & 8.5  & 0.73  & 0.42  & 1.7  & 0.061 & 0.002 & 0.003 & 0.75 & 2.5 \\ \hline
$\Delta R_{\tau\tau}^{min~(max)}$ & 8.2  & 0.34  & 0.26  & 1.6  & 0.048 & 0.002 & 0.002 & 0.79 & 2.5 \\ \hline
\end{tabular}}
\end{bigcenter}
\caption{\it Signal and background yields  in the $pp \to hh \to (h \to b\bar{b})(h \to aa \to 4\tau)$ channel, at each step of the cut-based analysis for the signal benchmarks $m_{a}= 20, 30, 40, 50,$ and $60$ GeV and the dominant backgrounds at the HL-LHC. Signal significance at the HL-LHC is also shown for $Br(h \to aa \to 4\tau) = 10\%$.}
\label{tab:cutflow_14tev}
\end{table}

The cut-flow information and the signal significance at the HL-LHC are presented in Tables~\ref{tab:cutflow_14tev}. Adopting a conservative approach, we consider $Br(h \to aa \to 4\tau) = 10\%$~\cite{Cai:2020lao, Curtin:2013fra} while computing the signal and $Zh$ background yields. We find that significance increases with $m_{a}$. At the HL-LHC, we obtain a significance of $\sim 1.1$ for $m_{a} = 20$ GeV which increases to $\sim 2.0, 2.2, 2.3,$ and $2.5$ at $m_{a} = 30, 40, 50,$ and $60$ GeV, respectively. 
In addition to the optimised cuts on $m_{4\tau}^{vis}$, $\Delta R_{\tau\tau}^{min}$ and $\Delta R_{\tau\tau}^{max}$, the selection cuts on $p_{T,\tau}$ plays a pivotal role towards improving the significance with $m_{a}$. For $m_{a} \sim 20$ GeV, the $\tau$s from the $a \to \tau\tau$ decay is produced rather softly compared to a scenario where $m_{a}$ is heavier. This behavior is illustrated in Fig. \ref{fig:2b4tau_parton}, Appendix~\ref{sec:appendixB} where we present the $p_{T,\tau}$ distributions at the parton-level for the five signal benchmarks at $\sqrt{s}=14$ TeV. At the basic selection stage, we require $p_{T,\tau_{h}} > 20$ GeV and $p_{T,\tau_{l}} >10$ GeV. The signal efficiency is the highest for $m_a = 60$ GeV and decreases with $m_a$. This hierarchy percolates down to the final stage of cut-based optimisation. The signal efficiency for smaller values of $m_{a}$ could be potentially improved through smaller and optimised trigger cuts on $p_{T,\tau_{h}}$ and $p_{T,\ell}$ at the HL-LHC. Such studies inspire a detailed analysis of the projected low-$p_{T}$ trigger efficiency at future colliders and are beyond the scope of the present work.

Owing to the complexity and limited statistics in the $pp \to hh \to 2b4\tau$ channel, it is of utmost importance to precisely explore all possible clues pertaining to beyond-SM interactions. In principle, new physics effects can potentially manifest through complex correlations between various input observables. The cut-based analysis, although robust, is susceptible to overlooking such multi-dimensional correlations. Accordingly, similar to Sec.~\ref{sec:hto4tau}, we adopt the XGBoost algorithm as a multivariate technique with several kinematic observables, 
\begin{equation}
\begin{split}
& ~m_{hh}^{vis},~p_{T,hh}^{vis},~\Delta R_{hh}^{vis},~m_{T2},~m_{T,h},~m_{eff},~ p_{T,bb},~m_{bb},~\Delta R_{bb},~m_{4\tau}^{vis},\\
& ~\Delta R_{\tau_1,\tau_2},~\Delta R_{\tau_1,\tau_3},~\Delta R_{\tau_1,\tau_4},~\Delta R_{\tau_2,\tau_3},~\Delta R_{\tau_2,\tau_4},~\Delta R_{\tau_3,\tau_4},~\Delta R_{\tau\tau}^{min},~\Delta R_{\tau\tau}^{max}.
\label{eqn:multivariate_nonres}
\end{split}
\end{equation}

Here, $m_{hh}^{vis}$ and $p_{T,hh}^{vis}$ are the invariant mass and transverse momentum, respectively, for the visible components of the di-Higgs system, $\Delta R_{hh}^{vis}$ is the distance in the $\eta-\phi$ plane between the reconstructed $h \to b\bar{b}$ and visible $h \to 4\tau$ system, $m_{eff}$ is the effective mass of the di-Higgs system, $m_{eff}~=~H_T~+~\met$, and $m_{T2}$ is motivated from Refs.~\cite{Lester:1999tx, Barr:2003rg}. The analysis in Refs.~\cite{Lester:1999tx, Barr:2003rg} illustrates the effectiveness of $m_{T2}$ in estimating the mass of pair-produced particles in scenarios where both decay into visible and invisible components. The decay topology of the exotic scalars in our signal corresponds to a similar scenario as they are pair-produced from the SM-like Higgs boson and undergo decay into visible and invisible candidates. The rest of the observables in Eq.~\ref{eqn:multivariate_nonres} have their usual meanings. 

The six most essential observables with the highest absolute SHAP values across all five signal benchmarks are $m_{4\tau}^{vis}$, $\Delta R_{\tau\tau}^{min}$, $\Delta R_{\tau\tau}^{max}$, $m_{T,h}$, $\Delta R_{b\bar{b}}$, $m_{b\bar{b}}$. We note that the cut-and-count analysis discussed earlier was performed by optimizing the selection cuts on the first three observables on this list. For illustrative purposes, we present the distributions for the other three most important observables $viz$ $\Delta R_{bb}$, $m_{T,h}$ and $m_{bb}$, in Fig. \ref{fig:BDT_obs_nonres}, Appendix~\ref{sec:appendixC}.

\begin{table}[htb!]
\centering
\scalebox{0.9}{%
\begin{tabular}{|c|c|c|c|c|c|}\hline
$\sqrt{s}$ & $m_a$ & Total Background & Signal Efficiency  & Signal & Significance \\

(TeV)      & (GeV) & Yield, B  & ($\times 10^{-4}$) & Yield, S  & (5$\%$ systematic) \\\hline

\multirow{5}{*}{14} & 20 & $1.8$ & $2.5$ & $3.2$ & $1.4$ ($1.4$) \\ \cline{2-6}

 & 30 & $2.7$ & $6.7$ & $8.6$  & $2.6$ ($2.6$) \\ \cline{2-6}
 
 & 40 & $3.8$ & $8.4$ & $10$  & $2.8$ ($2.8$)  \\ \cline{2-6}
 
 & 50 & $4.5$ & $9.3$ & $12$  & $2.9$ ($2.9$) \\ \cline{2-6}

 & 60 & $5.0$  & $11$ & $14$ & $3.2$ ($3.2$) \\ \hline
 
 \end{tabular}}
\caption{\it  Signal and background yields, and signal significance, at the HL-LHC from XGBoost analysis in the $pp \to hh \to (h \to b\bar{b})(h \to aa \to 4\tau)$ channel. The results shown here have been derived for Br($h\to aa\to 4\tau) = 10\%$.}
\label{tab:non-res_BDT_14TeV}
\end{table}

The optimised signal and background yields from the XGBoost analysis, alongside the signal significance at the HL-LHC (for $Br(h \to aa \to 4\tau) = 10\%$), without and with 5$\%$ systematic uncertainty, are presented in Table~\ref{tab:non-res_BDT_14TeV}. Compared to the cut-and-count analysis, the significance improves by $\mathcal{O}(25-45\%)$. Considering $Br(h \to aa \to 4\tau) = 10\%$, the significance goes beyond 2. However, single Higgs production at the HL-LHC would be able to constrain $Br(h \to aa \to 4\tau)$ up to $\sim 0.015\%$ for $m_{a}=60~$GeV~(see Sec.~\ref{sec:ggF}). Considering the projection described above for exotic Higgs branching fraction, null signal events would be observed in the non-resonant di-Higgs channel at the HL-LHC at $\gtrsim$ 2 s.d. 

We next focus on the $\sqrt{s}=100~$TeV hadron collider FCC-hh. The di-Higgs production cross-section is $\sim$ 30 times greater than that at the HL-LHC, which motivates us to investigate the prospects for exotic Higgs decay. We consider the same background processes as above. We use the FCC-hh detector card of \texttt{Delphes} \cite{Selvaggi:2717698} for the detector simulation with $b$-tagging, $c$, and light jet mistagging efficiencies as functions of $p_T$, as prescribed in Appendix A of \cite{Selvaggi:2717698}. For our analysis, the typical $p_T$ of a jet is much less than the TeV scale. Hence, the $b$-tagging, $c$-mistagging, and light jet mistagging efficiencies are approximately 85$\%$, 5$\%$, and 1$\%$ respectively. Jets are reconstructed from particle-flow objects using \texttt{Fastjet} with $R$ = 0.3 and $p_T > 30$ GeV. The basic trigger and selection cuts on the final state objects remain unchanged. 

\begin{figure}[htb!]
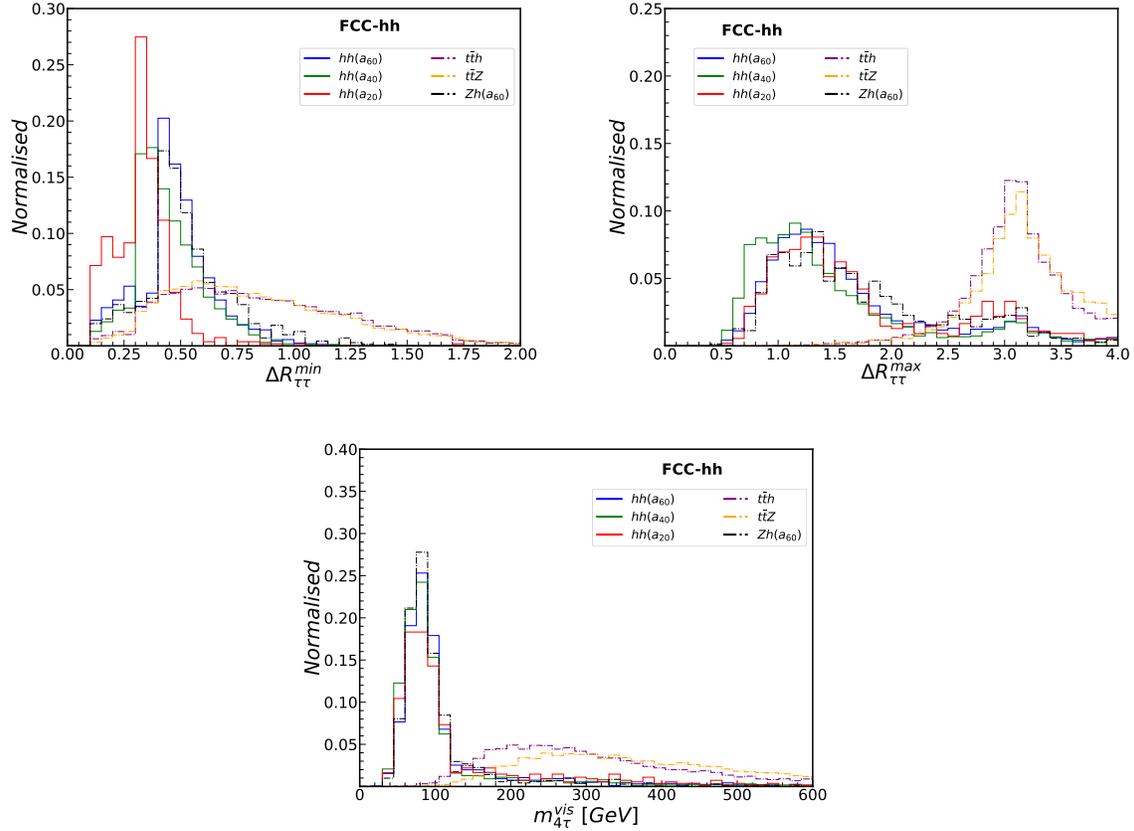

\centering
\includegraphics[width=0.5\textwidth]{./drmin_100_hh.pdf}~
\includegraphics[width=0.5\textwidth]{./drmax_100_hh.pdf}\\
\includegraphics[width=0.5\textwidth]{./mvis4tau_100_hh.pdf}
\caption{\it  Distributions of $\Delta R_{\tau\tau}^{min}$, $\Delta R_{\tau\tau}^{max}$, and $m_{4\tau}^{vis}$ for signal benchmarks $m_{a}=20,~40,~{60~\rm GeV}$, and dominant backgrounds, in the $pp \to hh \to (h \to b\bar{b})(h \to aa \to 4\tau)$ channel at $\sqrt{s}=100~${TeV} LHC.}
\label{fig:nonres_cutbased_optimised_100TeV}
\end{figure}

Fig. \ref{fig:nonres_cutbased_optimised_100TeV} shows the distributions of $\Delta R^{min},~\Delta R^{max}$, and $m^{vis}_{4\tau}$ variables. The peak in the $\Delta R^{min} < 0.3$ region is higher for $\sqrt{s}=100~$TeV because here the $\tau$ objects are more boosted, and the decay products are more collimated than at $\sqrt{s}=14~$TeV. Table \ref{tab:cuts_hh_100TeV} shows the optimised cuts on these variables.

\begin{table}[htb!]
\begin{bigcenter}\scalebox{0.9}{
\begin{tabular}{|c|}\hline 
Basic selection cuts \\ \hline
$p_{T,b_{1}(b_{2})} > 40 \; (30)~{\rm GeV}$, $0.4 < \Delta R_{bb} < 2.0$\\ 
$90~{\rm GeV} < m_{bb} < 130~{\rm GeV}$ \\
$p_{T,bb}>80$ GeV \\\hline \hline

Optimised cuts for 100 TeV analysis \\ \hline
$m_{4\tau}^{vis}< [200,200,170,170,150]$ GeV for $m_a = [20,30,40,50,60]$ GeV \\
$\Delta R_{\tau\tau}^{min}<[0.8,0.9,0.9,1.0,1.0]$ and $\Delta R_{\tau\tau}^{max}<[2.1,2.4,2.4,4.0,4.0]$ for $m_a = [20,30,40,50,60]$ GeV\\\hline
\end{tabular}}
\end{bigcenter}
\caption{\it Basic selection and optimised cuts imposed in the cut-and-count analysis for $pp \to hh \to (h \to b\bar{b})(h \to aa \to 4\tau)$ channel at the FCC-hh.}
\label{tab:cuts_hh_100TeV}
\end{table}

Using these cuts, we perform a cut-and-count analysis. The cut-flow information, along with the signal significance, is presented in Table \ref{tab:cutflow_100tev}. Assuming a 10$\%$ branching of $h\rightarrow 4\tau$, we observe a much higher signal yield and signal significance at the $\sqrt{s}=100~$TeV collider. At the FCC-hh, we obtain signal significance of $24, 48, 50, 54$ and $50$ for $m_{a} = 20, 30, 40, 50$ and $60$ GeV, respectively. We also apply the XGBoost analysis using the observables in Eq. \ref{eqn:multivariate_nonres}. Table \ref{tab:non-res_BDT_100TeV} shows the signal efficiency, signal and background yields, and signal significance with and without $5\%$ systematic uncertainty. As expected, the multivariate method resulted in $\mathcal{O}(30-45\%)$ improvement in the signal significance. \footnote{Appendix \ref{sec:appendixD} (see Table \ref{tab:non-res_BDT_100TeV_R04}) lists the results of XGBOOST analysis with jet radius R = 0.4 instead of 0.3.}

\begin{table}[htb!]
\begin{bigcenter}\scalebox{0.9}{
\begin{tabular}{|c||c|c|c|c|c|c|c|c|c|}
\hline
\multirow{3}{*}{Cut flow} & \multicolumn{7}{c|}{Event rates at $\sqrt{s}=100$ TeV with $\mathcal{L} = 30$ ab$^{-1}$} & \multirow{3}{*}{$\frac{S}{(S+B)}$} & \multirow{3}{*}{Significance}\\ \cline{2-8}

  & Signal & \multicolumn{6}{c|}{Backgrounds} &&\\\cline{3-8} 
 
 &  & $t\bar{t}h$ & $t\bar{t}Z$ & $Zh$ & $4\ell 2b$ & $t\bar{t}ZZ$ & $t\bar{t}WW$ &&\\\hline\hline

\multicolumn{10}{|c|}{$m_a=20$ GeV}\\\hline 

\hline
$p_{T,b}$, $\Delta R_{bb}$        & 1609 & 23454 & 39799 & 505 & 736 & 601 & 1497 & 0.024 & 6.2 \\ \hline
$m_{bb}$                          & 1215 & 7098  & 11089  & 67  & 195 & 163  & 374 & 0.060 & 8.6 \\ \hline
$p_{T,bb}$                        & 1208 & 6908  & 10683  & 67  & 188 & 158  & 364 & 0.062 & 8.6 \\ \hline
$m_{4\tau}^{vis}$                 & 972 & 1779   & 889   & 59  & 44  & 14   & 29  & 0.26 & 16  \\ \hline
$\Delta R_{\tau\tau}^{min~(max)}$ & 887 & 190   & 193   & 49  & 17   & 3.8   & 2.0   & 0.66 & 24 \\ \hline

\multicolumn{10}{|c|}{$m_a=30$ GeV}\\\hline 
 
\hline
$p_{T,b}$, $\Delta R_{bb}$        & 4362 & 23454 & 39799 & 973 & 736 & 601 & 1497 & 0.061 & 16 \\ \hline
$m_{bb}$                          & 3222  & 7098  & 11089  & 122  & 195 & 163  & 374 & 0.14 & 22  \\ \hline
$p_{T,bb}$                        & 3211  & 6908  & 10684  & 122  & 188 & 158  & 364 & 0.15 & 22  \\ \hline
$m_{4\tau}^{vis}$                 & 2814  & 915   & 425   & 118  & 29  & 8.1   & 13  & 0.65 & 43 \\ \hline
$\Delta R_{\tau\tau}^{min~(max)}$ & 2748  & 190   & 193   & 116  & 20   & 3.6   & 3.0   & 0.84 & 48 \\ \hline

\multicolumn{10}{|c|}{$m_a=40$ GeV}\\\hline

\hline
$p_{T,b}$, $\Delta R_{bb}$        & 4909 & 23454 & 39799 & 1201 & 736 & 601 & 1497 & 0.068 & 18  \\ \hline
$m_{bb}$                          & 3578 & 7098  & 11089  & 142  & 195 & 163  & 374 & 0.16 & 24  \\ \hline
$p_{T,bb}$                        & 3554 & 6908  & 10684  & 140  & 188 & 158  & 364 & 0.16 & 24  \\ \hline
$m_{4\tau}^{vis}$                 & 3112 & 915   & 425   & 126  & 29  & 8.1   & 13   & 0.67 & 46 \\ \hline
$\Delta R_{\tau\tau}^{min~(max)}$ & 2999 & 189   & 193   & 124  & 20   & 3.6   & 3.0 & 0.85 & 50 \\ \hline

\multicolumn{10}{|c|}{$m_a=50$ GeV}\\\hline

\hline
$p_{T,b}$, $\Delta R_{bb}$        & 5396 & 23454 & 39799 & 1223 & 736 & 601 & 1497 & 0.074 & 20  \\ \hline
$m_{bb}$                          & 4026 & 7098  & 11089  & 146  & 195 & 163  & 374 & 0.17 & 26 \\ \hline
$p_{T,bb}$                        & 3995 & 6908  & 10684  & 146  & 188 & 158  & 364 & 0.18 & 27 \\ \hline
$m_{4\tau}^{vis}$                 & 3454 & 363   & 193   & 124 & 19   & 4.3   & 6.0   & 0.83 & 54 \\ \hline
$\Delta R_{\tau\tau}^{min~(max)}$ & 3450 & 276   & 193   & 122 & 18   & 3.8   & 4.0   & 0.85 & 54 \\ \hline

\multicolumn{10}{|c|}{$m_a=60$ GeV}\\\hline

\hline
$p_{T,b}$, $\Delta R_{bb}$        & 6591 & 23454 & 39799 & 1608 & 736 & 601 & 1497 & 0.089 & 24 \\ \hline
$m_{bb}$                          & 4829 & 7098  & 11089  & 168  & 195 & 163  & 374 & 0.20 & 31 \\ \hline
$p_{T,bb}$                        & 4801 & 6908  & 10684  & 168  & 188 & 158  & 364 & 0.21 & 31 \\ \hline
$m_{4\tau}^{vis}$                 & 4083 & 363   & 193   & 158 & 19   & 4.3   & 6.0   & 0.85 & 59 \\ \hline
$\Delta R_{\tau\tau}^{min~(max)}$ & 4083 & 276   & 193   & 156 & 18   & 3.8   & 4.0 & 0.86 & 59 \\ \hline

\end{tabular}}
\end{bigcenter}
\caption{\it Signal and background yields in the $pp \to hh \to (h \to b\bar{b})(h \to aa \to 4\tau)$ channel, at each step of the cut-based analysis for the signal benchmarks $m_{a}= 20, 30, 40, 50,$ and $60$ GeV and the dominant backgrounds at the FCC-hh. Signal significance at the FCC-hh is also shown for $Br(h \to aa \to 4\tau) = 10\%$.}
\label{tab:cutflow_100tev}
\end{table}


\begin{table}[htb!]
\centering
\scalebox{0.9}{%
\begin{tabular}{|c|c|c|c|c|c|}\hline
$\sqrt{s}$ & $m_a$ & Total Background & Signal Efficiency  & Signal & Significance \\

(TeV)      & (GeV) & Yield, B  & ($\times 10^{-4}$) & Yield, S  & (5$\%$ systematic) \\\hline

 \multirow{5}{*}{100} & 20 & $716$ & $4.1$ & $1749$ & $35$ ($28$) \\ \cline{2-6}

 & 30 & $1125$ & $10.7$ & $4595$  & $61$ ($49$) \\ \cline{2-6}
 
 & 40 & $1247$ & $13$ & $5521$  & $67$ ($53$) \\ \cline{2-6}
 
 & 50 & $1384$ & $14$ & $5995$  & $70$ ($54$)  \\ \cline{2-6}

 & 60 & $1565$  & $16$ & $7015$ & $76$ ($58$) \\ \hline
 
 \end{tabular}}
\caption{\it  Signal and background yields, and signal significance, at the FCC-hh from XGBoost analysis in the $pp \to hh \to (h \to b\bar{b})(h \to aa \to 4\tau)$ channel. The results shown here have been derived for Br($h\to aa\to 4\tau) = 10\%$.}
\label{tab:non-res_BDT_100TeV}
\end{table}


The signal significance at FCC-hh is $\sim$ 76 s.d. for $m_{a} = 60~$GeV with $Br(h \to aa \to 4\tau) = 10\%$. The significance at FCC-hh drops by $\mathcal{O}(24\%)$ to $\sim$ 58 s.d. on introducing a $5\%$ systematic uncertainty. Adopting the HL-LHC reach for exotic Higgs branching ratio at $m_{a}=60~$GeV from searches in the single Higgs production channel~(see Fig.~\ref{fig:muh_14}), $Br(h \to aa \to 4\tau) \sim 0.015\%$, the significance at FCC-hh drops to $<$ 1 s.d. However, it must be noted that if a light scalar of mass $60~$GeV is first observed at the HL-LHC, there remains the possibility for a 5 s.d. discovery at the FCC-hh through the Higgs pair production channel, provided $Br(h \to aa \to 4\tau) \gtrsim ~1\%$. Overall, prospects at the $\sqrt{s}=100~$TeV collider look promising, and it might be possible to access exotic Higgs decays in $pp \to hh$ searches at the FCC-hh.

\begin{figure}[htb!]
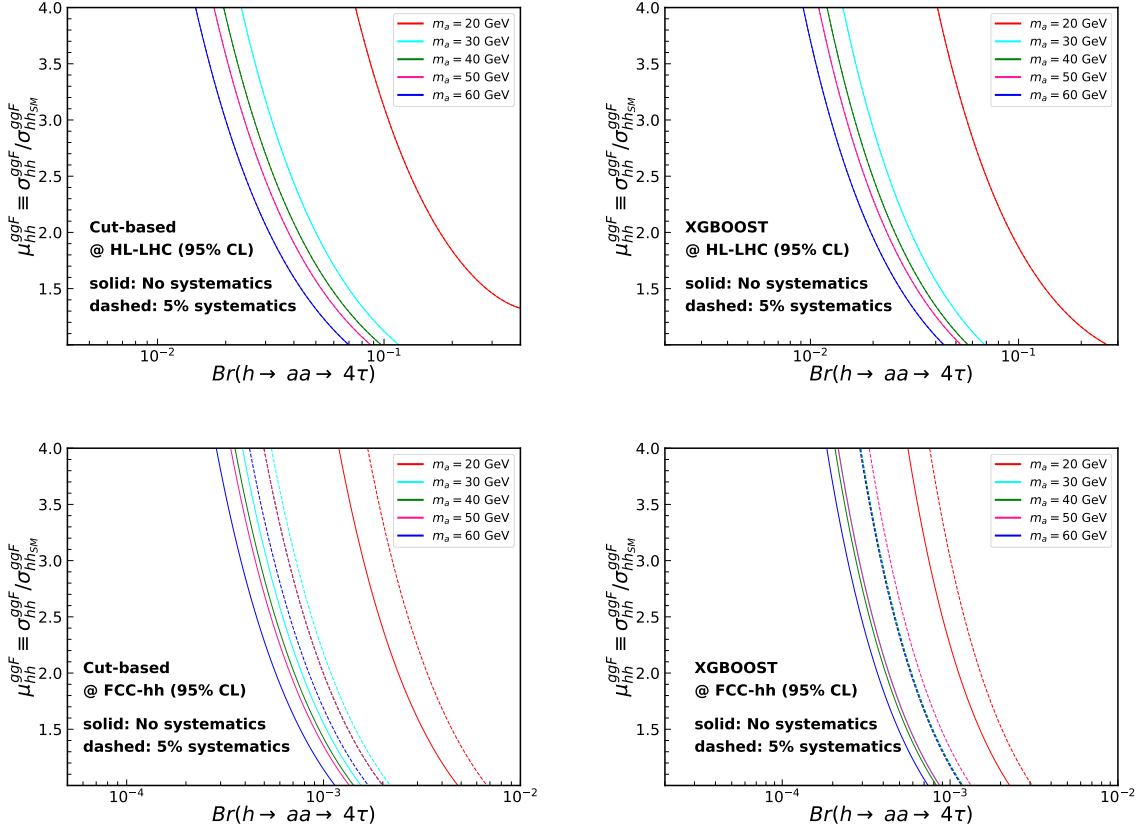

\centering
\includegraphics[width=0.5\textwidth]{./limit_14_cut.pdf}~
\includegraphics[width=0.5\textwidth]{./limit_14_xgb.pdf}\\
\includegraphics[width=0.5\textwidth]{./limit_100_cut.pdf}~
\includegraphics[width=0.5\textwidth]{./limit_100_xgb.pdf}
\caption{\it  Signal strength factor at $95\%$ CL for the non-resonant Higgs pair production with respect to $Br(h\to aa\to 4\tau)$ at $\sqrt{s}=14$ TeV with $\mathcal{L}=3~ab^{-1}$~(top panel) and $\sqrt{s}=100$ TeV with $\mathcal{L}=30~ab^{-1}$~(bottom panel). On the left, results from cut-based analysis are shown. The XGBoost results are shown on the right. 
}
\label{fig:muhh_14}
\end{figure}

Current measurements from CMS in the combined $b\bar{b}ZZ$, $multilepton$, $b\bar{b}\gamma\gamma$, $b\bar{b}\tau\tau$, and $b\bar{b}b\bar{b}$ channels~\cite{CMS:2022dwd} and ATLAS in the combined $b\bar{b}\gamma\gamma$, $b\bar{b}\tau\tau$, and $b\bar{b}b\bar{b}$ channels~\cite{ATLAS:2022jtk} have imposed constraints on the di-Higgs signal strength, limiting it to $\mu_{hh} < 3.4$ and $2.4$ at $95\%$ CL, respectively. For the sake of completeness, we translate the results from cut-and-count~(Table~\ref{tab:cutflow_14tev} and Table~\ref{tab:cutflow_100tev}) and XGBoost analyses~(Table~\ref{tab:non-res_BDT_14TeV} and Table~\ref{tab:non-res_BDT_100TeV}) into upper limit projections on the di-Higgs signal strength $\mu_{hh}^{ggF}$ as a function of $Br(h \to aa \to 4\tau)$, as shown in Fig.~\ref{fig:muhh_14}. The solid lines represent the projected upper limits when systematic uncertainties are not considered. We observe that the HL-LHC would be able to probe exotic Higgs decays up to $Br(h \to aa \to 4\tau) \sim 7\%$ for $m_a=60~$GeV through cut-and-count analysis in the non-resonant di-Higgs production channel, at $95\%$ CL. Upon using XGBoost, the projected sensitivity improves only to $Br(h \to aa \to 4\tau) \sim 5\%$. The potential reach at the FCC-hh is more than an order of magnitude stronger than at the HL-LHC. For $m_{a}=60~$GeV, FCC-hh would be able to probe $Br(h \to aa \to 4\tau)$ as small as $\sim 0.1\% ~$ and $0.07\% ~$ at $95\%$ CL through the cut-and-count and machine-learning based XGBoost analysis, respectively.

However, as discussed earlier, these limits are almost an order of magnitude weaker than the projections from single Higgs search channels considered in Sec.~\ref{sec:hto4tau}. Having exhausted the discovery prospects for exotic Higgs decay in the Higgs pair production channel, we focus on other possible Higgs production modes. The $b\bar{b}4\tau$ final state considered in the present section can also manifest in $Z$ associated Higgs production $pp \to (Z\to b\bar{b})h$, which is precisely the goal for our next section.

\section{Higgs-strahlung: $pp \to Zh \to (Z \to b\bar{b})(h \to aa \to 4\tau$)}
\label{sec:zh}

Higgs-strahlung production $pp \to Zh$, with $Z \to b\bar{b}$ and $h\to aa\to 4\tau$, served as a major background for the non-resonant di-Higgs signal considered in Sec.~\ref{sec:non-res}. An alternate point of view would be to consider the $Zh$ process with the $h$ decaying exotically as a signal by itself, and the di-Higgs process would be a potential background. We analyze such a case in the present section. The background processes are similar to those considered in Sec.~\ref{sec:non-res} barring $pp \to Zh$ which is adopted as the signal process here. The new addition to the list of backgrounds is non-resonant di-Higgs production $pp \to hh \to (h \to b\bar{b})(h \to aa \to 4\tau)$. 

\begin{table}[!h]
\centering
\begin{tabular}{|c|}\hline \hline
Cuts applied for 14 TeV analysis of $Zh\rightarrow b\bar{b}4\tau$\\ \hline \hline
$p_{T,b_{1}(b_{2})} > 30 \; (20)~{\rm GeV}$, $0.4 < \Delta R_{bb} < 2.0$\\ 
$50~{\rm GeV} < m_{bb} < 100~{\rm GeV}$ \\
$p_{T,bb}>50$ GeV \\\hline \hline
$m_{4\tau}^{vis}< [125,120,120,120,120]$ GeV for $m_a = [20,30,40,50,60]$ GeV \\
$\Delta R_{\tau\tau}^{min}<[0.7,0.9,1.2,1.2,1.2]$ and $\Delta R_{\tau\tau}^{max}<3.5$  for $m_a = [20,30,40,50,60]$ GeV\\\hline
\end{tabular}
\caption{\it Basic selection and optimized cuts imposed in the cut-and-count analysis for $pp \to Zh \to (Z \to b\bar{b})(h \to aa \to 4\tau)$ channel.}
\label{tab:cuts_zh}
\end{table}

\begin{table}[htb!]
\begin{bigcenter}\scalebox{0.9}{
\begin{tabular}{|c||c|c|c|c|c|c|c|c|c|}
\hline
\multirow{3}{*}{Cut flow} & \multicolumn{7}{c|}{Event rates at $\sqrt{s}=14$ TeV with $\mathcal{L} = 3$ ab$^{-1}$} & \multirow{3}{*}{$\frac{S}{(S+B)}$} & \multirow{3}{*}{Significance}\\ \cline{2-8}

  & Signal & \multicolumn{6}{c|}{Backgrounds} &&\\\cline{3-8} 
 
 &  & $t\bar{t}h$ & $t\bar{t}Z$ & $hh$ & $4\ell 2b$ & $t\bar{t}ZZ$ & $t\bar{t}WW$ &&\\\hline\hline

\multicolumn{10}{|c|}{$m_a=20$ GeV}\\\hline 

\hline
$p_{T,b}$, $\Delta R_{bb}$        & 2.3  & 34 & 82 & 2.3 & 2.8 & 0.73 & 1.4 & 0.022 & 0.25 \\ \hline
$m_{bb}$                          & 2.2 & 17 & 38 & 2.2 & 1.1 & 0.30 & 0.6 & 0.037 & 0.29 \\ \hline
$p_{T,bb}$                        & 2.2 & 17 & 38 & 2.2 & 1.1 & 0.30 & 0.59 & 0.037 & 0.29 \\ \hline
$m_{4\tau}^{vis}$                 & 2.1 & 0.85  & 0.55  & 0.94 & 0.09 & 0.003 & 0.010 & 0.46 & 0.99 \\ \hline
$\Delta R_{\tau\tau}^{min~(max)}$ & 2.1 & 0.49  & 0.28  & 0.94 & 0.08 & 0.005 & 0.006 & 0.54 & 1.1 \\ \hline

\multicolumn{10}{|c|}{$m_a=30$ GeV}\\\hline 
 
\hline
$p_{T,b}$, $\Delta R_{bb}$        & 6.2 & 35 & 82 & 8.4 & 2.1 & 0.73 & 1.4 & 0.046 & 0.53 \\ \hline
$m_{bb}$                          & 6.0 & 17 & 38 & 3.3 & 1.2 & 0.30 & 0.60 & 0.089 & 0.73 \\ \hline
$p_{T,bb}$                        & 6.0 & 17 & 38 & 3.3 & 1.2 & 0.29 & 0.59 & 0.090 & 0.73 \\ \hline
$m_{4\tau}^{vis}$                 & 5.6 & 0.72  & 0.48 & 2.9 & 0.090 & 0.001 & 0.008 & 0.57 & 1.8 \\ \hline
$\Delta R_{\tau\tau}^{min~(max)}$ & 5.5 & 0.56  & 0.33 & 2.9 & 0.090 & 0.001 & 0.006 & 0.59 & 1.8 \\ \hline

\multicolumn{10}{|c|}{$m_a=40$ GeV}\\\hline

\hline
$p_{T,b}$, $\Delta R_{bb}$        & 7.3 & 34 & 82 & 10 & 2.1 & 0.73 & 1.4 & 0.053 & 0.62 \\ \hline
$m_{bb}$                          & 7.0  & 17 & 38 & 3.9 & 1.2 & 0.30 & 0.60 & 0.10 & 0.84 \\ \hline
$p_{T,bb}$                        & 7.0  & 17 & 38 & 3.9 & 1.2 & 0.30 & 0.59 & 0.10 & 0.84 \\ \hline
$m_{4\tau}^{vis}$                 & 6.5  & 0.72  & 0.48  & 3.4 & 0.090 & 0.001 & 0.008 & 0.58 & 1.9 \\ \hline
$\Delta R_{\tau\tau}^{min~(max)}$ & 6.5  & 0.67  & 0.42  & 3.4 & 0.090 & 0.001 & 0.008 & 0.58 & 1.9 \\ \hline

\multicolumn{10}{|c|}{$m_a=50$ GeV}\\\hline

\hline
$p_{T,b}$, $\Delta R_{bb}$        & 8.5 & 34 & 82 & 11 & 2.1 & 0.73 & 1.45 & 0.06 & 0.72 \\ \hline
$m_{bb}$                          & 8.1  & 17 & 38 & 4.5 & 1.2 & 0.30 & 0.60 & 0.11 & 0.97 \\ \hline
$p_{T,bb}$                        & 8.1  & 17 & 38 & 4.5 & 1.2 & 0.30 & 0.59 & 0.12 & 0.97 \\ \hline
$m_{4\tau}^{vis}$                 & 7.6  & 0.72  & 0.48  & 4.1 & 0.090 & 0.001 & 0.008 & 0.58 & 2.1 \\ \hline
$\Delta R_{\tau\tau}^{min~(max)}$ & 7.5  & 0.67  & 0.42  & 4.1 & 0.090 & 0.001 & 0.008 & 0.59 & 2.1 \\ \hline

\multicolumn{10}{|c|}{$m_a=60$ GeV}\\\hline

\hline
$p_{T,b}$, $\Delta R_{bb}$        & 10 & 35 & 82 & 14 & 2.1 & 0.73 & 1.45 & 0.071 & 0.86 \\ \hline
$m_{bb}$                          & 9.9  & 17 & 38 & 5.2  & 1.2 & 0.30 & 0.60 & 0.14 & 1.16 \\ \hline
$p_{T,bb}$                        & 9.8  & 17 & 38 & 5.2  & 1.2 & 0.30 & 0.59 & 0.14 & 1.15 \\ \hline
$m_{4\tau}^{vis}$                 & 9.2  & 0.72  & 0.48  & 4.7  & 0.090 & 0.001 & 0.008 & 0.61 & 2.4 \\ \hline
$\Delta R_{\tau\tau}^{min~(max)}$ & 9.2  & 0.67  & 0.42  & 4.7  & 0.090 & 0.001 & 0.008 & 0.61 & 2.4 \\ \hline

\end{tabular}}
\end{bigcenter}
\caption{\it Signal and background yields in the $pp\to (Z\to b\bar{b})(h \to aa \to 4\tau) $ channel, at each step of the cut-based analysis for the signal benchmarks $m_{a}= 20, 30, 40, 50,$ and $60$ GeV and the dominant backgrounds at the HL-LHC. The signal significance at the HL-LHC is also shown for $Br(h \to aa \to 4\tau) = 10\%$.}
\label{tab:cutflow_zh_14tev}
\end{table}

\begin{table}[htbp!]
\centering
\scalebox{0.9}{%
\begin{tabular}{|c|c|c|c|c|c|}\hline
$\sqrt{s}$ & $m_a$ & Total Background & Signal Efficiency  & Signal & Significance \\

(TeV)      & (GeV) & Yield, B  & ($\times 10^{-4}$) & Yield, S  & (5$\%$ systematic)\\\hline

\multirow{5}{*}{14} & 20 & $1.5$ & $0.7$ & $2.7$ & $1.3$ ($1.3$)\\ \cline{2-6}

 & 30 & $3.0$ & $1.7$ & $6.4$  & $2.1$ ($2.1$) \\ \cline{2-6}
 
 & 40 & $3.3$ & $2.0$ & $7.4$  & $2.3$ ($2.3$)  \\ \cline{2-6}
 
 & 50 & $4.7$ & $2.4$ & $9.0$  & $2.4$ ($2.4$) \\ \cline{2-6}

 & 60 & $5.2$  & $3.0$ & $11$ & $2.8$ ($2.8$)\\ \hline
 
 \end{tabular}}
\caption{\it  Signal and background yields, and signal significance, at the HL-LHC from XGBoost analysis in the $pp \to Zh \to (Z \to b\bar{b})(h \to aa \to 4\tau)$ channel. The results assume Br($h\to aa\to 4\tau) = 10\%$.}
\label{tab:zh_BDT}
\end{table}

\begin{figure}[htb!]
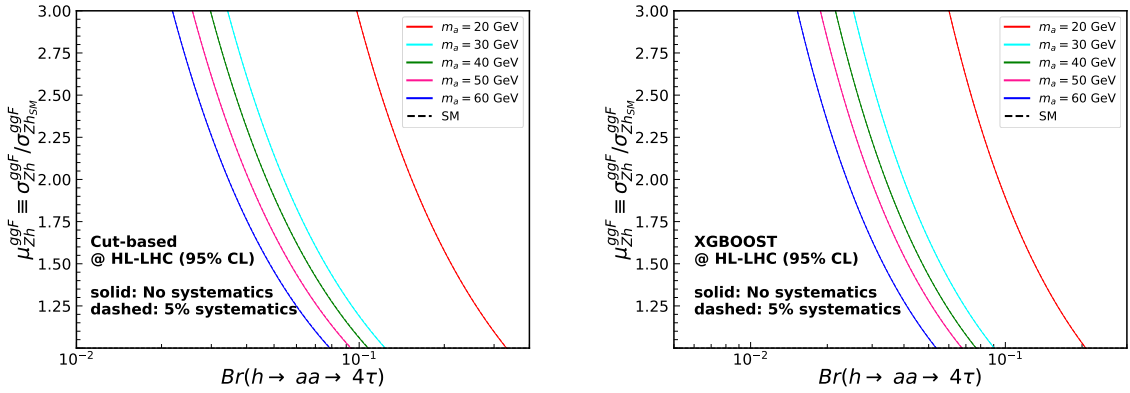

\centering
\includegraphics[width=0.5\textwidth]{./limit_14_cut_zh.pdf}~
\includegraphics[width=0.5\textwidth]{./limit_14_xgb_zh.pdf}\\
\caption{\it Signal strength factor in the $pp \to Zh \to (Z \to b\bar{b})(h \to aa \to 4\tau)$ channel at $95\%$ CL vs $Br(h\to aa\to 4\tau)$ at the HL-LHC for the cut-based~(left panel) and XGBoost~(right panel) analysis.}
\label{fig:muzh_14}
\end{figure}

We closely follow the analysis strategy adopted for the non-resonant di-Higgs channel considered in Sec.~\ref{sec:non-res}, with relatively smaller thresholds on the invariant mass and transverse momentum for the $b\bar{b}$ pairs. The $Z$ boson is reconstructed by constraining $m_{b\bar{b}} \in [50, 100]~\rm{GeV}$ and $p_{T,b\bar{b}}$ is required to be $p_{T,b\bar{b}} > 50~$GeV. We first perform a cut-and-count analysis to estimate the projected sensitivity at the HL-LHC by optimizing the selection cuts on $m_{4\tau}^{vis}$, $\Delta R_{\tau\tau}^{min}$ and $\Delta R_{\tau\tau}^{max}$. The optimization is performed for five signal benchmarks considered in Sec.~\ref{sec:hto4tau}. The basic selection cuts and the optimized cuts on $m_{4\tau}^{vis}$, $\Delta R_{\tau\tau}^{min}$ and $\Delta R_{\tau\tau}^{max}$ for different signal benchmarks are summarized in Table~\ref{tab:cuts_zh}. The cut-flow information and signal significance at the HL-LHC are shown in Table~\ref{tab:cutflow_zh_14tev}. Here again, we have assumed $Br(h \to aa \to 4\tau) = 10\%$. At $m_{a}=20~(60)~$GeV, we obtain significance of 1.06~(2.36) at 5 s.d. uncertainty. It must be noted that the significance values computed for the $Zh$ signal are roughly comparable with the results from searches in the non-resonant di-Higgs channel~(c.f. Sec.~\ref{sec:non-res}).

We also perform the XGBoost analysis using the kinematic observables in Eq.~\ref{eqn:multivariate_nonres}. The subset of training observables with the highest importance rankings is similar to that for the non-resonant di-Higgs channel. We present the corresponding signal and background yields, signal efficiency, and signal significance with and without systematic uncertainty~($\sigma_{unc} = 5\%$), at the HL-LHC, in Table~\ref{tab:zh_BDT}. Here again, we consider $Br(h \to aa \to 4\tau) = 10\%$ while computing the signal yields and the background yield for $pp \to hh$. Our results are also translated into projected upper limits on $\mu_{Zh} = \sigma_{Zh}/\sigma_{Zh}^{SM}$ and $Br(h \to aa \to 4\tau)$ as shown in Fig.~\ref{fig:muzh_14}. We observe that the HL-LHC would be able to probe $Br(h \to aa \to 4\tau)$ up to only $\sim 5\%$ at $95\%$ CL uncertainty for $m_a=60~$GeV through searches in the SM Higgs-strahlung production channel. We would like to note that the projected sensitivity for $Br(h \to aa \to 4\tau)$ from searches in the $Zh$ channel is comparable to that of the Higgs pair production channel considered in Sec.~\ref{sec:non-res}, and weaker compared to other single Higgs production channels in Sec.~\ref{sec:hto4tau}. 

\section{Resonant Higgs pair production: $pp \to H \to hh \to (h \to b\bar{b})(h \to aa \to 4\tau)$}
\label{sec:res}

The reach of non-resonant di-Higgs searches is limited mainly due to its smaller production rate. In general, the sensitivity in di-Higgs channels could benefit from an enhancement in production cross-section. One such possibility is presented by resonant Higgs pair production, where the SM-like Higgs boson pair is produced from the decay of a directly produced heavy scalar, $H$. Exotic Higgs searches in the resonant di-Higgs channel are also relevant in extending/complementing the coverage of the BSM landscape since the extended Higgs sector in several well-motivated BSM frameworks entail such a heavy Higgs boson alongside an exotic light Higgs \textit{viz.}, NMSSM~\cite{Ellwanger:2009dp, Djouadi:2008uj}, 2HDM+S~\cite{Chen:2013jvg}, among other models.  

\begin{figure}[htb!]
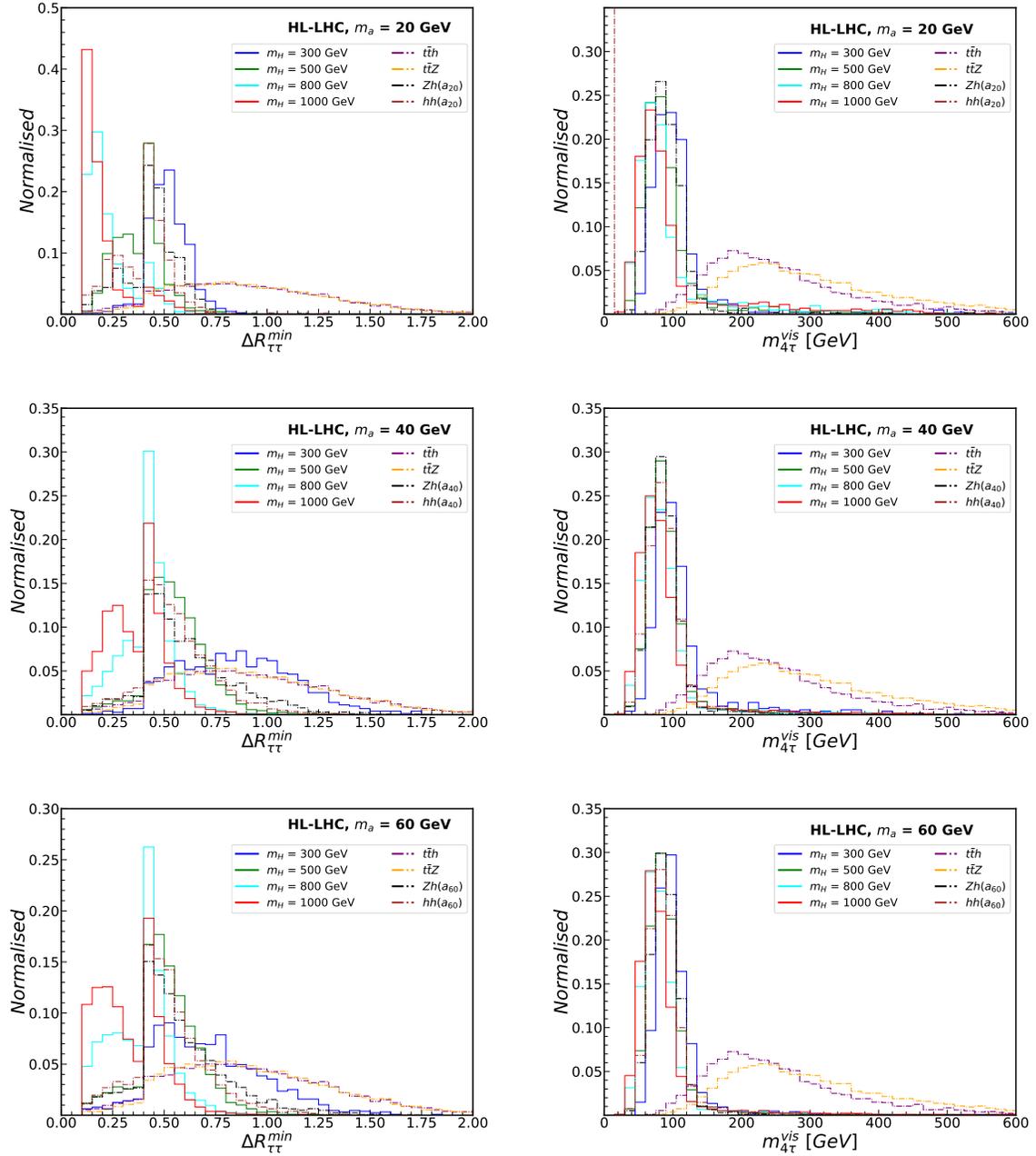

\centering
\includegraphics[width=0.5\textwidth]{./drmin_resh20_14.pdf}~
\includegraphics[width=0.5\textwidth]{./mvis4tau_resh20_14.pdf}\\
\includegraphics[width=0.5\textwidth]{./drmin_resh40_14.pdf}~
\includegraphics[width=0.5\textwidth]{./mvis4tau_resh40_14.pdf}\\
\includegraphics[width=0.5\textwidth]{./drmin_resh60_14.pdf}~
\includegraphics[width=0.5\textwidth]{./mvis4tau_resh60_14.pdf}
\caption{\it  Distributions of $\Delta R_{\tau\tau}^{min}$ (left), and $m_{4\tau}^{vis}$ (right) for signal benchmarks $m_{a}=20~GeV$ (top row), $m_{a}=40~GeV$ (middle row) and $m_{a}=60~GeV$ (bottom row), and dominant backgrounds, in the $pp \to H\to hh \to (h \to b\bar{b})(h \to aa \to 4\tau)$ channel, at $\sqrt{s}=14~${TeV} LHC. The mass of heavy Higgs boson $m_{H} = 300,~500,~800,~1000~GeV$.}
\label{fig:H_14}
\end{figure}

In this section, we explore the HL-LHC prospects of exotic Higgs searches in the resonant di-Higgs channel. We restrict to the final state considered in Sec.~\ref{sec:non-res} and analyze the signal process: $pp \to H \to hh \to (h \to b\bar{b})(h \to aa \to 4\tau)$. Along with the background processes in Sec.~\ref{sec:non-res}, we also consider the non-resonant Higgs pair production process: $pp \to hh \to (h \to b\bar{b})(h \to aa \to 4\tau)$ as a BSM background. The basic selection criteria imposed in Sec.~\ref{sec:non-res} are also applied here. We consider 4 benchmark values of heavy Higgs mass, $m_{H} = 300,~500,~800$ and 1000 GeV and three light scalar mass choices, $m_{a}=20,~40$, and 60 GeV, for each value of $m_{H}$. We perform an XGBoost analysis for these specific choices of $\{m_{H}, m_{a}\}$ using the observables in Eq.~\ref{eqn:multivariate_nonres} and translate the results into model-independent projected upper limits on $\sigma(pp \to H \to hh)$ as a function of $m_{H}$. Two important observables in the multivariate analysis are: $\Delta R_{\tau\tau}^{min}$ and $m_{4\tau}^{vis}$. In Fig.~\ref{fig:H_14}, we illustrate $\Delta R_{\tau\tau}^{min}$ and $m_{4\tau}^{vis}$ distributions for several signal benchmarks $\{m_{H} = 300,~500,~800,~1000~\text{GeV}, m_{a} = 20,~40,~60~\text{GeV}\}$ and dominant backgrounds, at $\sqrt{s}=14$ TeV LHC. For brevity, we are not showing the remaining important observables. The $\Delta R_{\tau\tau}^{min}$ distribution shifts towards smaller values with decreasing $m_{a}$. The exotic scalars are produced with a larger boost in case of low $m_a$, leading to more collimated decay products. For instance, the peak of $\Delta R_{\tau\tau}^{min}$ distribution is much higher in the $\Delta R < 0.4$ region for $m_a$=20 GeV, especially when $m_H$=1000 GeV.

\begin{table}[htb!]
\centering
\scalebox{0.9}{%
\begin{tabular}{|c|c|c|c|c|c|c|c|c|c|c|}\hline
\multicolumn{11}{|c|}{$pp\to H \to hh \to b\bar{b}aa \to 2b 4\tau$, $\sqrt{s}=14$ TeV} \\ \hline
 \multicolumn{2}{|c|}{Masses (GeV)} & \multicolumn{8}{c|}{Background yield at $3~ab^{-1}$ } & \multirow{2}{*}{\makecell{Signal Efficiency \\ ($\times 10^{-4}$) }} \\ \cline{1-10}
$m_H$  & $m_a$  & $t\bar{t}h$ & $t\bar{t}Z$ & $Zh$ & $ hh$ & $4\ell 2b$ & $t\bar{t}ZZ$ & $t\bar{t}WW$ & Total & \\ \hline 
 \multirow{3}{*}{300}  
 & 20 & 0.24 & 0.11 & 0.52 & 0.31 & 0.096 & 0.001 & 0.004 & 1.3 & 0.99 \\
 & 40 & 0.31 & 0.20 & 0.54 & 0.42 & 0.102 & 0.006 & 1.006 & 1.6 & 1.5\\
 & 60 & 0.34 & 0.11 & 0.85 & 0.65 & 0.12 & 0.001 & 0.006 & 2.1 & 2.8 \\ 
\hline  
 \multirow{3}{*}{500}  
 & 20 & 0.21 & 0.088 & 0.54 & 1.5 & 0.074 & 0.003 & 0.005 & 2.4 & 3.1  \\
 & 40 & 0.28 & 0.18 & 1.7 & 5.9 & 0.083 & 0.005 & 0.007 & 8.2 & 12 \\
 & 60 & 0.21 & 0.15 & 1.6 & 6.8 & 0.070 & 0.003 & 0.004 & 8.9 & 15 \\ 
\hline
 \multirow{3}{*}{800}  
 & 20 & 0.15 & 0.088 & 0.36 & 0.68 & 0.15 & 0.004 & 0.012 & 1.4 & 3.5  \\
 & 40 & 0.098 & 0.13 & 1.4 & 2.0 & 0.11 & 0.005 & 0.008 & 3.8 & 17 \\
 & 60 & 0.098 & 0.13 & 1.5 & 2.6 & 0.14 & 0.008 & 0.006 & 4.5 & 19 \\ 
\hline 
 \multirow{3}{*}{1000}  
 & 20 & 0.049 & 0.044 & 0.18 & 0.25 & 0.090 & 0.003 & 0.005 & 0.62 & 2.8  \\
 & 40 & 0.049 & 0.044 & 0.58 & 0.94 & 0.12 & 0.008 & 0.010 & 1.7 & 13 \\
 & 60 & 0.082 & 0.088 & 0.62 & 1.1 & 0.13 & 0.007 & 0.009 & 2.0 & 13  \\ \hline  
\end{tabular}}
\caption{\it  Signal efficiency and background yields from XGBoost analysis in the $pp \to H \to hh\to (h \to b\bar{b})(h \to aa \to 4\tau)$ channel at the HL-LHC. Results are derived for $Br(h \to aa \to 4\tau) = 10\%$.}
\label{tab:H14}
\end{table}  

In Table~\ref{tab:H14}, we present the signal efficiency and background yields at the HL-LHC, from the XGBoost analysis. As discussed earlier in Sec.~\ref{sec:ggF} and \ref{sec:non-res}, the signal efficiency falls with decreasing $m_{a}$. It must also be noted that for a given $m_{a}$, signal efficiency improves with increasing $m_{H}$, except when $m_H=1000$ GeV. This happens because the SM-like Higgs bosons are produced with a relatively more considerable boost for higher $m_{H}$, resulting in an overall shift towards the kinematic region farther away from the SM backgrounds. However, the decay products become too collimated for very high $m_H\sim 1000$ GeV, which becomes challenging to resolve.

We translate the optimised signal efficiencies, $\epsilon_{S}$, and background yields $B$, to model-independent projected upper limits on the resonant di-Higgs production cross-section $\sigma(pp \to H \to hh)_{UL}$,
\begin{equation}
    \sigma(pp \to H \to hh)_{UL} = \frac{\eta_{S} \cdot \sqrt{B}}{\epsilon_{S}\cdot \mathcal{L}\cdot Br(h \to b\bar{b})_{SM}\cdot Br(h \to aa \to 4\tau)\cdot 2}~. 
    \label{eqn:ul_res}
\end{equation}
Here $\eta_{S}$ represents the standard deviation from the background. In the present analysis, we derive the projected upper limits at $\eta_{S}=2$, which corresponds to a $95\%$ confidence interval, with $Br(h \to b\bar{b}) = 58\%$~\cite{Zyla:2020zbs} and $Br(h \to aa \to 4\tau) = 10\%$~\cite{Curtin:2013fra, Cai:2020lao}.

\begin{figure}[htb!]
\centering
\includegraphics[width=0.5\textwidth]{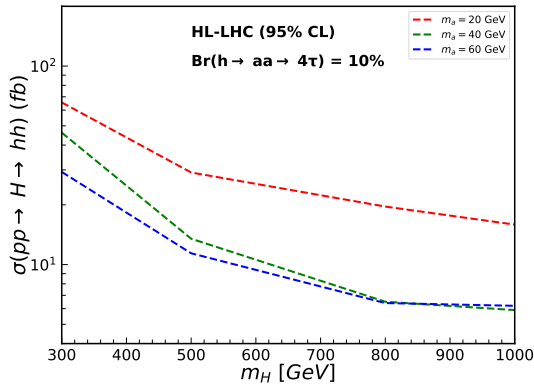}
\caption{\it  Projected upper limits at $95\%$ CL on $\sigma(pp\to H\to hh)$ as a function of $m_{H}$ from resonant di-Higgs searches in the $b\bar{b} 4\tau$ final state at the HL-LHC, assuming $Br(h \to aa \to 4\tau) = 10\%$. The red, green and blue lines correspond to different light scalar masses, $m_{a} = 20,~40$ and $60~$GeV, respectively.}
\label{fig:Hlimit}
\end{figure}

In Fig.~\ref{fig:Hlimit}, we present $\sigma(pp \to H \to hh)_{UL}$ as a function of $m_{H}$ for three light Higgs mass scenarios, $m_{a} = 20,~40$ and 60~GeV. These projections demonstrate the future sensitivity of resonant di-Higgs searches in the $pp \to H \to hh \to (h\to b\bar{b})(h \to aa \to 4\tau)$ channel at the HL-LHC. We observe that at $m_{H} = 300~$GeV, HL-LHC would be able to probe resonant di-Higgs production cross-sections up to $\sigma(pp \to H \to hh) \gtrsim 65, 46, 29~$fb for exotic scalar of mass, $m_{a} = 20, 40, 60~$GeV, respectively, for $Br(h \to aa \to 4\tau) = 10\%$, at $95\%$ CL. We observe that the projected sensitivity gets stronger with increasing $m_{H}$ up to $m_{h }\sim 800~$GeV, after which the upper limit remains almost unchanged. 

\begin{figure}[htb!]
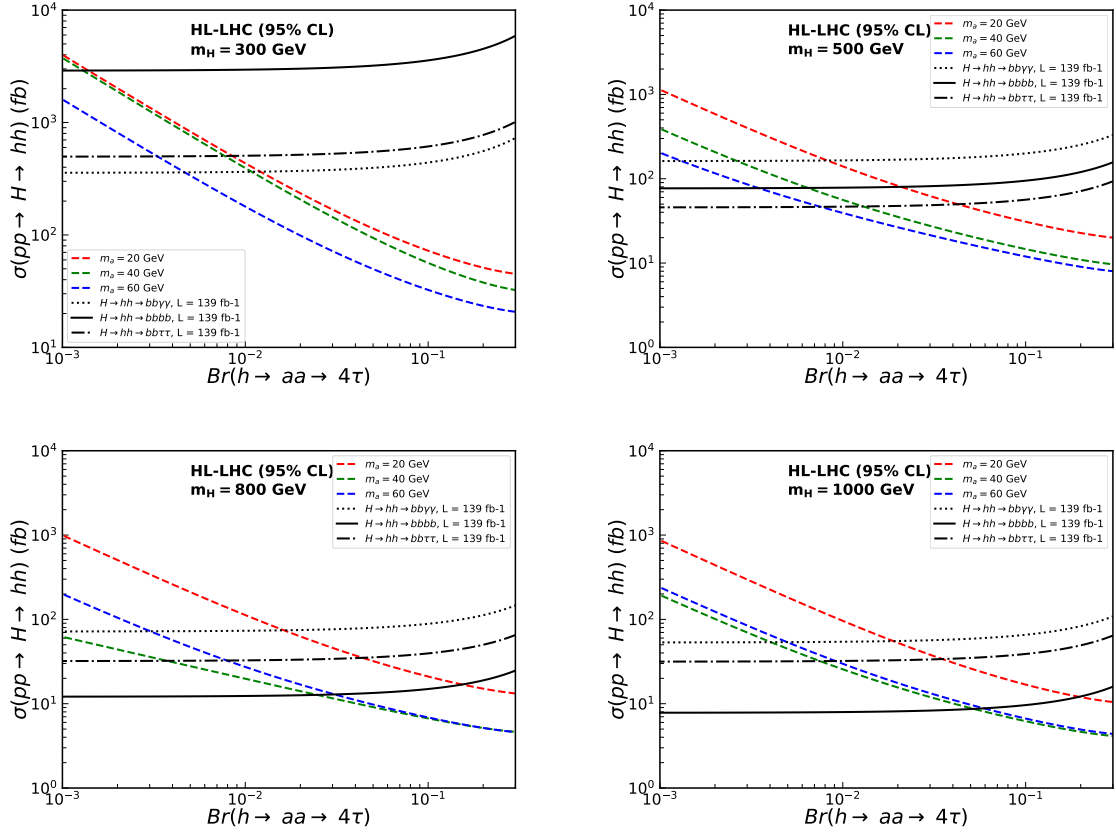

    \centering
    \includegraphics[width=0.5\textwidth]{./cs_br_MH300_res.pdf}~
    \includegraphics[width=0.5\textwidth]{./cs_br_MH500_res.pdf}\\
    \includegraphics[width=0.5\textwidth]{./cs_br_MH800_res.pdf}~
    \includegraphics[width=0.5\textwidth]{./cs_br_MH1000_res.pdf}
    \caption{\it Variation of the projected upper limits at $95\%$ CL on $\sigma(pp\to H\to hh)$ as a function of $Br(h \to aa \to 4\tau)$ for $m_H$ = 300, 500, 800, 1000 GeV at the HL-LHC. The red, green and blue lines correspond to different light scalar masses, $m_{a} = 20,~40$ and $60~$GeV, respectively. Variation of the current upper limits at $95\%$ CL on $\sigma(pp\to H\to hh)$, from resonant di-higgs searches in $b\bar{b}b\bar{b}$~\cite{ATLAS-CONF-2021-035}~(solid), $b\bar{b}\tau^{+}\tau^{-}$~\cite{ATLAS-CONF-2021-030}~(dash dot) and $b\bar{b}\gamma\gamma$~\cite{ATLAS-CONF-2021-016}~(dotted) final states, as a function of $Br(h \to aa \to 4\tau)$ are shown in black.}
    \label{fig:cs_v_br_res}
\end{figure}

Fig.~\ref{fig:cs_v_br_res} shows the $\sigma(pp \to H \to hh)_{UL}$ as a function of $Br(h \to aa \to 4\tau)$ following Eq. \ref{eqn:ul_res}. This section focuses on resonant di-Higgs production with an exotically decaying SM-like Higgs boson. The literature has also explored scenarios where SM-like Higgs bosons decay via SM modes. For example, both CMS and ATLAS collaborations have performed resonant di-Higgs searches in the $H \to hh \to b\bar{b}b\bar{b}$~\cite{ATLAS-CONF-2021-035}, $b\bar{b}\tau^{+}\tau^{-}$~\cite{ATLAS-CONF-2021-030} and $b\bar{b}\gamma\gamma$~\cite{ATLAS-CONF-2021-016} final states using the 13~TeV LHC data collected at $\mathcal{L}\sim 139~{\ifb}$, and derived upper bounds on $\sigma(pp \to H \to hh)$. A similar variation of these current upper limits at $95\%$ CL on $\sigma(pp\to H\to hh)$ with BSM $Br(h \to aa \to 4\tau)$ have been drawn in black.  

For a 300 GeV heavy Higgs, the current searches have ruled out up to $\sigma\sim$ 300 fb for $Br(h \to aa \to 4\tau) \sim 0.5\%$ at $95\%$ CL. With increasing mass of the heavy Higgs, the region becomes much more constrained by current limits. For $m_H$ = 1000 GeV, HL-LHC will be sensitive to $\sigma(pp\to H\to hh)$ up to 8 fb for $Br(h \to aa \to 4\tau)\sim$ 7$\%$.

 In Fig.~\ref{fig:limitcompare}~(left), we show the current upper limits from experiments for different masses of heavy Higgs with dashed lines. The projected sensitivity in the decay modes above at the HL-LHC has also been studied in Ref.~\cite{Adhikary:2018ise}. We display the projected upper limits at the HL-LHC in the $b\bar{b}b\bar{b}$ and $b\bar{b}\gamma\gamma$ channel in Fig.~\ref{fig:limitcompare}~(right) with dashed colored lines and contrast them with the findings of the present analysis where resonant di-Higgs production is probed in exotic final states. For this purpose, we consider the $m_{a}=60~$GeV scenario, which furnishes the most robust limits among other signal benchmarks considered in the present work. In Fig.~\ref{fig:limitcompare}~(right), we present the projected upper limits derived in the current analysis in solid black. The results indicate that the projected sensitivity for resonant di-Higgs searches in exotic final state assuming a saturated decay branching of $Br(h\rightarrow aa\rightarrow 4\tau)$ = 10$\%$ is almost comparable if not slightly stronger than the $4b$ and $2b2\gamma$ channel at the HL-LHC. We further compare the limits with the upper bounds from current LHC runs by rescaling our HL-LHC projections (see Fig.~\ref{fig:Hlimit}) to the current LHC luminosity~($\mathcal{L}=139~{fb}^{-1}$) in Fig.~\ref{fig:limitcompare}~(left). Below $m_{H} \lesssim 500~$GeV, upper limits from the $2b4\tau$ channel considered in the present section are almost comparable to the $2b2\tau$ channel. At higher heavy Higgs masses $m_{H}\gtrsim 650~$GeV, limits from the $4b$ channel begin to dominate. At $m_{H}\sim 1~$TeV, the upper limits from the present analysis are roughly 2-5 times weaker than the $4b$ channel.  

\begin{figure}[htb!]
\centering
\includegraphics[width=0.5\textwidth]{./limit_14_res_139_2.pdf}~
\includegraphics[width=0.5\textwidth]{./limit_14_res_proj.pdf}
\caption{\it  Left: Current upper limits on $\sigma(pp \to H \to hh)$ as a function of $m_{H}$ from resonant di-Higgs searches in $b\bar{b}b\bar{b}$~\cite{ATLAS-CONF-2021-035}~(blue), $b\bar{b}\tau^{+}\tau^{-}$~\cite{ATLAS-CONF-2021-030}~(green) and $b\bar{b}\gamma\gamma$~\cite{ATLAS-CONF-2021-016}~(red) final states. Right: Projected upper limits on $\sigma(pp \to H \to hh)$ as a function of $m_{H}$ from resonant di-Higgs searches in $b\bar{b}b\bar{b}$~(green) and $b\bar{b}\gamma\gamma$~(red) final states at the HL-LHC~\cite{Adhikary:2018ise}. The solid-black line in the left and right panels show the sensitivity of the present analysis.}
\label{fig:limitcompare}
\end{figure}

We briefly study the resonant production of a pair of exotic light scalars from a heavy Higgs. In this case, the final state consists of 4$\tau$s. Hence, we follow a similar analysis strategy to Sec. \ref{sec:ggF}.  We consider the resonant light Higgs pair production, $pp \to H \to aa \to 4\tau$, for two benchmark points, $m_H=300, 1000$ GeV, with $m_a$ = 20, 40, and 60 GeV, as the signal. The relevant background processes are the $gg \to h \to aa \to 4\tau$, the inclusive $4\ell$ and $h\rightarrow Z Z^* \rightarrow 4\ell$~($\ell = e,\mu,\tau$) processes. Subdominant contributions arise from QCD-QED $4\ell2\nu$, $4\ell2b$, $t\bar{t}Z$, $t\bar{t}h$, $t\bar{t}ZZ$ and $t\bar{t}WW$. Table \ref{tab:Haa14} lists the total background yield, the signal efficiency, and the upper limit on cross-section $\sigma(pp\to H\to aa)$ at $95\%$ CL when $Br(H \to aa \to 4\tau) = 10\%$.

\begin{table}[htb!]
\centering
\scalebox{0.9}{%
\begin{tabular}{|c|c|c|c|c|}\hline
$m_H$ & $m_a$ & Total Background & Signal Efficiency  & Upper limits on cross-section at 95$\%$ CL \\

(GeV)      & (GeV) & Yield, B  & ($\times 10^{-4}$) & (fb) \\\hline 
 \multirow{3}{*}{300}  
 & 20 & 31 & 12 & 32 \\
 & 40 & 158 & 146 & 5.7 \\
 & 60 & 270 & 144 & 7.6 \\
\hline  
 \multirow{3}{*}{1000}  
 & 20 & 6 & 1.7 & 93 \\
 & 40 & 24 & 19 & 17 \\
 & 60 & 86 & 113 & 5.5 \\ \hline  
\end{tabular}}
\caption{\it  Signal efficiency, total background yield, and upper limits at $95\%$ CL on $\sigma(pp\to H\to aa)$ from XGBoost analysis in the $pp \to H \to aa \to 4\tau$ channel at the HL-LHC when $Br(H \to aa \to 4\tau) = 10\%$.}
\label{tab:Haa14}
\end{table}  

Although we do not study heavy Higgs production at the FCC-hh, it would be interesting to see how the $t\bar{t}$ and $b\bar{b}$ initiated heavy Higgs production cross-sections compare to the ggF cross-section at high energy scales. We have used \texttt{MadGraph5\_aMC@NLO} for a simple estimation of the cross-sections at Leading Order (LO) when a heavy Higgs is produced from $b\bar{b}$ and $t\bar{t}$, and how they compare to the heavy Higgs production from $ggF$. For this, we consider two masses of SM-like heavy Higgs, $m_{H}$ = 500 GeV and 1000 GeV. We use the NNPDF2.3NNLO PDF set \cite{Ball:2012cx}. The cross-sections are listed in Table \ref{tab:tt_bb_flux}. 

\begin{table}[htb!]
    \centering
    \begin{tabular}{|c|c|c|}\hline 
        Process &   \multicolumn{2}{c|}{Cross-sections at LO (pb)} \\\cline{2-3}
        $\sqrt{s}$ = 100 TeV  &  $m_H$ = 500 GeV & $m_H$ = 1000 GeV \\\hline 
        $b\bar{b} \rightarrow H$ & 0.61  & 0.055  \\\hline 
        $t\bar{t} \rightarrow H$& 24 & 5.1  \\\hline 
        $g\bar{g} \rightarrow H$& 81 &  5.7 \\\hline 
    \end{tabular}
    \caption{Cross-sections of processes: $b\bar{b}\rightarrow H$, $t\bar{t} \rightarrow H$, and $g\bar{g} \rightarrow H$ for $m_H$ = 500 GeV and 1000 GeV, as calculated by \texttt{MadGraph5\_aMC@NLO}.}
    \label{tab:tt_bb_flux}
\end{table}
The LO cross-sections of $b\bar{b} \rightarrow H$, $t\bar{t} \rightarrow H$ and $g\bar{g} \rightarrow H$ at $\sqrt{s}$ = 100 TeV are 51 (148), 70 (201), 43 (108) times greater than the respective LO cross-sections at $\sqrt{s}$ = 14 TeV for $m_H$ = 500 (1000) GeV. At the 100 TeV collider, the top quark, along with the bottom quark, can be effectively treated as a light particle \cite{Mangano:2016jyj}, having a significant contribution to the heavy Higgs production. In such a scenario, a dedicated study involving the $t\bar{t}$ initiated process would be interesting.
\section{Summary}
\label{sec:summary}
In this paper, we have investigated the prospects for probing the exotic Higgs decay mode $h \rightarrow a a \rightarrow 4 \tau$ in single Higgs and resonant di-Higgs searches at the HL-LHC, and in non-resonant di-Higgs searches at the HL-LHC as well as at the FCC-hh.
Searches for exotic Higgs cascade decays to multiple SM particles via new intermediate light (pseudo-)scalars are particularly relevant in extending the coverage of BSM models with an extended Higgs sector. We first focused on the ggF $gg \to h\to aa$ and VBF $pp\to ( h \to aa)jj$ channels, where the light exotic Higgs boson $a$ subsequently decays to $\tau\bar\tau$. We considered five signal benchmarks $m_a = 20,~30,~40,~50$ and 60 GeV and performed a machine-learning-based analysis using the XGBoost algorithm to estimate the projected sensitivity at the HL-LHC. Searches in the $ggF$-induced single Higgs production channel at the HL-LHC lead to signal significance values of 6.0, 5.4, 5.2, 5.8, and 10.3 in the five signal benchmarks, respectively, for $Br(h \to aa \to 4\tau) = 0.1\%$. We also derived upper limits on the exotic Higgs branching fraction assuming SM production rates for the Higgs boson, finding that the HL-LHC will be able to probe $Br(h\to aa\to 4\tau)$ up to $\sim 0.015\%$ at $95\%$ CL for $m_{a}=60~$GeV. The VBF production mode yields smaller signal significances at the HL-LHC, the signal significance of 1.5, 2.6, 2.9, 3.1, and 3.7, for $m_a = 20,~30,~40,~50$ and 60 GeV, respectively, mainly due to the smaller cross section. Through searches in the VBF channel, the HL-LHC could exclude $Br(h\to aa\to 4\tau)$ up to $0.043\%$ at $95\%$ CL for $m_a$ = 60 GeV. The combined projected upper limit on $Br(h\to aa\to 4\tau)$ from searches in the ggF and VBF channels is $0.014\%$ at $95\%$ CL.

Di-Higgs searches in exotic channels provide additional search modes besides the usual SM decay channels. 
We have analyzed the prospects for probing exotic Higgs decays in the $gg \to hh\to ( h \to b\bar{b})(h \to aa \to 4\tau)$ channel at the HL-LHC as well as FCC-hh, for several light scalar masses. The analysis is performed using the traditional cut-and-count approach and the XGBoost algorithm. The cut-based analysis results in the signal significance of 1.1~(24), 2.0~(48), 2.2~(50), 2.3~(54), and 2.5~(59) for the five signal benchmarks, respectively, at the HL-LHC~(FCC-hh), for $Br(h \to aa \to 4\tau) = 10\%$. The XGBoost analysis performs better and leads to a signal significance of 1.4~(35), 2.6~(61), 2.8(67), 2.9(70), and 3.2~(76) for $m_{a} = 20,~30,~40,~50$ and $50~$GeV, respectively, at the HL-LHC~(FCC-hh).  
Overall, the prospects for probing exotic Higgs decays in the di-Higgs channel at the HL-LHC is comparable with the current LHC limits on $Br(h \to aa \to 4\tau)$ from single Higgs searches~\cite{Cai:2020lao}.
We also investigate the HL-LHC prospects for the $Z$ associated Higgs production channel, $pp\to Zh \to (Z\to b\bar{b})(h \to aa \to 4\tau)$. Searches in this channel furnish comparable results to the non-resonant di-Higgs production. 

We finally examine the case of resonant di-Higgs production $pp \to H \to hh \to (h \to b\bar{b})(h \to aa \to 4\tau)$ at the HL-LHC, for several combinations of $\{m_{H},m_{a}\}$. We derive model-independent projected upper limits for $\sigma(pp\to H\to hh)$ as a function of $m_{H}$, which are also translated to the current LHC through luminosity scaling. At the current LHC, the $2b4\tau$ channel considered in the present work performs almost comparably with $4b$, $2b2\tau$ and $2b2\gamma$ channels, provided $Br(h \to aa \to 4\tau) = 10\%$. The same holds even at the HL-LHC. Current searches put a strong constraint on the resonant di-Higgs production cross-section for large masses of heavy Higgs.

This work mainly focused on the $4\tau$ and $2b 4\tau$ final states. Possible improvements in the search potential for exotic Higgs decays at future colliders can be expected through combined searches in other final states. Furthermore, the projected sensitivity might benefit through better background modeling, including higher-order signal and background generation effects. Another critical aspect of the HL-LHC and FCC-hh would be to study the implications of systematic uncertainties and devise effective techniques to mitigate their effects. We defer examining these aspects to future work.

\acknowledgments
A.A. acknowledges partial financial support from the Polish National Science Center under the Beethoven series grant numberDEC-2016/23/G/ST2/04301. In the final stage of this project, the research of A.A. has received funding from the Norwegian Financial Mechanism for years 2014-2021, grant nr DEC-2019/34/H/ST2/00707. RKB thanks the U.S.~Department of Energy for the financial support under grant number DE-SC0016013. The work of Brian Batell is supported by the U.S. Department of Energy under grant No. DE- SC0007914. Biplob Bhattacherjee and C.B. thank Rhitaja Sengupta and Prabhat Solanki for useful discussion. The work of Zhuoni Qian is supported by the Helmholtz-OCPC fellowship program. The work of Michael Spannowsky is supported by the STFC under grant ST/P001246/1. 

\newpage 
\appendix

\section{Summary of the generation cuts and production cross sections for the signal and backgrounds in single Higgs and non-resonant di-Higgs production}
\label{sec:appendixA}

\begin{table}[htb!]
\centering
\begin{bigcenter}
\scalebox{0.8}{%
\begin{tabular}{|c|c|c|c|}

\hline
\multirow{2}{*}{Process} & \multirow{2}{*}{Signal and Backgrounds} & Generation-level cuts ($\ell=e^\pm,\mu^\pm,\tau^\pm$) & \multicolumn{1}{c|}{Cross section (fb) at $\sqrt{s}=$} \\ \cline{4-4}

&& (NA : Not Applied) & $14$ TeV \\ \hline
\multirow{11}{*}{$4\tau$}          
                                           & \multicolumn{1}{l|}{Signal ($gg\to h\to aa \to 4\tau$)} & NA  & $49.68\times 10^3$ \\\cline{2-4}
                                           
                                           & \multicolumn{1}{l|}{$gg\to h\to ZZ \to 4l$} & NA  & $10.46$ \\\cline{2-4}
                                           
                                           & \multicolumn{1}{l|}{$4l$} & {\makecell{ $p_{T,\ell}>3~\text{GeV}$,  $|\eta_{\ell}|<3.0$, $\Delta R_{\ell\ell}>0.1$}}  & $442.5$    \\\cline{2-4}
                                           
                                           & \multicolumn{1}{l|}{$4l2\nu$} & \multirow{1}{*}{$p_{T,\ell}>10~\text{GeV}$,  $|\eta_{\ell}|<3.0$, $\Delta R_{\ell\ell}>0.1$}  & $1.36$    \\\cline{2-4}
                                           
                                           & \multicolumn{1}{l|}{Signal ($pp\to (h\to aa \to 4\tau)jj$)} &{\makecell{$p_{T,j}>20~\text{GeV}$, $|\eta_{j}|<5.0$, $\Delta R_{jj}>0.2$, \\ $m_{jj}>500$ GeV}}  & $4.26\times 10^3$ \\\cline{2-4}

                                           & \multicolumn{1}{l|}{$4l2j$} &{\makecell{$p_{T,j}>20~\text{GeV}$, $p_{T,\ell}>3~\text{GeV}$, $|\eta_{j}|<5.0$,\\ $\Delta R_{jj}>0.2$, $\Delta R_{\ell\ell}>0.1$,\\ $m_{jj}>500$ GeV}}  & $6.2$ \\\cline{2-4}
                                      
                                           & \multicolumn{1}{l|}{$t\bar{t}h$ ($\geq 3\ell$ final states)} & \multicolumn{1}{l|}{\makecell{$p_{T,j/b}>15~\text{GeV}$, $p_{T,\ell}>3~\text{GeV}$, $|\eta_{j}|<4.0$,\\ $|\eta_{b/\ell}|<3.0$, $\Delta R_{j,b,\ell}>0.2$ except $\Delta R_{\ell\ell}>0.1$,\\ $m_{bb}>50$ GeV}}  & $12.4$  \\\cline{2-4}
                                                                                      
                                           & \multicolumn{1}{l|}{$t\bar{t}Z$, $Z\to \ell\ell$} & \multirow{4}{*}{same as $t\bar{t}h$}  & $29.2$    \\\cline{2-2}\cline{4-4}                                 
                                           
                                           & \multicolumn{1}{l|}{$4\ell 2b$} &   & $1.77$   \\\cline{2-2}\cline{4-4}
                                           
                                           & \multicolumn{1}{l|}{$t\bar{t}ZZ$ ($\geq 3\ell$ final states)} &    & $0.15$   \\\cline{2-2}\cline{4-4}
                                           
                                           & \multicolumn{1}{l|}{$t\bar{t}WW$ ($\geq 3\ell$ final states)} &    & $0.6$  \\\hline 
                                                                                      
\end{tabular}}
\end{bigcenter}
\caption{\it Generation level cuts and cross-sections for the various backgrounds used in the analyses. The branching ratio for $h\to aa\to 4\tau$ is assumed to be 0.1\%.}
\label{app1:1}
\end{table}


\begin{table}[htb!]
\centering
\begin{bigcenter}
\scalebox{0.8}{%
\begin{tabular}{|c|c|c|c|c|c|}

\hline
\multirow{2}{*}{Process} & \multirow{2}{*}{Signal and Backgrounds} & Generation-level cuts ($\ell=e^\pm,\mu^\pm,\tau^\pm$) & \multicolumn{2}{c|}{Cross section (fb) at $\sqrt{s}=$} \\ \cline{4-5}

&& (NA : Not Applied) & $14$ TeV & $100$ TeV \\ \hline
\multirow{6}{*}{$b \bar{b} 4\tau$}          
                                           & \multicolumn{1}{l|}{Signal ($pp\to hh\to b\bar{b}aa \to 2b4\tau$)} & NA  & $4.27$ & $142.6$ \\\cline{2-5}
                                           
                                           & \multicolumn{1}{l|}{$t\bar{t}h$ ($\geq 3\ell$ final states)} & \multicolumn{1}{l|}{\makecell{$p_{T,j/b}>15~\text{GeV}$, $p_{T,\ell}>3~\text{GeV}$, $|\eta_{j}|<4.0$,\\ $|\eta_{b/\ell}|<3.0$, $\Delta R_{j,b,\ell}>0.2$ except $\Delta R_{\ell\ell}>0.1$,\\ $m_{bb}>50$ GeV}}  & $12.4$ & $575.7$ \\\cline{2-5}
                                                                                      
                                           & \multicolumn{1}{l|}{$t\bar{t}Z$, $Z\to \ell\ell$} & \multirow{4}{*}{same as $t\bar{t}h$}  & $29.2$ & $1288$   \\\cline{2-2}\cline{4-5}                                 
                                           
                                           & \multicolumn{1}{l|}{$4\ell 2b$} &   & $1.77$ & $31.02$  \\\cline{2-2}\cline{4-5}
                                           
                                           & \multicolumn{1}{l|}{$t\bar{t}ZZ$ ($\geq 3\ell$ final states)} &    & $0.15$ & $7.93$  \\\cline{2-2}\cline{4-5}
                                           
                                           & \multicolumn{1}{l|}{$t\bar{t}WW$ ($\geq 3\ell$ final states)} &    & $0.6$ & $33.2$  \\\cline{2-5}
                                           
                                                                                      & \multicolumn{1}{l|}{$Zh,~Z\to b\bar{b},~h\to aa\to 4\tau$} & NA & $12.64$ & $169.01$  \\\hline 
\end{tabular}}
\end{bigcenter}
\caption{\it Generation level cuts and cross-sections for the various backgrounds used in the analyses. The branching ratio for $h\to aa\to 4\tau$ is assumed to be 10\%.}
\label{app1:2}
\end{table}

\section{Truth level transverse momentum distributions of $\tau$'s in VBF single Higgs and di-Higgs production}
\label{sec:appendixB}

\begin{figure}[htb!]
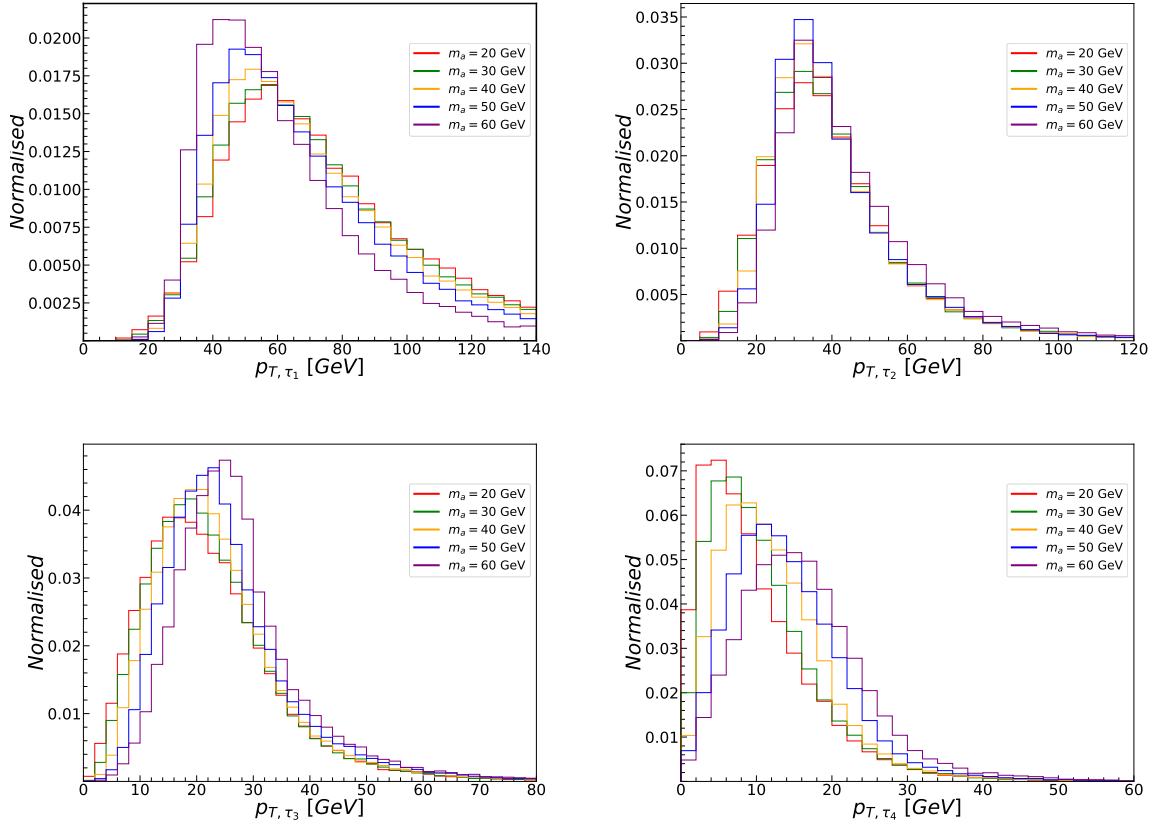

    \centering
    \includegraphics[width=0.5\textwidth]{./pt_tau1_VBF.pdf}~
    \includegraphics[width=0.5\textwidth]{./pt_tau2_VBF.pdf}\\
    \includegraphics[width=0.5\textwidth]{./pt_tau3_VBF.pdf}~
    \includegraphics[width=0.5\textwidth]{./pt_tau4_VBF.pdf}
    \caption{\it Distributions for the transverse momentum $p_T$ of the $\tau$ leptons at the truth level in the $VBF$ production channel  $pp \to (h \to aa \to 4\tau)jj$. Here we assume $\sqrt{s}=14~$TeV LHC.}
    \label{fig:pt_4tau_vbf}
\end{figure}

\begin{figure}[htb!]
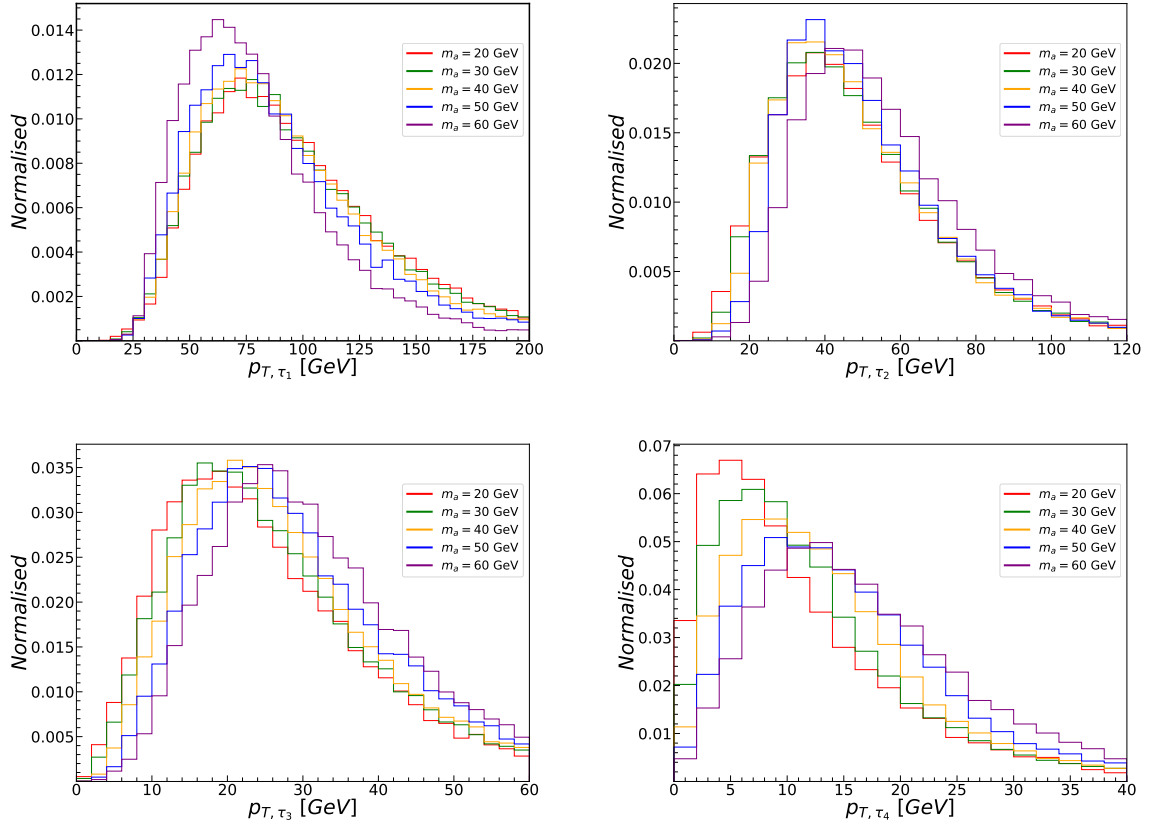

\centering
\includegraphics[width=0.5\textwidth]{./pt_tau1_hh_14.pdf}~
\includegraphics[width=0.5\textwidth]{./pt_tau2_hh_14.pdf}\\
\includegraphics[width=0.5\textwidth]{./pt_tau3_hh_14.pdf}~
\includegraphics[width=0.5\textwidth]{./pt_tau4_hh_14.pdf}
\caption{\it  Distributions for the transverse momentum $p_{T}$ for the four $\tau$ leptons in the non-resonant di-Higgs production channel $pp \to hh \to (h \to b\bar{b})(h \to aa \to 4\tau)$ at the truth level for several exotic scalar masses $m_a=~20,~30,~40,~50$ and $60$ GeV. The centre of mass energy is fixed at $\sqrt{s}=14$ TeV.}
\label{fig:2b4tau_parton}
\end{figure}

\pagebreak

\section{Normalised distribution of kinematic observables in non-resonant di-Higgs production at the HL-LHC and FCC-hh}
\label{sec:appendixC}

\begin{figure}[htb!]
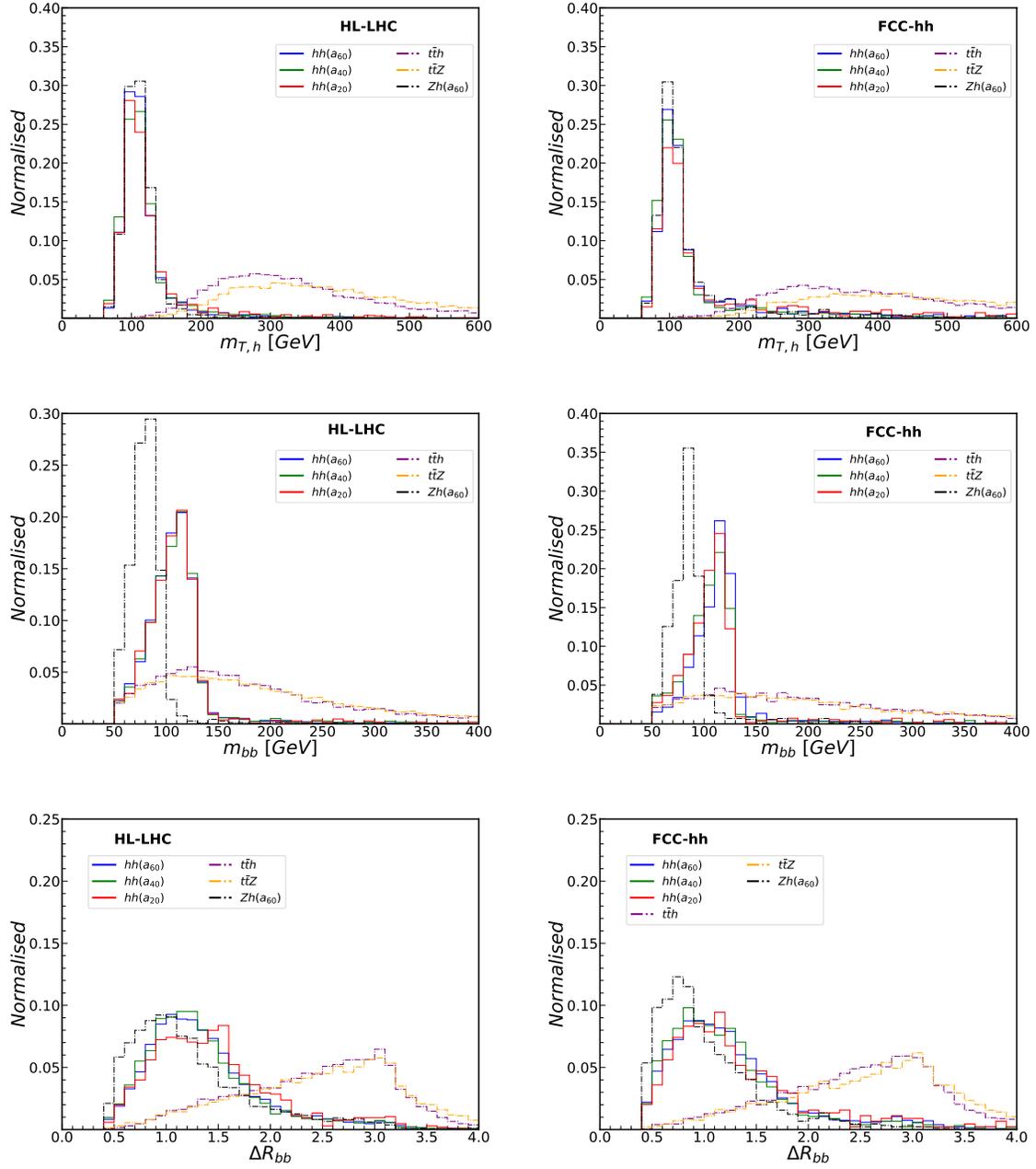

\centering
\includegraphics[width=0.5\textwidth]{./MTh_14_hh.pdf}\includegraphics[width=0.5\textwidth]{./MTh_100_hh.pdf}
\includegraphics[width=0.5\textwidth]{./mbb_14_hh.pdf}\includegraphics[width=0.5\textwidth]{./mbb_100_hh.pdf}
\includegraphics[width=0.5\textwidth]{./drbb_14_hh.pdf}\includegraphics[width=0.5\textwidth]{./drbb_100_hh.pdf}
\caption{\it Distributions of $m_{T,h}$, $m_{bb}$, and $\Delta R_{bb}$ for signal benchmarks $m_{a}=20,~40,~60~{\rm GeV}$, and dominant backgrounds, in the $pp \to hh \to (h \to b\bar{b})(h \to aa \to 4\tau)$ channel. The left and right panels show the distributions at $\sqrt{s}=14$ and $100$ TeV, respectively.}
\label{fig:BDT_obs_nonres}
\end{figure}

\pagebreak

\section{XGBoost analysis of non-resonant di-Higgs production at FCC-hh with jet radius $R$ = 0.4}
\label{sec:appendixD}

We list the results for the case where $R$ = 0.4 is used to reconstruct jets at the FCC-hh. Compared with Table \ref{tab:non-res_BDT_100TeV}, the signal efficiency in the case of R = 0.3 jets is 1.2-1.8 times more than in the case of R = 0.4 jets.

\begin{table}[htb!]
\centering
\scalebox{0.9}{%
\begin{tabular}{|c|c|c|c|c|c|}\hline
$\sqrt{s}$ & $m_a$ & Total Background & Signal Efficiency  & Signal & Significance \\

(TeV)      & (GeV) & Yield, B  & ($\times 10^{-4}$) & Yield, S  & (5$\%$ systematic) \\\hline

 \multirow{5}{*}{100} & 20 & $514$ & $2.2$ & $965$ & $25$ ($21$) \\ \cline{2-6}

 & 30 & $959$ & $6.7$ & $2866$  & $46$ ($37$) \\ \cline{2-6}
 
 & 40 & $993$ & $9.9$ & $4240$  & $59$ ($48$) \\ \cline{2-6}
 
 & 50 & $1046$ & $10$ & $4383$  & $59$ ($48$)  \\ \cline{2-6}

 & 60 & $1146$  & $12$ & $5108$ & $64$ ($52$) \\ \hline
 
 \end{tabular}}
\caption{\it  Signal and background yields, and signal significance, at the FCC-hh from XGBoost analysis in the $pp \to hh \to (h \to b\bar{b})(h \to aa \to 4\tau)$ channel. The results shown here have been derived for Br($h\to aa\to 4\tau)\sim 10\%$.}
\label{tab:non-res_BDT_100TeV_R04}
\end{table}

\newpage 


\begin{thebibliography}{100}

\bibitem{ATLAS-CONF-2020-027}
{\scshape ATLAS Collaboration} collaboration, \emph{{A combination of
  measurements of Higgs boson production and decay using up to $139$ fb$^{-1}$
  of proton--proton collision data at $\sqrt{s}=$ 13 TeV collected with the
  ATLAS experiment}},  Technical Report ATLAS-CONF-2020-027, CERN, Geneva, Aug, 2020.

\bibitem{Curtin:2013fra}
D.~Curtin et~al., \emph{{Exotic decays of the 125 GeV Higgs boson}},
  \href{http://dx.doi.org/10.1103/PhysRevD.90.075004}{\emph{Phys. Rev. D} {\bf
  90} (2014) 075004}, [\href{https://arxiv.org/abs/1312.4992}{{\tt
  1312.4992}}].

\bibitem{Cepeda:2021rql}
M.~Cepeda, S.~Gori, V.~M. Outschoorn and J.~Shelton, \emph{{Exotic Higgs
  Decays}},  \href{https://arxiv.org/abs/2111.12751}{{\tt 2111.12751}}.

\bibitem{Silveira:1985rk}
V.~Silveira and A.~Zee, \emph{{SCALAR PHANTOMS}},
  \href{http://dx.doi.org/10.1016/0370-2693(85)90624-0}{\emph{Phys. Lett. B}
  {\bf 161} (1985) 136--140}.

\bibitem{Burgess:2000yq}
C.~P. Burgess, M.~Pospelov and T.~ter Veldhuis, \emph{{The Minimal model of
  nonbaryonic dark matter: A Singlet scalar}},
  \href{http://dx.doi.org/10.1016/S0550-3213(01)00513-2}{\emph{Nucl. Phys. B}
  {\bf 619} (2001) 709--728}, [\href{https://arxiv.org/abs/hep-ph/0011335}{{\tt
  hep-ph/0011335}}].

\bibitem{Draper:2010ew}
P.~Draper, T.~Liu, C.~E.~M. Wagner, L.-T. Wang and H.~Zhang, \emph{{Dark Light
  Higgs}}, \href{http://dx.doi.org/10.1103/PhysRevLett.106.121805}{\emph{Phys.
  Rev. Lett.} {\bf 106} (2011) 121805},
  [\href{https://arxiv.org/abs/1009.3963}{{\tt 1009.3963}}].

\bibitem{Englert:2011yb}
C.~Englert, T.~Plehn, D.~Zerwas and P.~M. Zerwas, \emph{{Exploring the Higgs
  portal}}, \href{http://dx.doi.org/10.1016/j.physletb.2011.08.002}{\emph{Phys.
  Lett. B} {\bf 703} (2011) 298--305},
  [\href{https://arxiv.org/abs/1106.3097}{{\tt 1106.3097}}].

\bibitem{Bhattacherjee:2013jca}
B.~Bhattacherjee, S.~Matsumoto, S.~Mukhopadhyay and M.~M. Nojiri,
  \emph{{Phenomenology of light fermionic asymmetric dark matter}},
  \href{http://dx.doi.org/10.1007/JHEP10(2013)032}{\emph{JHEP} {\bf 10} (2013)
  032}, [\href{https://arxiv.org/abs/1306.5878}{{\tt 1306.5878}}].

\bibitem{Robens_2015}
T.~Robens and T.~Stefaniak, \emph{Status of the higgs singlet extension of the
  standard model after lhc run 1},
  \href{http://dx.doi.org/10.1140/epjc/s10052-015-3323-y}{\emph{The European
  Physical Journal C} {\bf 75} (Mar, 2015) }.

\bibitem{Robens_2016}
T.~Robens and T.~Stefaniak, \emph{Lhc benchmark scenarios for the real higgs
  singlet extension of the standard model},
  \href{http://dx.doi.org/10.1140/epjc/s10052-016-4115-8}{\emph{The European
  Physical Journal C} {\bf 76} (May, 2016) }.

\bibitem{Robens_2020}
T.~Robens, T.~Stefaniak and J.~Wittbrodt, \emph{Two-real-scalar-singlet
  extension of the sm: Lhc phenomenology and benchmark scenarios},
  \href{http://dx.doi.org/10.1140/epjc/s10052-020-7655-x}{\emph{The European
  Physical Journal C} {\bf 80} (Feb, 2020) }.

\bibitem{Bauer_2017}
M.~Bauer, M.~Neubert and A.~Thamm, \emph{Collider probes of axion-like
  particles}, \href{http://dx.doi.org/10.1007/jhep12(2017)044}{\emph{Journal of
  High Energy Physics} {\bf 2017} (Dec, 2017) }.

\bibitem{Alves:2021puo}
A.~Alves, A.~G. Dias and D.~D. Lopes, \emph{{Jets and photons spectroscopy of
  Higgs-ALP interactions}},
  \href{http://dx.doi.org/10.1007/JHEP10(2021)012}{\emph{JHEP} {\bf 10} (2021)
  012}, [\href{https://arxiv.org/abs/2105.01095}{{\tt 2105.01095}}].

\bibitem{Ellwanger_2010}
U.~Ellwanger, C.~Hugonie and A.~M. Teixeira, \emph{The next-to-minimal
  supersymmetric standard model},
  \href{http://dx.doi.org/10.1016/j.physrep.2010.07.001}{\emph{Physics Reports}
  {\bf 496} (Nov, 2010) 1–77}.

\bibitem{NILLES1983346}
H.~Nilles, M.~Srednicki and D.~Wyler, \emph{Weak interaction breakdown induced
  by supergravity},
  \href{http://dx.doi.org/https://doi.org/10.1016/0370-2693(83)90460-4}{\emph{Physics
  Letters B} {\bf 120} (1983) 346--348}.

\bibitem{PhysRevD.39.844}
J.~Ellis, J.~F. Gunion, H.~E. Haber, L.~Roszkowski and F.~Zwirner, \emph{Higgs
  bosons in a nonminimal supersymmetric model},
  \href{http://dx.doi.org/10.1103/PhysRevD.39.844}{\emph{Phys. Rev. D} {\bf 39}
  (Feb, 1989) 844--869}.

\bibitem{ATLAS:2018pvw}
{\scshape ATLAS} collaboration, M.~Aaboud et~al., \emph{{Search for the Higgs
  boson produced in association with a vector boson and decaying into two
  spin-zero particles in the $H \rightarrow aa \rightarrow 4b$ channel in $pp$
  collisions at $\sqrt{s} = 13$ TeV with the ATLAS detector}},
  \href{http://dx.doi.org/10.1007/JHEP10(2018)031}{\emph{JHEP} {\bf 10} (2018)
  031}, [\href{https://arxiv.org/abs/1806.07355}{{\tt 1806.07355}}].

\bibitem{ATLAS:2020ahi}
{\scshape ATLAS} collaboration, G.~Aad et~al., \emph{{Search for Higgs boson
  decays into two new low-mass spin-0 particles in the 4$b$ channel with the
  ATLAS detector using $pp$ collisions at $\sqrt{s}= 13$ TeV}},
  \href{http://dx.doi.org/10.1103/PhysRevD.102.112006}{\emph{Phys. Rev. D} {\bf
  102} (2020) 112006}, [\href{https://arxiv.org/abs/2005.12236}{{\tt
  2005.12236}}].

\bibitem{Sirunyan_2018}
A.~Sirunyan, A.~Tumasyan, W.~Adam, F.~Ambrogi, E.~Asilar, T.~Bergauer et~al.,
  \emph{Search for an exotic decay of the higgs boson to a pair of light
  pseudoscalars in the final state with two b quarks and two $\tau$ leptons in
  proton–proton collisions at $\sqrt{s} = 13$ tev},
  \href{http://dx.doi.org/10.1016/j.physletb.2018.08.057}{\emph{Physics Letters
  B} {\bf 785} (Oct, 2018) 462–488}.

\bibitem{ATLAS:2018emt}
{\scshape ATLAS} collaboration, M.~Aaboud et~al., \emph{{Search for Higgs boson
  decays into a pair of light bosons in the $bb\mu\mu$ final state in $pp$
  collision at $\sqrt{s} = $13 TeV with the ATLAS detector}},
  \href{http://dx.doi.org/10.1016/j.physletb.2018.10.073}{\emph{Phys. Lett. B}
  {\bf 790} (2019) 1--21}, [\href{https://arxiv.org/abs/1807.00539}{{\tt
  1807.00539}}].

\bibitem{CMS:2018nsh}
{\scshape CMS} collaboration, A.~M. Sirunyan et~al., \emph{{Search for an
  exotic decay of the Higgs boson to a pair of light pseudoscalars in the final
  state with two muons and two b quarks in pp collisions at 13 TeV}},
  \href{http://dx.doi.org/10.1016/j.physletb.2019.06.021}{\emph{Phys. Lett. B}
  {\bf 795} (2019) 398--423}, [\href{https://arxiv.org/abs/1812.06359}{{\tt
  1812.06359}}].

\bibitem{ATLAS:2021hbr}
{\scshape ATLAS} collaboration, G.~Aad et~al., \emph{{Search for Higgs boson
  decays into a pair of pseudoscalar particles in the $bb\mu\mu$ final state
  with the ATLAS detector in $pp$ collisions at $\sqrt{s}=13$ TeV}},
  \href{https://arxiv.org/abs/2110.00313}{{\tt 2110.00313}}.

\bibitem{Sirunyan_2019}
A.~Sirunyan, A.~Tumasyan, W.~Adam, F.~Ambrogi, E.~Asilar, T.~Bergauer et~al.,
  \emph{A search for pair production of new light bosons decaying into muons in
  proton-proton collisions at 13 tev},
  \href{http://dx.doi.org/10.1016/j.physletb.2019.07.013}{\emph{Physics Letters
  B} {\bf 796} (Sep, 2019) 131–154}.

\bibitem{ATLAS:2018coo}
{\scshape ATLAS} collaboration, M.~Aaboud et~al., \emph{{Search for Higgs boson
  decays to beyond-the-Standard-Model light bosons in four-lepton events with
  the ATLAS detector at $\sqrt{s}=13$ TeV}},
  \href{http://dx.doi.org/10.1007/JHEP06(2018)166}{\emph{JHEP} {\bf 06} (2018)
  166}, [\href{https://arxiv.org/abs/1802.03388}{{\tt 1802.03388}}].

\bibitem{Sirunyan_2020}
A.~M. Sirunyan, A.~Tumasyan, W.~Adam, F.~Ambrogi, T.~Bergauer, M.~Dragicevic
  et~al., \emph{Search for a light pseudoscalar higgs boson in the boosted
  $\mu\mu\tau\tau$ final state in proton-proton collisions at $\sqrt{s} = 13$
  tev}, \href{http://dx.doi.org/10.1007/jhep08(2020)139}{\emph{Journal of High
  Energy Physics} {\bf 2020} (Aug, 2020) }.

\bibitem{Sirunyan_2020_2}
A.~Sirunyan, A.~Tumasyan, W.~Adam, F.~Ambrogi, E.~Asilar, T.~Bergauer et~al.,
  \emph{Search for light pseudoscalar boson pairs produced from decays of the
  125 gev higgs boson in final states with two muons and two nearby tracks in
  pp collisions at $\sqrt{s} = $13 tev},
  \href{http://dx.doi.org/10.1016/j.physletb.2019.135087}{\emph{Physics Letters
  B} {\bf 800} (Jan, 2020) 135087}.

\bibitem{CMS:2018qvj}
{\scshape CMS} collaboration, A.~M. Sirunyan et~al., \emph{{Search for an
  exotic decay of the Higgs boson to a pair of light pseudoscalars in the final
  state of two muons and two $\tau$ leptons in proton-proton collisions at $
  \sqrt{s}=13 $ TeV}},
  \href{http://dx.doi.org/10.1007/JHEP11(2018)018}{\emph{JHEP} {\bf 11} (2018)
  018}, [\href{https://arxiv.org/abs/1805.04865}{{\tt 1805.04865}}].

\bibitem{CMS:2017dmg}
{\scshape CMS} collaboration, V.~Khachatryan et~al., \emph{{Search for light
  bosons in decays of the 125 GeV Higgs boson in proton-proton collisions at $
  \sqrt{s}=8 $ TeV}},
  \href{http://dx.doi.org/10.1007/JHEP10(2017)076}{\emph{JHEP} {\bf 10} (2017)
  076}, [\href{https://arxiv.org/abs/1701.02032}{{\tt 1701.02032}}].

\bibitem{ATLAS:2015unc}
{\scshape ATLAS} collaboration, G.~Aad et~al., \emph{{Search for Higgs bosons
  decaying to $aa$ in the $\mu\mu\tau\tau$ final state in $pp$ collisions at
  $\sqrt{s} = $ 8 TeV with the ATLAS experiment}},
  \href{http://dx.doi.org/10.1103/PhysRevD.92.052002}{\emph{Phys. Rev. D} {\bf
  92} (2015) 052002}, [\href{https://arxiv.org/abs/1505.01609}{{\tt
  1505.01609}}].

\bibitem{ATLAS:2015rsn}
{\scshape ATLAS} collaboration, G.~Aad et~al., \emph{{Search for new phenomena
  in events with at least three photons collected in $pp$ collisions at
  $\sqrt{s}$ = 8 TeV with the ATLAS detector}},
  \href{http://dx.doi.org/10.1140/epjc/s10052-016-4034-8}{\emph{Eur. Phys. J.
  C} {\bf 76} (2016) 210}, [\href{https://arxiv.org/abs/1509.05051}{{\tt
  1509.05051}}].

\bibitem{ATLAS:2018jnf}
{\scshape ATLAS} collaboration, M.~Aaboud et~al., \emph{{Search for Higgs boson
  decays into pairs of light (pseudo)scalar particles in the $\gamma\gamma jj$
  final state in $pp$ collisions at $\sqrt{s}=13$ TeV with the ATLAS
  detector}},
  \href{http://dx.doi.org/10.1016/j.physletb.2018.06.011}{\emph{Phys. Lett. B}
  {\bf 782} (2018) 750--767}, [\href{https://arxiv.org/abs/1803.11145}{{\tt
  1803.11145}}].

\bibitem{Cai:2020lao}
H.~Cai, \emph{{Search for exotic Higgs boson decays to four leptons with the
  ATLAS detector}}.
\newblock PhD thesis, Illinois U., Urbana, 2020.

\bibitem{CMS:2012pxt}
{\scshape CMS} collaboration, \emph{{A Search for Anomalous Production of
  Events with three or more leptons using 9.2 fb $^{-1}$ of $\sqrt{s}$ = 8 TeV
  CMS data}}, Technical Report CMS-PAS-SUS-12-026, CERN, Geneva, 2012.

\bibitem{CMS-PAS-SUS-13-010}
{\scshape CMS Collaboration} collaboration, \emph{{Search for RPV SUSY in the
  four-lepton final state}},  Technical Report CMS-PAS-SUS-13-010, CERN, Geneva, 2013.

\bibitem{Banerjee:2016nzb}
S.~Banerjee, B.~Batell and M.~Spannowsky, \emph{{Invisible decays in Higgs
  boson pair production}},
  \href{http://dx.doi.org/10.1103/PhysRevD.95.035009}{\emph{Phys. Rev.} {\bf
  D95} (2017) 035009}, [\href{https://arxiv.org/abs/1608.08601}{{\tt
  1608.08601}}].

\bibitem{Arganda:2017wjh}
E.~Arganda, J.~L. D\'\i{}az-Cruz, N.~Mileo, R.~A. Morales and A.~Szynkman,
  \emph{{Search strategies for pair production of heavy Higgs bosons decaying
  invisibly at the LHC}},
  \href{http://dx.doi.org/10.1016/j.nuclphysb.2018.02.004}{\emph{Nucl. Phys. B}
  {\bf 929} (2018) 171--192}, [\href{https://arxiv.org/abs/1710.07254}{{\tt
  1710.07254}}].

\bibitem{Alves:2019emf}
A.~Alves, T.~Ghosh and F.~S. Queiroz, \emph{{Dark and bright signatures of
  di-Higgs boson production}},
  \href{http://dx.doi.org/10.1103/PhysRevD.100.036012}{\emph{Phys. Rev. D} {\bf
  100} (2019) 036012}, [\href{https://arxiv.org/abs/1905.03271}{{\tt
  1905.03271}}].

\bibitem{FCC:2018byv}
{\scshape FCC} collaboration, A.~Abada et~al., \emph{{FCC Physics
  Opportunities}: {Future Circular Collider Conceptual Design Report Volume
  1}}, \href{http://dx.doi.org/10.1140/epjc/s10052-019-6904-3}{\emph{Eur. Phys.
  J. C} {\bf 79} (2019) 474}.

\bibitem{Dawson:1998py}
S.~Dawson, S.~Dittmaier and M.~Spira, \emph{{Neutral Higgs boson pair
  production at hadron colliders: QCD corrections}},
  \href{http://dx.doi.org/10.1103/PhysRevD.58.115012}{\emph{Phys. Rev.} {\bf
  D58} (1998) 115012}, [\href{https://arxiv.org/abs/hep-ph/9805244}{{\tt
  hep-ph/9805244}}].

\bibitem{Borowka:2016ehy}
S.~Borowka, N.~Greiner, G.~Heinrich, S.~Jones, M.~Kerner, J.~Schlenk et~al.,
  \emph{{Higgs Boson Pair Production in Gluon Fusion at Next-to-Leading Order
  with Full Top-Quark Mass Dependence}},
  \href{http://dx.doi.org/10.1103/PhysRevLett.117.079901}{\emph{Phys. Rev.
  Lett.} {\bf 117} (2016) 012001},
  [\href{https://arxiv.org/abs/1604.06447}{{\tt 1604.06447}}].

\bibitem{Baglio:2018lrj}
J.~Baglio, F.~Campanario, S.~Glaus, M.~Mühlleitner, M.~Spira and J.~Streicher,
  \emph{{Gluon fusion into Higgs pairs at NLO QCD and the top mass scheme}},
  \href{http://dx.doi.org/10.1140/epjc/s10052-019-6973-3}{\emph{Eur. Phys. J.
  C} {\bf 79} (2019) 459}, [\href{https://arxiv.org/abs/1811.05692}{{\tt
  1811.05692}}].

\bibitem{deFlorian:2013jea}
D.~de~Florian and J.~Mazzitelli, \emph{{Higgs Boson Pair Production at
  Next-to-Next-to-Leading Order in QCD}},
  \href{http://dx.doi.org/10.1103/PhysRevLett.111.201801}{\emph{Phys. Rev.
  Lett.} {\bf 111} (2013) 201801}, [\href{https://arxiv.org/abs/1309.6594}{{\tt
  1309.6594}}].

\bibitem{Shao:2013bz}
D.~Y. Shao, C.~S. Li, H.~T. Li and J.~Wang, \emph{{Threshold resummation
  effects in Higgs boson pair production at the LHC}},
  \href{http://dx.doi.org/10.1007/JHEP07(2013)169}{\emph{JHEP} {\bf 07} (2013)
  169}, [\href{https://arxiv.org/abs/1301.1245}{{\tt 1301.1245}}].

\bibitem{deFlorian:2015moa}
D.~de~Florian and J.~Mazzitelli, \emph{{Higgs pair production at
  next-to-next-to-leading logarithmic accuracy at the LHC}},
  \href{http://dx.doi.org/10.1007/JHEP09(2015)053}{\emph{JHEP} {\bf 09} (2015)
  053}, [\href{https://arxiv.org/abs/1505.07122}{{\tt 1505.07122}}].

\bibitem{Grazzini:2018bsd}
M.~Grazzini, G.~Heinrich, S.~Jones, S.~Kallweit, M.~Kerner, J.~M. Lindert
  et~al., \emph{{Higgs boson pair production at NNLO with top quark mass
  effects}}, \href{http://dx.doi.org/10.1007/JHEP05(2018)059}{\emph{JHEP} {\bf
  05} (2018) 059}, [\href{https://arxiv.org/abs/1803.02463}{{\tt 1803.02463}}].

\bibitem{Baglio:2020ini}
J.~Baglio, F.~Campanario, S.~Glaus, M.~Mühlleitner, J.~Ronca, M.~Spira et~al.,
  \emph{{Higgs-Pair Production via Gluon Fusion at Hadron Colliders: NLO QCD
  Corrections}}, \href{http://dx.doi.org/10.1007/JHEP04(2020)181}{\emph{JHEP}
  {\bf 04} (2020) 181}, [\href{https://arxiv.org/abs/2003.03227}{{\tt
  2003.03227}}].

\bibitem{Aaboud:2018knk}
{\scshape ATLAS} collaboration, M.~Aaboud et~al., \emph{{Search for pair
  production of Higgs bosons in the $b\bar{b}b\bar{b}$ final state using
  proton-proton collisions at $\sqrt{s} = 13$ TeV with the ATLAS detector}},
  \href{http://dx.doi.org/10.1007/JHEP01(2019)030}{\emph{JHEP} {\bf 01} (2019)
  030}, [\href{https://arxiv.org/abs/1804.06174}{{\tt 1804.06174}}].

\bibitem{Aaboud:2018sfw}
{\scshape ATLAS} collaboration, M.~Aaboud et~al., \emph{{Search for resonant
  and non-resonant Higgs boson pair production in the ${b\bar{b}\tau^+\tau^-}$
  decay channel in $pp$ collisions at $\sqrt{s}=13$ TeV with the ATLAS
  detector}},
  \href{http://dx.doi.org/10.1103/PhysRevLett.121.191801}{\emph{Phys. Rev.
  Lett.} {\bf 121} (2018) 191801},
  [\href{https://arxiv.org/abs/1808.00336}{{\tt 1808.00336}}].

\bibitem{Sirunyan:2017djm}
{\scshape CMS} collaboration, A.~M. Sirunyan et~al., \emph{{Search for Higgs
  boson pair production in events with two bottom quarks and two tau leptons in
  proton--proton collisions at $\sqrt s$ =13TeV}},
  \href{http://dx.doi.org/10.1016/j.physletb.2018.01.001}{\emph{Phys. Lett. B}
  {\bf 778} (2018) 101--127}, [\href{https://arxiv.org/abs/1707.02909}{{\tt
  1707.02909}}].

\bibitem{Aaboud:2018ftw}
{\scshape ATLAS} collaboration, M.~Aaboud et~al., \emph{{Search for Higgs boson
  pair production in the $\gamma\gamma b\bar{b}$ final state with 13 TeV $pp$
  collision data collected by the ATLAS experiment}},
  \href{http://dx.doi.org/10.1007/JHEP11(2018)040}{\emph{JHEP} {\bf 11} (2018)
  040}, [\href{https://arxiv.org/abs/1807.04873}{{\tt 1807.04873}}].

\bibitem{CMS:2017ihs}
{\scshape CMS} collaboration, \emph{{Search for Higgs boson pair production in
  the final state containing two photons and two bottom quarks in proton-proton
  collisions at $\sqrt{s}=13~\mathrm{TeV}$}}, .

\bibitem{Aaboud:2018zhh}
{\scshape ATLAS} collaboration, M.~Aaboud et~al., \emph{{Search for Higgs boson
  pair production in the $b\bar{b}WW^{*}$ decay mode at $\sqrt{s}=13$ TeV with
  the ATLAS detector}},
  \href{http://dx.doi.org/10.1007/JHEP04(2019)092}{\emph{JHEP} {\bf 04} (2019)
  092}, [\href{https://arxiv.org/abs/1811.04671}{{\tt 1811.04671}}].

\bibitem{Aaboud:2018ewm}
{\scshape ATLAS} collaboration, M.~Aaboud et~al., \emph{{Search for Higgs boson
  pair production in the $\gamma\gamma WW^{*}$ channel using $pp$ collision
  data recorded at $\sqrt{s} = 13$ TeV with the ATLAS detector}},
  \href{http://dx.doi.org/10.1140/epjc/s10052-018-6457-x}{\emph{Eur. Phys. J.
  C} {\bf 78} (2018) 1007}, [\href{https://arxiv.org/abs/1807.08567}{{\tt
  1807.08567}}].

\bibitem{Aaboud:2018ksn}
{\scshape ATLAS} collaboration, M.~Aaboud et~al., \emph{{Search for Higgs boson
  pair production in the $WW^{(*)}WW^{(*)}$ decay channel using ATLAS data
  recorded at $\sqrt{s}=13$ TeV}},
  \href{http://dx.doi.org/10.1007/JHEP05(2019)124}{\emph{JHEP} {\bf 05} (2019)
  124}, [\href{https://arxiv.org/abs/1811.11028}{{\tt 1811.11028}}].

\bibitem{deFlorian:2227475}
{\scshape LHCHiggsCrossSectionWorkingGroup} collaboration, D.~de~Florian,
  C.~Grojean, F.~Maltoni, C.~Mariotti, A.~Nikitenko, M.~Pieri et~al.,
  \emph{{Handbook of LHC Higgs Cross Sections: 4. Deciphering the Nature of the
  Higgs Sector}}.
\newblock CERN Yellow Reports: Monographs. CERN, Geneva, 2017,
  \href{http://dx.doi.org/10.23731/CYRM-2017-002}{10.23731/CYRM-2017-002}.

\bibitem{Durieux:2022hbu}
G.~Durieux, G.~Durieux, M.~McCullough, M.~McCullough, E.~Salvioni and
  E.~Salvioni, \emph{{Charting the Higgs self-coupling boundaries}},
  \href{http://dx.doi.org/10.1007/JHEP12(2022)148}{\emph{JHEP} {\bf 12} (2022)
  148}, [\href{https://arxiv.org/abs/2209.00666}{{\tt 2209.00666}}].

\bibitem{CMS:2022dwd}
{\scshape CMS} collaboration, A.~Tumasyan et~al., \emph{{A portrait of the
  Higgs boson by the CMS experiment ten years after the discovery.}},
  \href{http://dx.doi.org/10.1038/s41586-022-04892-x}{\emph{Nature} {\bf 607}
  (2022) 60--68}, [\href{https://arxiv.org/abs/2207.00043}{{\tt 2207.00043}}].

\bibitem{ATLAS:2022jtk}
{\scshape ATLAS} collaboration, G.~Aad et~al., \emph{{Constraints on the Higgs
  boson self-coupling from single- and double-Higgs production with the ATLAS
  detector using pp collisions at s=13 TeV}},
  \href{http://dx.doi.org/10.1016/j.physletb.2023.137745}{\emph{Phys. Lett. B}
  {\bf 843} (2023) 137745}, [\href{https://arxiv.org/abs/2211.01216}{{\tt
  2211.01216}}].

\bibitem{Dolan:2012rv}
M.~J. Dolan, C.~Englert and M.~Spannowsky, \emph{{Higgs self-coupling
  measurements at the LHC}},
  \href{http://dx.doi.org/10.1007/JHEP10(2012)112}{\emph{JHEP} {\bf 10} (2012)
  112}, [\href{https://arxiv.org/abs/1206.5001}{{\tt 1206.5001}}].

\bibitem{Kim:2018cxf}
J.~H. Kim, K.~Kong, K.~T. Matchev and M.~Park, \emph{{Probing the Triple Higgs
  Self-Interaction at the Large Hadron Collider}},
  \href{http://dx.doi.org/10.1103/PhysRevLett.122.091801}{\emph{Phys. Rev.
  Lett.} {\bf 122} (2019) 091801},
  [\href{https://arxiv.org/abs/1807.11498}{{\tt 1807.11498}}].

\bibitem{Kim:2019wns}
J.~H. Kim, M.~Kim, K.~Kong, K.~T. Matchev and M.~Park, \emph{{Portraying Double
  Higgs at the Large Hadron Collider}},
  \href{http://dx.doi.org/10.1007/JHEP09(2019)047}{\emph{JHEP} {\bf 09} (2019)
  047}, [\href{https://arxiv.org/abs/1904.08549}{{\tt 1904.08549}}].

\bibitem{Barr:2013tda}
A.~J. Barr, M.~J. Dolan, C.~Englert and M.~Spannowsky, \emph{{Di-Higgs final
  states augMT2ed -- selecting $hh$ events at the high luminosity LHC}},
  \href{http://dx.doi.org/10.1016/j.physletb.2013.12.011}{\emph{Phys. Lett.}
  {\bf B728} (2014) 308--313}, [\href{https://arxiv.org/abs/1309.6318}{{\tt
  1309.6318}}].

\bibitem{Barger:2013jfa}
V.~Barger, L.~L. Everett, C.~B. Jackson and G.~Shaughnessy, \emph{{Higgs-Pair
  Production and Measurement of the Triscalar Coupling at LHC(8,14)}},
  \href{http://dx.doi.org/10.1016/j.physletb.2013.12.013}{\emph{Phys. Lett.}
  {\bf B728} (2014) 433--436}, [\href{https://arxiv.org/abs/1311.2931}{{\tt
  1311.2931}}].

\bibitem{Kling:2016lay}
F.~Kling, T.~Plehn and P.~Schichtel, \emph{{Maximizing the significance in
  Higgs boson pair analyses}},
  \href{http://dx.doi.org/10.1103/PhysRevD.95.035026}{\emph{Phys. Rev.} {\bf
  D95} (2017) 035026}, [\href{https://arxiv.org/abs/1607.07441}{{\tt
  1607.07441}}].

\bibitem{Alves:2017ued}
A.~Alves, T.~Ghosh and K.~Sinha, \emph{{Can We Discover Double Higgs Production
  at the LHC?}},
  \href{http://dx.doi.org/10.1103/PhysRevD.96.035022}{\emph{Phys. Rev.} {\bf
  D96} (2017) 035022}, [\href{https://arxiv.org/abs/1704.07395}{{\tt
  1704.07395}}].

\bibitem{Adhikary:2017jtu}
A.~Adhikary, S.~Banerjee, R.~K. Barman, B.~Bhattacherjee and S.~Niyogi,
  \emph{{Revisiting the non-resonant Higgs pair production at the HL-LHC}},
  \href{http://dx.doi.org/10.1007/JHEP07(2018)116}{\emph{JHEP} {\bf 07} (2018)
  116}, [\href{https://arxiv.org/abs/1712.05346}{{\tt 1712.05346}}].

\bibitem{Amacker:2020bmn}
J.~Amacker et~al., \emph{{Higgs self-coupling measurements using deep learning
  and jet substructure in the $b\bar{b}b\bar{b}$ final state}},
  \href{https://arxiv.org/abs/2004.04240}{{\tt 2004.04240}}.

\bibitem{Abdughani:2020xfo}
M.~Abdughani, D.~Wang, L.~Wu, J.~M. Yang and J.~Zhao, \emph{{Probing triple
  Higgs coupling with machine learning at the LHC}},
  \href{https://arxiv.org/abs/2005.11086}{{\tt 2005.11086}}.

\bibitem{Heinrich:2019bkc}
G.~Heinrich, S.~Jones, M.~Kerner, G.~Luisoni and L.~Scyboz, \emph{{Probing the
  trilinear Higgs boson coupling in di-Higgs production at NLO QCD including
  parton shower effects}},
  \href{http://dx.doi.org/10.1007/JHEP06(2019)066}{\emph{JHEP} {\bf 06} (2019)
  066}, [\href{https://arxiv.org/abs/1903.08137}{{\tt 1903.08137}}].

\bibitem{Arganda:2018ftn}
E.~Arganda, C.~Garcia-Garcia and M.~J. Herrero, \emph{{Probing the Higgs
  self-coupling through double Higgs production in vector boson scattering at
  the LHC}},
  \href{http://dx.doi.org/10.1016/j.nuclphysb.2019.114687}{\emph{Nucl. Phys. B}
  {\bf 945} (2019) 114687}, [\href{https://arxiv.org/abs/1807.09736}{{\tt
  1807.09736}}].

\bibitem{Chang:2018uwu}
J.~Chang, K.~Cheung, J.~S. Lee, C.-T. Lu and J.~Park, \emph{{Higgs-boson-pair
  production H($\rightarrow$b$\bar{b}$)H($\rightarrow\gamma\gamma$) from gluon
  fusion at the HL-LHC and HL-100 TeV hadron collider}},
  \href{http://dx.doi.org/10.1103/PhysRevD.100.096001}{\emph{Phys. Rev. D} {\bf
  100} (2019) 096001}, [\href{https://arxiv.org/abs/1804.07130}{{\tt
  1804.07130}}].

\bibitem{Cao:2015oxx}
Q.-H. Cao, Y.~Liu and B.~Yan, \emph{{Measuring trilinear Higgs coupling in WHH
  and ZHH productions at the high-luminosity LHC}},
  \href{http://dx.doi.org/10.1103/PhysRevD.95.073006}{\emph{Phys. Rev.} {\bf
  D95} (2017) 073006}, [\href{https://arxiv.org/abs/1511.03311}{{\tt
  1511.03311}}].

\bibitem{Mangano:2020sao}
M.~L. Mangano, G.~Ortona and M.~Selvaggi, \emph{{Measuring the Higgs
  self-coupling via Higgs-pair production at a 100 TeV p-p collider}},
  \href{https://arxiv.org/abs/2004.03505}{{\tt 2004.03505}}.

\bibitem{Banerjee:2019jys}
S.~Banerjee, F.~Krauss and M.~Spannowsky, \emph{{Revisiting the $t\bar{t}hh$
  channel at the FCC-hh}},
  \href{http://dx.doi.org/10.1103/PhysRevD.100.073012}{\emph{Phys. Rev. D} {\bf
  100} (2019) 073012}, [\href{https://arxiv.org/abs/1904.07886}{{\tt
  1904.07886}}].

\bibitem{Banerjee:2018yxy}
S.~Banerjee, C.~Englert, M.~L. Mangano, M.~Selvaggi and M.~Spannowsky,
  \emph{{$hh+\text{jet}$ production at 100 TeV}},
  \href{http://dx.doi.org/10.1140/epjc/s10052-018-5788-y}{\emph{Eur. Phys. J.
  C} {\bf 78} (2018) 322}, [\href{https://arxiv.org/abs/1802.01607}{{\tt
  1802.01607}}].

\bibitem{Bizon:2018syu}
W.~Bizo\'n, U.~Haisch and L.~Rottoli, \emph{{Constraints on the quartic Higgs
  self-coupling from double-Higgs production at future hadron colliders}},
  \href{http://dx.doi.org/10.1007/JHEP10(2019)267}{\emph{JHEP} {\bf 10} (2019)
  267}, [\href{https://arxiv.org/abs/1810.04665}{{\tt 1810.04665}}].

\bibitem{Goncalves:2018qas}
D.~Gonçalves, T.~Han, F.~Kling, T.~Plehn and M.~Takeuchi, \emph{{Higgs boson
  pair production at future hadron colliders: From kinematics to dynamics}},
  \href{http://dx.doi.org/10.1103/PhysRevD.97.113004}{\emph{Phys. Rev. D} {\bf
  97} (2018) 113004}, [\href{https://arxiv.org/abs/1802.04319}{{\tt
  1802.04319}}].

\bibitem{Barr:2014sga}
A.~J. Barr, M.~J. Dolan, C.~Englert, D.~E. Ferreira~de Lima and M.~Spannowsky,
  \emph{{Higgs Self-Coupling Measurements at a 100 TeV Hadron Collider}},
  \href{http://dx.doi.org/10.1007/JHEP02(2015)016}{\emph{JHEP} {\bf 02} (2015)
  016}, [\href{https://arxiv.org/abs/1412.7154}{{\tt 1412.7154}}].

\bibitem{Contino:2016spe}
R.~Contino et~al., \emph{{Physics at a 100 TeV pp collider: Higgs and EW
  symmetry breaking studies}},
  \href{http://dx.doi.org/10.23731/CYRM-2017-003.255}{\emph{CERN Yellow Rep.}
  (2017) 255--440}, [\href{https://arxiv.org/abs/1606.09408}{{\tt
  1606.09408}}].

\bibitem{Park:2020yps}
J.~Park, J.~Chang, K.~Cheung and J.~S. Lee, \emph{{Measuring the trilinear
  Higgs boson self--coupling at the 100 TeV hadron collider via multivariate
  analysis}},  \href{https://arxiv.org/abs/2003.12281}{{\tt 2003.12281}}.

\bibitem{Adhikary:2020fqf}
A.~Adhikary, R.~K. Barman and B.~Bhattacherjee, \emph{{Prospects of
  non-resonant di-Higgs searches and Higgs boson self-coupling measurement at
  the HE-LHC using machine learning techniques}},
  \href{http://dx.doi.org/10.1007/JHEP12(2020)179}{\emph{JHEP} {\bf 12} (2020)
  179}, [\href{https://arxiv.org/abs/2006.11879}{{\tt 2006.11879}}].

\bibitem{Liu:2004pv}
J.-J. Liu, W.-G. Ma, G.~Li, R.-Y. Zhang and H.-S. Hou, \emph{{Higgs boson pair
  production in the little Higgs model at hadron collider}},
  \href{http://dx.doi.org/10.1103/PhysRevD.70.015001}{\emph{Phys. Rev.} {\bf
  D70} (2004) 015001}, [\href{https://arxiv.org/abs/hep-ph/0404171}{{\tt
  hep-ph/0404171}}].

\bibitem{Dib:2005re}
C.~O. Dib, R.~Rosenfeld and A.~Zerwekh, \emph{{Double Higgs production and
  quadratic divergence cancellation in little Higgs models with T parity}},
  \href{http://dx.doi.org/10.1088/1126-6708/2006/05/074}{\emph{JHEP} {\bf 05}
  (2006) 074}, [\href{https://arxiv.org/abs/hep-ph/0509179}{{\tt
  hep-ph/0509179}}].

\bibitem{Pierce:2006dh}
A.~Pierce, J.~Thaler and L.-T. Wang, \emph{{Disentangling Dimension Six
  Operators through Di-Higgs Boson Production}},
  \href{http://dx.doi.org/10.1088/1126-6708/2007/05/070}{\emph{JHEP} {\bf 05}
  (2007) 070}, [\href{https://arxiv.org/abs/hep-ph/0609049}{{\tt
  hep-ph/0609049}}].

\bibitem{Wang:2007zx}
L.~Wang, W.~Wang, J.~M. Yang and H.~Zhang, \emph{{Higgs-pair production in
  littlest Higgs model with T-parity}},
  \href{http://dx.doi.org/10.1103/PhysRevD.76.017702}{\emph{Phys. Rev.} {\bf
  D76} (2007) 017702}, [\href{https://arxiv.org/abs/0705.3392}{{\tt
  0705.3392}}].

\bibitem{Kanemura:2008ub}
S.~Kanemura and K.~Tsumura, \emph{{Effects of the anomalous Higgs couplings on
  the Higgs boson production at the Large Hadron Collider}},
  \href{http://dx.doi.org/10.1140/epjc/s10052-009-1077-0}{\emph{Eur. Phys. J.}
  {\bf C63} (2009) 11--21}, [\href{https://arxiv.org/abs/0810.0433}{{\tt
  0810.0433}}].

\bibitem{Contino:2010mh}
R.~Contino, C.~Grojean, M.~Moretti, F.~Piccinini and R.~Rattazzi, \emph{{Strong
  Double Higgs Production at the LHC}},
  \href{http://dx.doi.org/10.1007/JHEP05(2010)089}{\emph{JHEP} {\bf 05} (2010)
  089}, [\href{https://arxiv.org/abs/1002.1011}{{\tt 1002.1011}}].

\bibitem{Grober:2010yv}
R.~Grober and M.~Muhlleitner, \emph{{Composite Higgs Boson Pair Production at
  the LHC}}, \href{http://dx.doi.org/10.1007/JHEP06(2011)020}{\emph{JHEP} {\bf
  06} (2011) 020}, [\href{https://arxiv.org/abs/1012.1562}{{\tt 1012.1562}}].

\bibitem{Sun:2012zzm}
H.~Sun, Y.-J. Zhou and H.~Chen, \emph{{Constraints on large-extra-dimensions
  model through 125-GeV Higgs pair production at the LHC}},
  \href{http://dx.doi.org/10.1140/epjc/s10052-012-2011-4}{\emph{Eur. Phys. J.}
  {\bf C72} (2012) 2011}, [\href{https://arxiv.org/abs/1211.5197}{{\tt
  1211.5197}}].

\bibitem{Contino:2012xk}
R.~Contino, M.~Ghezzi, M.~Moretti, G.~Panico, F.~Piccinini and A.~Wulzer,
  \emph{{Anomalous Couplings in Double Higgs Production}},
  \href{http://dx.doi.org/10.1007/JHEP08(2012)154}{\emph{JHEP} {\bf 08} (2012)
  154}, [\href{https://arxiv.org/abs/1205.5444}{{\tt 1205.5444}}].

\bibitem{Kribs:2012kz}
G.~D. Kribs and A.~Martin, \emph{{Enhanced di-Higgs Production through Light
  Colored Scalars}},
  \href{http://dx.doi.org/10.1103/PhysRevD.86.095023}{\emph{Phys. Rev.} {\bf
  D86} (2012) 095023}, [\href{https://arxiv.org/abs/1207.4496}{{\tt
  1207.4496}}].

\bibitem{Dolan:2012ac}
M.~J. Dolan, C.~Englert and M.~Spannowsky, \emph{{New Physics in LHC Higgs
  boson pair production}},
  \href{http://dx.doi.org/10.1103/PhysRevD.87.055002}{\emph{Phys. Rev.} {\bf
  D87} (2013) 055002}, [\href{https://arxiv.org/abs/1210.8166}{{\tt
  1210.8166}}].

\bibitem{Nishiwaki:2013cma}
K.~Nishiwaki, S.~Niyogi and A.~Shivaji, \emph{{$ttH$ Anomalous Coupling in
  Double Higgs Production}},
  \href{http://dx.doi.org/10.1007/JHEP04(2014)011}{\emph{JHEP} {\bf 04} (2014)
  011}, [\href{https://arxiv.org/abs/1309.6907}{{\tt 1309.6907}}].

\bibitem{Ellwanger:2013ova}
U.~Ellwanger, \emph{{Higgs pair production in the NMSSM at the LHC}},
  \href{http://dx.doi.org/10.1007/JHEP08(2013)077}{\emph{JHEP} {\bf 08} (2013)
  077}, [\href{https://arxiv.org/abs/1306.5541}{{\tt 1306.5541}}].

\bibitem{Cao:2013si}
J.~Cao, Z.~Heng, L.~Shang, P.~Wan and J.~M. Yang, \emph{{Pair Production of a
  125 GeV Higgs Boson in MSSM and NMSSM at the LHC}},
  \href{http://dx.doi.org/10.1007/JHEP04(2013)134}{\emph{JHEP} {\bf 04} (2013)
  134}, [\href{https://arxiv.org/abs/1301.6437}{{\tt 1301.6437}}].

\bibitem{Dawson:2012mk}
S.~Dawson, E.~Furlan and I.~Lewis, \emph{{Unravelling an extended quark sector
  through multiple Higgs production?}},
  \href{http://dx.doi.org/10.1103/PhysRevD.87.014007}{\emph{Phys. Rev.} {\bf
  D87} (2013) 014007}, [\href{https://arxiv.org/abs/1210.6663}{{\tt
  1210.6663}}].

\bibitem{No:2013wsa}
J.~M. No and M.~Ramsey-Musolf, \emph{{Probing the Higgs Portal at the LHC
  Through Resonant di-Higgs Production}},
  \href{http://dx.doi.org/10.1103/PhysRevD.89.095031}{\emph{Phys. Rev.} {\bf
  D89} (2014) 095031}, [\href{https://arxiv.org/abs/1310.6035}{{\tt
  1310.6035}}].

\bibitem{Goertz:2014qta}
F.~Goertz, A.~Papaefstathiou, L.~L. Yang and J.~Zurita, \emph{{Higgs boson pair
  production in the D=6 extension of the SM}},
  \href{http://dx.doi.org/10.1007/JHEP04(2015)167}{\emph{JHEP} {\bf 04} (2015)
  167}, [\href{https://arxiv.org/abs/1410.3471}{{\tt 1410.3471}}].

\bibitem{Liu:2014rba}
N.~Liu, S.~Hu, B.~Yang and J.~Han, \emph{{Impact of top-Higgs couplings on
  Di-Higgs production at future colliders}},
  \href{http://dx.doi.org/10.1007/JHEP01(2015)008}{\emph{JHEP} {\bf 01} (2015)
  008}, [\href{https://arxiv.org/abs/1408.4191}{{\tt 1408.4191}}].

\bibitem{Chen:2014xra}
C.-R. Chen and I.~Low, \emph{{Double take on new physics in double Higgs boson
  production}}, \href{http://dx.doi.org/10.1103/PhysRevD.90.013018}{\emph{Phys.
  Rev.} {\bf D90} (2014) 013018}, [\href{https://arxiv.org/abs/1405.7040}{{\tt
  1405.7040}}].

\bibitem{Baglio:2014nea}
J.~Baglio, O.~Eberhardt, U.~Nierste and M.~Wiebusch, \emph{{Benchmarks for
  Higgs Pair Production and Heavy Higgs boson Searches in the Two-Higgs-Doublet
  Model of Type II}},
  \href{http://dx.doi.org/10.1103/PhysRevD.90.015008}{\emph{Phys. Rev.} {\bf
  D90} (2014) 015008}, [\href{https://arxiv.org/abs/1403.1264}{{\tt
  1403.1264}}].

\bibitem{Hespel:2014sla}
B.~Hespel, D.~Lopez-Val and E.~Vryonidou, \emph{{Higgs pair production via
  gluon fusion in the Two-Higgs-Doublet Model}},
  \href{http://dx.doi.org/10.1007/JHEP09(2014)124}{\emph{JHEP} {\bf 09} (2014)
  124}, [\href{https://arxiv.org/abs/1407.0281}{{\tt 1407.0281}}].

\bibitem{Barger:2014taa}
V.~Barger, L.~L. Everett, C.~B. Jackson, A.~D. Peterson and G.~Shaughnessy,
  \emph{{New physics in resonant production of Higgs boson pairs}},
  \href{http://dx.doi.org/10.1103/PhysRevLett.114.011801}{\emph{Phys. Rev.
  Lett.} {\bf 114} (2015) 011801}, [\href{https://arxiv.org/abs/1408.0003}{{\tt
  1408.0003}}].

\bibitem{Dawson:2015oha}
S.~Dawson, A.~Ismail and I.~Low, \emph{{What\textquoteright{}s in the loop? The
  anatomy of double Higgs production}},
  \href{http://dx.doi.org/10.1103/PhysRevD.91.115008}{\emph{Phys. Rev. D} {\bf
  91} (2015) 115008}, [\href{https://arxiv.org/abs/1504.05596}{{\tt
  1504.05596}}].

\bibitem{Azatov:2015oxa}
A.~Azatov, R.~Contino, G.~Panico and M.~Son, \emph{{Effective field theory
  analysis of double Higgs boson production via gluon fusion}},
  \href{http://dx.doi.org/10.1103/PhysRevD.92.035001}{\emph{Phys. Rev.} {\bf
  D92} (2015) 035001}, [\href{https://arxiv.org/abs/1502.00539}{{\tt
  1502.00539}}].

\bibitem{Lu:2015qqa}
L.-C. Lü, C.~Du, Y.~Fang, H.-J. He and H.~Zhang, \emph{{Searching heavier
  Higgs boson via di-Higgs production at LHC Run-2}},
  \href{http://dx.doi.org/10.1016/j.physletb.2016.02.026}{\emph{Phys. Lett.}
  {\bf B755} (2016) 509--522}, [\href{https://arxiv.org/abs/1507.02644}{{\tt
  1507.02644}}].

\bibitem{Lu:2015jza}
C.-T. Lu, J.~Chang, K.~Cheung and J.~S. Lee, \emph{{An exploratory study of
  Higgs-boson pair production}},
  \href{http://dx.doi.org/10.1007/JHEP08(2015)133}{\emph{JHEP} {\bf 08} (2015)
  133}, [\href{https://arxiv.org/abs/1505.00957}{{\tt 1505.00957}}].

\bibitem{Carvalho:2015ttv}
A.~Carvalho, M.~Dall'Osso, T.~Dorigo, F.~Goertz, C.~A. Gottardo and M.~Tosi,
  \emph{{Higgs Pair Production: Choosing Benchmarks With Cluster Analysis}},
  \href{http://dx.doi.org/10.1007/JHEP04(2016)126}{\emph{JHEP} {\bf 04} (2016)
  126}, [\href{https://arxiv.org/abs/1507.02245}{{\tt 1507.02245}}].

\bibitem{Cao:2015oaa}
Q.-H. Cao, B.~Yan, D.-M. Zhang and H.~Zhang, \emph{{Resolving the Degeneracy in
  Single Higgs Production with Higgs Pair Production}},
  \href{http://dx.doi.org/10.1016/j.physletb.2015.11.045}{\emph{Phys. Lett.}
  {\bf B752} (2016) 285--290}, [\href{https://arxiv.org/abs/1508.06512}{{\tt
  1508.06512}}].

\bibitem{Costa:2015llh}
R.~Costa, M.~Mühlleitner, M.~O.~P. Sampaio and R.~Santos, \emph{{Singlet
  Extensions of the Standard Model at LHC Run 2: Benchmarks and Comparison with
  the NMSSM}}, \href{http://dx.doi.org/10.1007/JHEP06(2016)034}{\emph{JHEP}
  {\bf 06} (2016) 034}, [\href{https://arxiv.org/abs/1512.05355}{{\tt
  1512.05355}}].

\bibitem{Batell:2015koa}
B.~Batell, M.~McCullough, D.~Stolarski and C.~B. Verhaaren, \emph{{Putting a
  Stop to di-Higgs Modifications}},
  \href{http://dx.doi.org/10.1007/JHEP09(2015)216}{\emph{JHEP} {\bf 09} (2015)
  216}, [\href{https://arxiv.org/abs/1508.01208}{{\tt 1508.01208}}].

\bibitem{Cao:2016zob}
Q.-H. Cao, G.~Li, B.~Yan, D.-M. Zhang and H.~Zhang, \emph{{Double Higgs
  production at the 14 TeV LHC and the 100 TeV pp-collider}},
  \href{https://arxiv.org/abs/1611.09336}{{\tt 1611.09336}}.

\bibitem{Kotwal:2016tex}
A.~V. Kotwal, M.~J. Ramsey-Musolf, J.~M. No and P.~Winslow,
  \emph{{Singlet-catalyzed electroweak phase transitions in the 100 TeV
  frontier}}, \href{http://dx.doi.org/10.1103/PhysRevD.94.035022}{\emph{Phys.
  Rev.} {\bf D94} (2016) 035022}, [\href{https://arxiv.org/abs/1605.06123}{{\tt
  1605.06123}}].

\bibitem{Grober:2016wmf}
R.~Grober, M.~Muhlleitner and M.~Spira, \emph{{Signs of Composite Higgs Pair
  Production at Next-to-Leading Order}},
  \href{http://dx.doi.org/10.1007/JHEP06(2016)080}{\emph{JHEP} {\bf 06} (2016)
  080}, [\href{https://arxiv.org/abs/1602.05851}{{\tt 1602.05851}}].

\bibitem{Bian:2016awe}
L.~Bian and N.~Chen, \emph{{Higgs pair productions in the CP-violating
  two-Higgs-doublet model}},
  \href{http://dx.doi.org/10.1007/JHEP09(2016)069}{\emph{JHEP} {\bf 09} (2016)
  069}, [\href{https://arxiv.org/abs/1607.02703}{{\tt 1607.02703}}].

\bibitem{Crivellin:2016ihg}
A.~Crivellin, M.~Ghezzi and M.~Procura, \emph{{Effective Field Theory with Two
  Higgs Doublets}},
  \href{http://dx.doi.org/10.1007/JHEP09(2016)160}{\emph{JHEP} {\bf 09} (2016)
  160}, [\href{https://arxiv.org/abs/1608.00975}{{\tt 1608.00975}}].

\bibitem{Gorbahn:2016uoy}
M.~Gorbahn and U.~Haisch, \emph{{Indirect probes of the trilinear Higgs
  coupling: $gg \to h$ and $h \to \gamma \gamma$}},
  \href{http://dx.doi.org/10.1007/JHEP10(2016)094}{\emph{JHEP} {\bf 10} (2016)
  094}, [\href{https://arxiv.org/abs/1607.03773}{{\tt 1607.03773}}].

\bibitem{Carvalho:2016rys}
A.~Carvalho, M.~Dall'Osso, P.~De~Castro~Manzano, T.~Dorigo, F.~Goertz,
  M.~Gouzevich et~al., \emph{{Analytical parametrization and shape
  classification of anomalous HH production in the EFT approach}},
  \href{https://arxiv.org/abs/1608.06578}{{\tt 1608.06578}}.

\bibitem{Huang:2017nnw}
P.~Huang, A.~Joglekar, M.~Li and C.~E.~M. Wagner, \emph{{Corrections to
  di-Higgs boson production with light stops and modified Higgs couplings}},
  \href{http://dx.doi.org/10.1103/PhysRevD.97.075001}{\emph{Phys. Rev. D} {\bf
  97} (2018) 075001}, [\href{https://arxiv.org/abs/1711.05743}{{\tt
  1711.05743}}].

\bibitem{Nakamura:2017irk}
K.~Nakamura, K.~Nishiwaki, K.-y. Oda, S.~C. Park and Y.~Yamamoto,
  \emph{{Di-higgs enhancement by neutral scalar as probe of new colored
  sector}}, \href{http://dx.doi.org/10.1140/epjc/s10052-017-4835-4}{\emph{Eur.
  Phys. J. C} {\bf 77} (2017) 273},
  [\href{https://arxiv.org/abs/1701.06137}{{\tt 1701.06137}}].

\bibitem{Gao:2019uco}
Y.~Gao and N.~A. Neill, \emph{{Probing Exotic Triple Higgs Couplings for Almost
  Inert Higgs Bosons at the LHC}},
  \href{http://dx.doi.org/10.1007/JHEP05(2020)087}{\emph{JHEP} {\bf 05} (2020)
  087}, [\href{https://arxiv.org/abs/2001.00069}{{\tt 2001.00069}}].

\bibitem{Huang:2019bcs}
P.~Huang and Y.~H. Ng, \emph{{Di-Higgs Production in SUSY models at the LHC}},
  \href{https://arxiv.org/abs/1910.13968}{{\tt 1910.13968}}.

\bibitem{Barducci:2019xkq}
D.~Barducci, K.~Mimasu, J.~No, C.~Vernieri and J.~Zurita, \emph{{Enlarging the
  scope of resonant di-Higgs searches: Hunting for Higgs-to-Higgs cascades in
  $4b$ final states at the LHC and future colliders}},
  \href{http://dx.doi.org/10.1007/JHEP02(2020)002}{\emph{JHEP} {\bf 02} (2020)
  002}, [\href{https://arxiv.org/abs/1910.08574}{{\tt 1910.08574}}].

\bibitem{Basler:2019nas}
P.~Basler, S.~Dawson, C.~Englert and M.~Mühlleitner, \emph{{Di-Higgs boson
  peaks and top valleys: Interference effects in Higgs sector extensions}},
  \href{http://dx.doi.org/10.1103/PhysRevD.101.015019}{\emph{Phys. Rev. D} {\bf
  101} (2020) 015019}, [\href{https://arxiv.org/abs/1909.09987}{{\tt
  1909.09987}}].

\bibitem{Alves:2019igs}
A.~Alves, D.~Gonçalves, T.~Ghosh, H.-K. Guo and K.~Sinha, \emph{{Di-Higgs
  Production in the $4b$ Channel and Gravitational Wave Complementarity}},
  \href{http://dx.doi.org/10.1007/JHEP03(2020)053}{\emph{JHEP} {\bf 03} (2020)
  053}, [\href{https://arxiv.org/abs/1909.05268}{{\tt 1909.05268}}].

\bibitem{Englert:2019eyl}
C.~Englert and J.~Jaeckel, \emph{{Probing the Symmetric Higgs Portal with
  Di-Higgs Boson Production}},
  \href{http://dx.doi.org/10.1103/PhysRevD.100.095017}{\emph{Phys. Rev. D} {\bf
  100} (2019) 095017}, [\href{https://arxiv.org/abs/1908.10615}{{\tt
  1908.10615}}].

\bibitem{Bauer:2017cov}
M.~Bauer, M.~Carena and A.~Carmona, \emph{{Higgs Pair Production as a Signal of
  Enhanced Yukawa Couplings}},
  \href{http://dx.doi.org/10.1103/PhysRevLett.121.021801}{\emph{Phys. Rev.
  Lett.} {\bf 121} (2018) 021801},
  [\href{https://arxiv.org/abs/1801.00363}{{\tt 1801.00363}}].

\bibitem{Babu:2018uik}
K.~Babu and S.~Jana, \emph{{Enhanced Di-Higgs Production in the Two Higgs
  Doublet Model}}, \href{http://dx.doi.org/10.1007/JHEP02(2019)193}{\emph{JHEP}
  {\bf 02} (2019) 193}, [\href{https://arxiv.org/abs/1812.11943}{{\tt
  1812.11943}}].

\bibitem{Basler:2018dac}
P.~Basler, S.~Dawson, C.~Englert and M.~Mühlleitner, \emph{{Showcasing HH
  production: Benchmarks for the LHC and HL-LHC}},
  \href{http://dx.doi.org/10.1103/PhysRevD.99.055048}{\emph{Phys. Rev. D} {\bf
  99} (2019) 055048}, [\href{https://arxiv.org/abs/1812.03542}{{\tt
  1812.03542}}].

\bibitem{Alves:2018jsw}
A.~Alves, T.~Ghosh, H.-K. Guo, K.~Sinha and D.~Vagie, \emph{{Collider and
  Gravitational Wave Complementarity in Exploring the Singlet Extension of the
  Standard Model}},
  \href{http://dx.doi.org/10.1007/JHEP04(2019)052}{\emph{JHEP} {\bf 04} (2019)
  052}, [\href{https://arxiv.org/abs/1812.09333}{{\tt 1812.09333}}].

\bibitem{Adhikary:2018ise}
A.~Adhikary, S.~Banerjee, R.~Kumar~Barman and B.~Bhattacherjee, \emph{{Resonant
  heavy Higgs searches at the HL-LHC}},
  \href{http://dx.doi.org/10.1007/JHEP09(2019)068}{\emph{JHEP} {\bf 09} (2019)
  068}, [\href{https://arxiv.org/abs/1812.05640}{{\tt 1812.05640}}].

\bibitem{Borowka:2018pxx}
S.~Borowka, C.~Duhr, F.~Maltoni, D.~Pagani, A.~Shivaji and X.~Zhao,
  \emph{{Probing the scalar potential via double Higgs boson production at
  hadron colliders}},
  \href{http://dx.doi.org/10.1007/JHEP04(2019)016}{\emph{JHEP} {\bf 04} (2019)
  016}, [\href{https://arxiv.org/abs/1811.12366}{{\tt 1811.12366}}].

\bibitem{Chen:2018uim}
N.~Chen, C.~Du, Y.~Wu and X.-J. Xu, \emph{{Further study of the global minimum
  constraint on the two-Higgs-doublet models: LHC searches for heavy Higgs
  bosons}}, \href{http://dx.doi.org/10.1103/PhysRevD.99.035011}{\emph{Phys.
  Rev. D} {\bf 99} (2019) 035011},
  [\href{https://arxiv.org/abs/1810.04689}{{\tt 1810.04689}}].

\bibitem{Alves:2018oct}
A.~Alves, T.~Ghosh, H.-K. Guo and K.~Sinha, \emph{{Resonant Di-Higgs Production
  at Gravitational Wave Benchmarks: A Collider Study using Machine Learning}},
  \href{http://dx.doi.org/10.1007/JHEP12(2018)070}{\emph{JHEP} {\bf 12} (2018)
  070}, [\href{https://arxiv.org/abs/1808.08974}{{\tt 1808.08974}}].

\bibitem{Buchalla:2018yce}
G.~Buchalla, M.~Capozi, A.~Celis, G.~Heinrich and L.~Scyboz, \emph{{Higgs boson
  pair production in non-linear Effective Field Theory with full
  $m_t$-dependence at NLO QCD}},
  \href{http://dx.doi.org/10.1007/JHEP09(2018)057}{\emph{JHEP} {\bf 09} (2018)
  057}, [\href{https://arxiv.org/abs/1806.05162}{{\tt 1806.05162}}].

\bibitem{Heng:2018kyd}
Z.~Heng, X.~Gong and H.~Zhou, \emph{{Pair production of Higgs boson in NMSSM at
  the LHC with the next-to-lightest CP-even Higgs boson being SM-like}},
  \href{http://dx.doi.org/10.1088/1674-1137/42/7/073103}{\emph{Chin. Phys. C}
  {\bf 42} (2018) 073103}, [\href{https://arxiv.org/abs/1805.01598}{{\tt
  1805.01598}}].

\bibitem{Kim:2018uty}
J.~H. Kim, Y.~Sakaki and M.~Son, \emph{{Combined analysis of double Higgs
  production via gluon fusion at the HL-LHC in the effective field theory
  approach}}, \href{http://dx.doi.org/10.1103/PhysRevD.98.015016}{\emph{Phys.
  Rev. D} {\bf 98} (2018) 015016},
  [\href{https://arxiv.org/abs/1801.06093}{{\tt 1801.06093}}].

\bibitem{Flores:2019hcf}
M.~Flores, C.~Gross, J.~S. Kim, O.~Lebedev and S.~Mondal, \emph{{Multi-Higgs
  Probes of the Dark Sector}},  \href{https://arxiv.org/abs/1912.02204}{{\tt
  1912.02204}}.

\bibitem{Englert:2019xhz}
C.~Englert, D.~J. Miller and D.~D. Smaranda, \emph{{Phenomenology of
  GUT-inspired gauge-Higgs unification}},
  \href{https://arxiv.org/abs/1911.05527}{{\tt 1911.05527}}.

\bibitem{DiMicco:2019ngk}
J.~Alison et~al., \emph{{Higgs boson potential at colliders: Status and
  perspectives}},
  \href{http://dx.doi.org/10.1016/j.revip.2020.100045}{\emph{Rev. Phys.} {\bf
  5} (2020) 100045}, [\href{https://arxiv.org/abs/1910.00012}{{\tt
  1910.00012}}].

\bibitem{Alasfar:2019pmn}
L.~Alasfar, R.~Corral~Lopez and R.~Gröber, \emph{{Probing Higgs couplings to
  light quarks via Higgs pair production}},
  \href{http://dx.doi.org/10.1007/JHEP11(2019)088}{\emph{JHEP} {\bf 11} (2019)
  088}, [\href{https://arxiv.org/abs/1909.05279}{{\tt 1909.05279}}].

\bibitem{Capozi:2019xsi}
M.~Capozi and G.~Heinrich, \emph{{Exploring anomalous couplings in Higgs boson
  pair production through shape analysis}},
  \href{http://dx.doi.org/10.1007/JHEP03(2020)091}{\emph{JHEP} {\bf 03} (2020)
  091}, [\href{https://arxiv.org/abs/1908.08923}{{\tt 1908.08923}}].

\bibitem{Li:2019uyy}
G.~Li, L.-X. Xu, B.~Yan and C.-P. Yuan, \emph{{Resolving the degeneracy in top
  quark Yukawa coupling with Higgs pair production}},
  \href{http://dx.doi.org/10.1016/j.physletb.2019.135070}{\emph{Phys. Lett. B}
  {\bf 800} (2020) 135070}, [\href{https://arxiv.org/abs/1904.12006}{{\tt
  1904.12006}}].

\bibitem{Cheung:2020xij}
K.~Cheung, A.~Jueid, C.-T. Lu, J.~Song and Y.~W. Yoon, \emph{{Disentangling new
  physics effects on non-resonant Higgs boson pair production from gluon
  fusion}},  \href{https://arxiv.org/abs/2003.11043}{{\tt 2003.11043}}.

\bibitem{Kon:2018vmv}
T.~Kon, T.~Nagura, T.~Ueda and K.~Yagyu, \emph{{Double Higgs boson production
  at $e^+e^-$ colliders in the two-Higgs-doublet model}},
  \href{http://dx.doi.org/10.1103/PhysRevD.99.095027}{\emph{Phys. Rev. D} {\bf
  99} (2019) 095027}, [\href{https://arxiv.org/abs/1812.09843}{{\tt
  1812.09843}}].

\bibitem{Arhrib:2018qmw}
A.~Arhrib, R.~Benbrik, M.~El~Kacimi, L.~Rahili and S.~Semlali, \emph{{Extended
  Higgs sector of 2HDM with real singlet facing LHC data}},
  \href{http://dx.doi.org/10.1140/epjc/s10052-019-7472-2}{\emph{Eur. Phys. J.
  C} {\bf 80} (2020) 13}, [\href{https://arxiv.org/abs/1811.12431}{{\tt
  1811.12431}}].

\bibitem{Bahl:2020kwe}
H.~Bahl, P.~Bechtle, S.~Heinemeyer, S.~Liebler, T.~Stefaniak and G.~Weiglein,
  \emph{{HL-LHC and ILC sensitivities in the hunt for heavy Higgs bosons}},
  \href{http://dx.doi.org/10.1140/epjc/s10052-020-08472-z}{\emph{Eur. Phys. J.
  C} {\bf 80} (2020) 916}, [\href{https://arxiv.org/abs/2005.14536}{{\tt
  2005.14536}}].

\bibitem{Barman:2020ulr}
R.~K. Barman, C.~Englert, D.~Gon\c{c}alves and M.~Spannowsky, \emph{{Di-Higgs
  resonance searches in weak boson fusion}},
  \href{http://dx.doi.org/10.1103/PhysRevD.102.055014}{\emph{Phys. Rev. D} {\bf
  102} (2020) 055014}, [\href{https://arxiv.org/abs/2007.07295}{{\tt
  2007.07295}}].

\bibitem{Cheung:2022bhx}
K.~Cheung, Y.-L. Chung and S.-C. Hsu, \emph{{Sensitivity on Two-Higgs-Doublet
  Models from Higgs-Pair Production via $b\bar{b}b\bar{b}$ Final State}},
  \href{https://arxiv.org/abs/2207.09602}{{\tt 2207.09602}}.

\bibitem{Arroyo-Urena:2022oft}
M.~A. Arroyo-Ure\~na, A.~Chakraborty, J.~L. D\'\i{}az-Cruz, D.~K. Ghosh,
  N.~Khan and S.~Moretti, \emph{{Higgs Pair Production at the LHC through the
  Flavon}},  \href{https://arxiv.org/abs/2205.12641}{{\tt 2205.12641}}.

\bibitem{Bhaskar:2022ygp}
A.~Bhaskar, D.~Das, B.~De, S.~Mitra, A.~K. Nayak and C.~Neeraj,
  \emph{{Leptoquark-assisted singlet-mediated di-Higgs production at the LHC}},
  \href{http://dx.doi.org/10.1016/j.physletb.2022.137341}{\emph{Phys. Lett. B}
  {\bf 833} (2022) 137341}, [\href{https://arxiv.org/abs/2205.12210}{{\tt
  2205.12210}}].

\bibitem{Kanemura:2022ldq}
S.~Kanemura, M.~Kikuchi and K.~Yagyu, \emph{{Next-to-leading order corrections
  to decays of the heavier CP-even Higgs boson in the two Higgs doublet
  model}}, \href{http://dx.doi.org/10.1016/j.nuclphysb.2022.115906}{\emph{Nucl.
  Phys. B} {\bf 983} (2022) 115906},
  [\href{https://arxiv.org/abs/2203.08337}{{\tt 2203.08337}}].

\bibitem{htwiki}
\url{https://twiki.cern.ch/twiki/bin/view/LHCPhysics/CERNYellowReportPageAt14TeV}.

\bibitem{ATLAS:2020bhl}
{\scshape ATLAS} collaboration, G.~Aad et~al., \emph{{Measurements of Higgs
  bosons decaying to bottom quarks from vector boson fusion production with the
  ATLAS experiment at $\sqrt{s}=13\,\text {TeV}$}},
  \href{http://dx.doi.org/10.1140/epjc/s10052-021-09192-8}{\emph{Eur. Phys. J.
  C} {\bf 81} (2021) 537}, [\href{https://arxiv.org/abs/2011.08280}{{\tt
  2011.08280}}].

\bibitem{ATLAS:2021tbi}
{\scshape ATLAS} collaboration, G.~Aad et~al., \emph{{Constraints on Higgs
  boson production with large transverse momentum using $H\rightarrow b\bar{b}$
  decays in the ATLAS detector}},
  \href{http://dx.doi.org/10.1103/PhysRevD.105.092003}{\emph{Phys. Rev. D} {\bf
  105} (2022) 092003}, [\href{https://arxiv.org/abs/2111.08340}{{\tt
  2111.08340}}].

\bibitem{ATLAS:2022tnm}
{\scshape ATLAS} collaboration, \emph{{Measurement of the properties of Higgs
  boson production at $\sqrt{s} = 13$ TeV in the $H\to\gamma\gamma$ channel
  using $139$ fb$^{-1}$ of $pp$ collision data with the ATLAS experiment}},
  \href{https://arxiv.org/abs/2207.00348}{{\tt 2207.00348}}.

\bibitem{ATLAS:2022qef}
{\scshape ATLAS} collaboration, \emph{{Measurement of the total and
  differential Higgs boson production cross-sections at $\sqrt{s} = 13$ TeV
  with the ATLAS detector by combining the $H \rightarrow ZZ^* \rightarrow
  4\ell$ and $H \rightarrow \gamma \gamma$ decay channels}},
  \href{https://arxiv.org/abs/2207.08615}{{\tt 2207.08615}}.

\bibitem{ATLAS:2022yrq}
{\scshape ATLAS} collaboration, G.~Aad et~al., \emph{{Measurements of Higgs
  boson production cross-sections in the $H\to\tau^{+}\tau^{-}$ decay channel
  in $pp$ collisions at $\sqrt{s}=13\,\text{TeV}$ with the ATLAS detector}},
  \href{https://arxiv.org/abs/2201.08269}{{\tt 2201.08269}}.

\bibitem{ATLAS:2022net}
{\scshape ATLAS} collaboration, \emph{{Measurement of the Higgs boson mass in
  the $H \rightarrow ZZ^* \rightarrow 4\ell$ decay channel using 139 fb$^{-1}$
  of $\sqrt{s}=13$ TeV $pp$ collisions recorded by the ATLAS detector at the
  LHC}},  \href{https://arxiv.org/abs/2207.00320}{{\tt 2207.00320}}.

\bibitem{ATLAS:2022ooq}
{\scshape ATLAS} collaboration, \emph{{Measurements of Higgs boson production
  by gluon$-$gluon fusion and vector-boson fusion using $H\rightarrow W W^*
  \rightarrow e\nu \mu\nu$ decays in $pp$ collisions at $\sqrt{s}=13$ TeV with
  the ATLAS detector}},  \href{https://arxiv.org/abs/2207.00338}{{\tt
  2207.00338}}.

\bibitem{ATLAS:2021pkb}
{\scshape ATLAS} collaboration, G.~Aad et~al., \emph{{Constraints on Higgs
  boson properties using $WW^{*}(\rightarrow e\nu\mu\nu) jj$ production in 36.1
  fb$^{-1}$ of $\sqrt{s}$=13 TeV $pp$ collisions with the ATLAS detector}},
  \href{https://arxiv.org/abs/2109.13808}{{\tt 2109.13808}}.

\bibitem{ATLAS:2020fzp}
{\scshape ATLAS} collaboration, G.~Aad et~al., \emph{{A search for the dimuon
  decay of the Standard Model Higgs boson with the ATLAS detector}},
  \href{http://dx.doi.org/10.1016/j.physletb.2020.135980}{\emph{Phys. Lett. B}
  {\bf 812} (2021) 135980}, [\href{https://arxiv.org/abs/2007.07830}{{\tt
  2007.07830}}].

\bibitem{Alwall:2014hca}
J.~Alwall, R.~Frederix, S.~Frixione, V.~Hirschi, F.~Maltoni, O.~Mattelaer
  et~al., \emph{{The automated computation of tree-level and next-to-leading
  order differential cross sections, and their matching to parton shower
  simulations}}, \href{http://dx.doi.org/10.1007/JHEP07(2014)079}{\emph{JHEP}
  {\bf 07} (2014) 079}, [\href{https://arxiv.org/abs/1405.0301}{{\tt
  1405.0301}}].

\bibitem{Sjostrand:2001yu}
T.~Sjostrand, L.~Lonnblad and S.~Mrenna, \emph{{PYTHIA 6.2: Physics and
  manual}},  \href{https://arxiv.org/abs/hep-ph/0108264}{{\tt hep-ph/0108264}}.

\bibitem{Sjostrand:2014zea}
T.~Sjöstrand, S.~Ask, J.~R. Christiansen, R.~Corke, N.~Desai, P.~Ilten et~al.,
  \emph{{An Introduction to PYTHIA 8.2}},
  \href{http://dx.doi.org/10.1016/j.cpc.2015.01.024}{\emph{Comput. Phys.
  Commun.} {\bf 191} (2015) 159--177},
  [\href{https://arxiv.org/abs/1410.3012}{{\tt 1410.3012}}].

\bibitem{Cacciari:2008gp}
M.~Cacciari, G.~P. Salam and G.~Soyez, \emph{{The Anti-k(t) jet clustering
  algorithm}},
  \href{http://dx.doi.org/10.1088/1126-6708/2008/04/063}{\emph{JHEP} {\bf 04}
  (2008) 063}, [\href{https://arxiv.org/abs/0802.1189}{{\tt 0802.1189}}].

\bibitem{Cacciari:2011ma}
M.~Cacciari, G.~P. Salam and G.~Soyez, \emph{{FastJet User Manual}},
  \href{http://dx.doi.org/10.1140/epjc/s10052-012-1896-2}{\emph{Eur. Phys. J.}
  {\bf C72} (2012) 1896}, [\href{https://arxiv.org/abs/1111.6097}{{\tt
  1111.6097}}].

\bibitem{deFavereau:2013fsa}
{\scshape DELPHES 3} collaboration, J.~de~Favereau, C.~Delaere, P.~Demin,
  A.~Giammanco, V.~Lemaître, A.~Mertens et~al., \emph{{DELPHES 3, A modular
  framework for fast simulation of a generic collider experiment}},
  \href{http://dx.doi.org/10.1007/JHEP02(2014)057}{\emph{JHEP} {\bf 02} (2014)
  057}, [\href{https://arxiv.org/abs/1307.6346}{{\tt 1307.6346}}].

\bibitem{Sirunyan:2017ezt}
{\scshape CMS} collaboration, A.~M. Sirunyan et~al., \emph{{Identification of
  heavy-flavour jets with the CMS detector in pp collisions at 13 TeV}},
  \href{http://dx.doi.org/10.1088/1748-0221/13/05/P05011}{\emph{JINST} {\bf 13}
  (2018) P05011}, [\href{https://arxiv.org/abs/1712.07158}{{\tt 1712.07158}}].

\bibitem{CMS-PAS-TAU-16-002}
{\scshape CMS Collaboration} collaboration, \emph{{Performance of
  reconstruction and identification of tau leptons in their decays to hadrons
  and tau neutrino in LHC Run-2}},  Technical Report CMS-PAS-TAU-16-002, CERN,
  Geneva, 2016.

\bibitem{Chen:2016:XST:2939672.2939785}
T.~Chen and C.~Guestrin, \emph{{XGBoost}: A scalable tree boosting system},  in
  \emph{Proceedings of the 22nd ACM SIGKDD International Conference on
  Knowledge Discovery and Data Mining}, KDD '16, (New York, NY, USA),
  pp.~785--794, ACM, 2016.
\newblock \href{http://dx.doi.org/10.1145/2939672.2939785}{DOI}.

\bibitem{Lundberg17}
S.~Lundberg and S.-I. Lee, \emph{A unified approach to interpreting model
  predictions},  2017.
\newblock 10.48550/ARXIV.1705.07874.

\bibitem{Lundberg18}
S.~M. Lundberg, G.~G. Erion and S.-I. Lee, \emph{Consistent individualized
  feature attribution for tree ensembles},  2018.
\newblock 10.48550/ARXIV.1802.03888.

\bibitem{Shapley+2016+307+318}
L.~S. Shapley, \emph{17. A Value for n-Person Games}, pp.~307--318.
\newblock Princeton University Press, 2016.
\newblock doi:10.1515/9781400881970-018.

\bibitem{Cornell_2022}
A.~S. Cornell, W.~Doorsamy, B.~Fuks, G.~Harmsen and L.~Mason, \emph{Boosted
  decision trees in the era of new physics: a smuon analysis case study},
  \href{http://dx.doi.org/10.1007/jhep04(2022)015}{\emph{Journal of High Energy
  Physics} {\bf 2022} (apr, 2022) }.

\bibitem{Alvestad:2021sje}
D.~Alvestad, N.~Fomin, J.~Kersten, S.~Maeland and I.~Str\"umke, \emph{{Beyond
  Cuts in Small Signal Scenarios -- Enhanced Sneutrino Detectability Using
  Machine Learning}},  \href{https://arxiv.org/abs/2108.03125}{{\tt
  2108.03125}}.

\bibitem{Grojean:2020ech}
C.~Grojean, A.~Paul and Z.~Qian, \emph{{Resurrecting $ b\overline{b}h $ with
  kinematic shapes}},
  \href{http://dx.doi.org/10.1007/JHEP04(2021)139}{\emph{JHEP} {\bf 04} (2021)
  139}, [\href{https://arxiv.org/abs/2011.13945}{{\tt 2011.13945}}].

\bibitem{CMS:2021qzz}
{\scshape CMS} collaboration, A.~Tumasyan et~al., \emph{{Evidence for WW/WZ
  vector boson scattering in the decay channel
  \ensuremath{\ell}\ensuremath{\nu}qq produced in association with two jets in
  proton-proton collisions at s=13 TeV}},
  \href{http://dx.doi.org/10.1016/j.physletb.2022.137438}{\emph{Phys. Lett. B}
  {\bf 834} (2022) 137438}, [\href{https://arxiv.org/abs/2112.05259}{{\tt
  2112.05259}}].

\bibitem{Grojean:2022mef}
C.~Grojean, A.~Paul, Z.~Qian and I.~Str\"umke, \emph{{Lessons on interpretable
  machine learning from particle physics}},
  \href{http://dx.doi.org/10.1038/s42254-022-00456-0}{\emph{Nature Rev. Phys.}
  {\bf 4} (2022) 284--286}, [\href{https://arxiv.org/abs/2203.08021}{{\tt
  2203.08021}}].

\bibitem{ATLAS:2022vkf}
{\scshape ATLAS} collaboration, G.~Aad et~al., \emph{{A detailed map of Higgs
  boson interactions by the ATLAS experiment ten years after the discovery}},
  \href{http://dx.doi.org/10.1038/s41586-022-04893-w}{\emph{Nature} {\bf 607}
  (2022) 52--59}, [\href{https://arxiv.org/abs/2207.00092}{{\tt 2207.00092}}].

\bibitem{ATL-PHYS-PUB-2015-045}
{\scshape ATLAS} collaboration, \emph{{Reconstruction, Energy Calibration, and
  Identification of Hadronically Decaying Tau Leptons in the ATLAS Experiment
  for Run-2 of the LHC}},  Technical Report ATL-PHYS-PUB-2015-045, CERN, Geneva, 2015.

\bibitem{Tumasyan_2022}
A.~Tumasyan, W.~Adam, J.~Andrejkovic, T.~Bergauer, S.~Chatterjee, M.~Dragicevic
  et~al., \emph{Identification of hadronic tau lepton decays using a deep
  neural network},
  \href{http://dx.doi.org/10.1088/1748-0221/17/07/p07023}{\emph{Journal of
  Instrumentation} {\bf 17} (jul, 2022) P07023}.

\bibitem{Elagin:2010aw}
A.~Elagin, P.~Murat, A.~Pranko and A.~Safonov, \emph{{A New Mass Reconstruction
  Technique for Resonances Decaying to di-tau}},
  \href{http://dx.doi.org/10.1016/j.nima.2011.07.009}{\emph{Nucl. Instrum.
  Meth.} {\bf A654} (2011) 481--489},
  [\href{https://arxiv.org/abs/1012.4686}{{\tt 1012.4686}}].

\bibitem{ATLAS:2018rnh}
{\scshape ATLAS} collaboration, M.~Aaboud et~al., \emph{{Search for pair
  production of Higgs bosons in the $b\bar{b}b\bar{b}$ final state using
  proton-proton collisions at $\sqrt{s} = 13$ TeV with the ATLAS detector}},
  \href{http://dx.doi.org/10.1007/JHEP01(2019)030}{\emph{JHEP} {\bf 01} (2019)
  030}, [\href{https://arxiv.org/abs/1804.06174}{{\tt 1804.06174}}].

\bibitem{Lester:1999tx}
C.~G. Lester and D.~J. Summers, \emph{{Measuring masses of semiinvisibly
  decaying particles pair produced at hadron colliders}},
  \href{http://dx.doi.org/10.1016/S0370-2693(99)00945-4}{\emph{Phys. Lett.}
  {\bf B463} (1999) 99--103}, [\href{https://arxiv.org/abs/hep-ph/9906349}{{\tt
  hep-ph/9906349}}].

\bibitem{Barr:2003rg}
A.~Barr, C.~Lester and P.~Stephens, \emph{{m(T2): The Truth behind the
  glamour}}, \href{http://dx.doi.org/10.1088/0954-3899/29/10/304}{\emph{J.
  Phys.} {\bf G29} (2003) 2343--2363},
  [\href{https://arxiv.org/abs/hep-ph/0304226}{{\tt hep-ph/0304226}}].

\bibitem{Selvaggi:2717698}
M.~Selvaggi, \emph{{A Delphes parameterisation of the FCC-hh detector}},  
  \href{https://cds.cern.ch/record/2717698}{\emph{CERN-FCC-PHYS-2020-0003}},
  Technical Report, CERN, Geneva, 2020.

\bibitem{Mangano:2016jyj}
 M. L.~Mangano and others, \emph{{Physics at a 100 TeV pp Collider: 
 Standard Model Processes}}, 
 \href{http://dx.doi.org/10.23731/CYRM-2017-003.1}{\emph{CERN-TH-2016-112} 
 (2016)},
  [\href{https://arxiv.org/abs/1607.01831}{{\tt 1607.01831}}]

\bibitem{Ellwanger:2009dp}
U.~Ellwanger, C.~Hugonie and A.~M. Teixeira, \emph{{The Next-to-Minimal
  Supersymmetric Standard Model}},
  \href{http://dx.doi.org/10.1016/j.physrep.2010.07.001}{\emph{Phys. Rept.}
  {\bf 496} (2010) 1--77}, [\href{https://arxiv.org/abs/0910.1785}{{\tt
  0910.1785}}].

\bibitem{Djouadi:2008uj}
A.~Djouadi, U.~Ellwanger and A.~M. Teixeira, \emph{{Phenomenology of the
  constrained NMSSM}},
  \href{http://dx.doi.org/10.1088/1126-6708/2009/04/031}{\emph{JHEP} {\bf 04}
  (2009) 031}, [\href{https://arxiv.org/abs/0811.2699}{{\tt 0811.2699}}].

\bibitem{Chen:2013jvg}
C.-Y. Chen, M.~Freid and M.~Sher, \emph{{Next-to-minimal two Higgs doublet
  model}}, \href{http://dx.doi.org/10.1103/PhysRevD.89.075009}{\emph{Phys. Rev.
  D} {\bf 89} (2014) 075009}, [\href{https://arxiv.org/abs/1312.3949}{{\tt
  1312.3949}}].

\bibitem{Zyla:2020zbs}
{\scshape Particle Data Group} collaboration, P.~Zyla et~al., \emph{{Review of
  Particle Physics}}, \href{http://dx.doi.org/10.1093/ptep/ptaa104}{\emph{PTEP}
  {\bf 2020} (2020) 083C01}.

\bibitem{ATLAS-CONF-2021-035}
{\scshape ATLAS Collaboration} collaboration, \emph{{Search for resonant pair
  production of Higgs bosons in the $b\bar{b}b\bar{b}$ final state using $pp$
  collisions at $\sqrt{s}$ = 13 TeV with the ATLAS detector}},  
  Technical Report ATLAS-CONF-2021-035, CERN, Geneva, Aug, 2021.

\bibitem{ATLAS-CONF-2021-030}
{\scshape ATLAS Collaboration} collaboration, \emph{{Search for resonant and
  non-resonant Higgs boson pair production in the $b\bar b\tau^+\tau^-$ decay
  channel using 13 TeV $pp$ collision data from the ATLAS detector}},  
  Technical report  ATLAS-CONF-2021-030, CERN, Geneva, Jul, 2021.

\bibitem{ATLAS-CONF-2021-016}
{\scshape ATLAS Collaboration} collaboration, \emph{{Search for Higgs boson
  pair production in the two bottom quarks plus two photons final state in $pp$
  collisions at $\sqrt{s}=13$ TeV with the ATLAS detector}},  
  Technical Report ATLAS-CONF-2021-016, CERN,  Geneva, Mar, 2021.

\bibitem{Ball:2012cx}
R.~D. Ball et~al., \emph{{Parton distributions with LHC data}},
  \href{http://dx.doi.org/10.1016/j.nuclphysb.2012.10.003}{\emph{Nucl. Phys.}
  {\bf B867} (2013) 244--289}, [\href{https://arxiv.org/abs/1207.1303}{{\tt
  1207.1303}}].


\end{thebibliography}

\providecommand{\href}[2]{#2}\begingroup\raggedright\endgroup


\end{document}